\documentclass[12pt,preprint]{aastex}
\usepackage{graphicx}
\usepackage{natbib}
\usepackage{amssymb}
\usepackage{longtable}
\def\hii{\mbox{H\,{\sc ii}}}
\def\checkmark{\tikz\fill[scale=0.4](0,.35) -- (.25,0) -- (1,.7) -- (.25,.15) -- cycle;} 
\usepackage{tikz}
\usepackage{graphicx}
\usepackage{times}
\usepackage{natbib}
\usepackage{subfigure}
\usepackage{url}
\usepackage{float,lscape}
\usepackage{amsmath}
\usepackage{multirow}
\usepackage[titletoc,page]{appendix}
\usepackage{rotating}

\shorttitle{Radio and infrared study of G351.6-1.3}
\shortauthors{Vig et al.}

\begin{document}


\title{ Star forming activity in the \hii~regions associated with IRAS~17160--3707 complex
}

\author{G. Nandakumar$^1$, V. S. Veena$^1$, S. Vig$^{1}$, A. Tej$^1$, S. K. Ghosh$^{2,3}$, D. K. Ojha$^2$}
\affil{$^1$ Indian Institute of Space Science and Technology, Thiruvananthapuram
 695 547, India \\
$^2$Tata Institute of Fundamental Research, Mumbai (Bombay) 400 005, India\\
$^3$National Centre for Radio Astrophysics, Pune, 411 007, India\\}

\begin{abstract}
We present a multiwavelength investigation of star formation activity towards the southern \hii~regions associated with IRAS~17160--3707, located at a distance of 6.2~kpc with a bolometric luminosity of $8.3\times10^5$~L$_\odot$. The ionised gas distribution and dust clumps in the parental molecular cloud are examined in detail using  measurements at infrared, submillimeter and radio wavelengths. The radio continuum images at 1280 and 610~MHz obtained using Giant Metrewave Radio Telescope reveal the presence of multiple compact sources as well as nebulous emission. At submillimeter wavelengths, we identify seven dust clumps and estimate their physical properties like temperature: $24-30$~K, mass: $300 - 4800$~M$_\odot$ and luminosity: $9 - 317\times10^2$~L$_\odot$
using modified blackbody fits to the spectral energy distributions between 70 and 870~$\mu$m. We find 24 young stellar objects in the mid-infrared, with few of them coincident with the compact radio sources. The spectral energy distributions of young stellar objects  have been fitted by the Robitaille models and the results indicate that those having radio compact sources as counterparts host massive objects in early evolutionary stages with best fit age $\le 0.2$~Myr. We compare the relative evolutionary stages of clumps using various signposts such as masers, ionised gas, presence of young stellar objects and infrared nebulosity and find six massive star forming clumps and one quiescent clump. Of the former, five are in a relatively advanced stage and one in an earlier stage.  

\end{abstract}



\section{Introduction}

\hii~regions are the first reliable pointers announcing the formation of massive O and B stars embedded in their natal molecular clouds. The hypercompact (size $<0.01$~pc) and ultracompact (size $<0.1$~pc) \hii~region phases with large electron densities ($n_e\ge 10^4$~cm$^{-3}$) are indicative of the early evolutionary stages comprising young massive stars in the accretion phase, and these young \hii~regions evolve with time into compact and classical \hii~regions \citep{1999PASP..111.1049G,{2007prpl.conf..181H}}. 

These young \hii~regions are bright in radio and infrared and possess a wide range of  morphologies  such as cometary, core-halo, shell, bipolar, and irregular \citep{{2002ARA&A..40...27C},{wood1989morphologies}}. While the morphology of a \hii~region provides clues to the density of the surrounding interstellar medium, it also depends on the scale of the region observed: a large scale morphology could lead to another at alternate angular resolution \citep{{wood1989morphologies}, {2004ApJ...605..285S}}. Ultracompact \hii~regions are usually found close to other ultracompact and compact \hii~regions within complexes  \citep{{1993ApJ...412..684M},{2001ApJ...549..979K},{sanchez2013deciphering}}. 

\par The host massive stars (mass greater than $\sim$8~M$_\odot$) remain in an embedded phase that lasts about 15\% of their lifetime before evolving into the optically visible main-sequence phase  \citep{{2002ARA&A..40...27C}, {2007ARA&A..45..481Z}}. These initial phases remain observationally elusive as massive stars are rare, evolve fast compared to their lower mass counterparts, and are at relatively large distances ($> 1$~kpc) from us, apart from being born in a clustered environment. The circumstellar dust in the vicinity of a \hii~region absorbs nearly all the stellar radiation and re-emit in the mid and far-infrared. The relatively cooler dust in the envelope absorbs infrared radiation and shows thermal emission in the submillimeter regime, that is optically thin.  The far-infrared and submillimeter maps of such regions therefore show warm and cold dust clumps as well as filamentary emission, that can be used as a potential tool for studying the physical properties of the associated molecular clouds \citep[e.g.][]{{2011A&A...533A..94H}, {2012A&A...542A..10A}, {2013A&A...554A..42R}}. By comparing the emission from radio (ionised gas) as well as  infrared and submillimeter (warm and cold dust), various stages of evolutionary sequence of massive star forming clumps in molecular clouds can be examined in detail \citep{ {2013A&A...556A..16G},{2010ApJ...721..222B}, {2013A&A...550A..21S}}. These studies highlight the importance of multiwavelength investigation of star forming regions. 

\par In this paper, we investigate a southern Galactic massive star forming region IRAS~17160--3707.  IRAS~17160-3707, also known as G350.10+0.08, harbours a number of compact and ultracompact \hii~regions  \citep{2003A&A...407..957M} with a far-infrared (IRAS)  luminosity of 3.6$\times$10$^{5}$~L$_{\odot}$ \citep{2002A&A...381..571P}. The near distance to this region ranges from 5.7 to 7.3~kpc \citep{2002A&A...381..571P, 2004A&A...426...97F, 1987A&A...171..261C}. We adopt a distance of 6.2~kpc based on sensitive radio recombination line kinematics that resolve the distance ambiguity using measurement of \mbox{H\,{\sc i}} line absorption towards the region \citep{2006ApJ...653.1226Q}. Being bright in radio, this \hii~complex has been detected in many Galactic plane surveys \citep{{1989ApJS...71..469L},{1990ApJS...74..181Z},{1994ApJS...91..347B},{1998MNRAS.301..640W}, {2006ApJS..165..338Q}}. Among these surveys, \citet{1994ApJS...91..347B} have detected a radio source of size $12''$ in their snapshot observations at 1.4 and 5 GHz. High resolution maps of this region  at 4.8 and 8.6~GHz are presented in \citet{2003A&A...407..957M} where they detected five compact \hii~regions. IRAS 17160--3707 also encompasses a number of maser spots that have been detected as a part of maser surveys, such as methanol masers \citep[e.g.][]{1998MNRAS.301..640W}, water masers \citep[e.g.][]{1989A&A...213..339F}, and hydroxyl masers \citep[e.g.][]{2001MNRAS.326..805C}.
Emission from Polycyclic Aromatic Hydrocarbons (PAH) as well as  silicate absorption were reported in this region by \cite{2002A&A...381..571P}. 

\par In the present work, we examine the star formation activity in this region using measurements at infrared, submillimeter and radio wavelengths. In particular, our interferometric observations using the Giant Metrewave Radio Telescope (GMRT) enable us to probe the ionised gas at different angular scales: compact
objects requiring high resolution as well as the broad diffuse emission.  The far-infrared and submillimeter wavebands provide a wide coverage of thermal dust emission, useful for investigating the physical properties of the associated molecular cloud. We attempt to locate the mid-infrared young stellar objects and clumps of gas and dust in this \hii~region complex and ascertain their properties. The details of observations, data reduction and archival data are given in Section 2. In Section 3, we discuss the results. The radio morphology and multiwavelength scenario are presented in Section 4, and in Section 5, we summarize the results of this work.

\section{Observations and Data Reduction}
\label{obvs.sect}
\subsection{Radio continuum observations using GMRT}

The radio continuum observations of the region around IRAS~17160--3707 were carried out 
using the GMRT \citep[][]{1991CuSc...60...95S}, India. GMRT consists of 30 antennas arranged in a Y-shaped hybrid configuration, each having a diameter of 45~m. There are three arms (east, west and south) along which six antennas each are placed. The remaining twelve antennas are located at the center forming a compact but random pattern within $1\times1$~km$^{2}$ area. The shortest baseline ($ \sim 100$~m) aids in the detection of large-scale diffuse emission while the longest baseline ($\sim 25$~km) provides high angular resolution. 
 The largest angular scale of diffuse emission observable with GMRT is  $7'$ at 1280 MHz and $17'$ at 610 MHz.
 Observations towards IRAS 17160--3707 were carried out at two frequency bands: 1280 and 610 MHz. The details of observations are listed in Table~\ref{radio_tab}. The radio quasars 3C286 and 3C48 were used as the primary flux calibrators, while the sources 1822-096 and 1626-298 were used for calibration of  phases at 610 and 1280 MHz, respectively.

\begin{table*}[hbt!]
\begin{center}
\caption{Details of the radio continuum observations.}
\label{radio_tab}
\vspace{0.2cm}
\begin{tabular}{l c c}
\hline \hline 
    Frequency (MHz) & 610 & 1280\\
 \hline 
 Observation date & 2002 May 19 & 2002 Sep 27 \\
 On source time (min) & 156 & 169 \\
 Bandwidth (MHz) & 16 & 16\\
 Primary Beam & $45'.8$ & $21'.8$\\
 Synthesized beam & $8''.1\times 5''.2$ & $4''.5\times2''.3$\\
 Position angle ($ ^\circ $) & 9.5 & -0.2\\
 Noise (mJy/beam) & 1.4 & 0.3\\
Flux Calibrator & 3C48, 3C286& 3C286\\
Phase Calibrator & 1822-096 & 1626-298\\
 \hline 
 \end{tabular}
\end{center}
 \end{table*}

\par Data reduction was carried out using the NRAO Astronomical Image Processing System (AIPS). The visibilities affected by radio frequency interference, and inoperative antennas were flagged using the tasks {\tt TVFLG} and {\tt UVFLG}. The task {\tt SPLIT} was used to create the target source file, on which the calibrations were applied. The calibrated data were cleaned and deconvolved using the task {\tt IMAGR}. The atmospheric and ionospheric phase corruptions were minimized after several iterations of self calibration. 
At the time of these observations, noise switching to measure the system temperature ($T_\mathrm{sys}$)
was not available leading to a difference in system temperature  between the
flux calibrator and the target source positions. To account for the contribution of emission from the Galactic plane ($T_\mathrm{gal}$) to the system temperature, we used a correction factor given by ($T_\mathrm{gal}$ + $T_\mathrm{sys}$)/$T_\mathrm{sys}$. This is particularly significant at the lower frequency of 610 MHz. $T_\mathrm{gal}$ is estimated using the 408 MHz map of \cite{1982A&AS...47....1H} and extrapolated to 610 MHz using a spectral index of -2.6 \citep{{2011A&A...525A.138G}, {1999A&AS..137....7R}}. The correction factor obtained this way is used to scale the flux densities in the final radio continuum map at 610 MHz. The maps are finally corrected for the primary beam using AIPS task {\tt PBCOR} to obtain the final images. 

\subsection{Archival Data}

In addition to the low frequency radio continuum observations, dust emission from the molecular cloud is investigated using 
submillimeter and far-infrared images from ATLASGAL and \textit{Herschel} Hi-GAL surveys. The \textit{Spitzer}-IRAC GLIMPSE survey images and point sources from the archive have been used for probing the warm dust as well as for identification of young stellar objects (YSOs). 

\subsection{ATLASGAL}
The Apex Telescope Large Area Survey of the GALaxy (ATLASGAL) using the Large APEX Bolometer Camera (LABOCA) is a systematic survey of the Galaxy carried out at 870 $\mu$m with a resolution of $18''.2$ and a pixel size of $6''$ \citep{2009A&A...504..415S}. The images available from the archive are tiles of size $5'\times5'$, that have been mosaiced using the Montage software \citep{2010ascl.soft10036J} to obtain the image covering our region of interest.

\subsection{\textit{Herschel}}
The images from the \textit{Herschel} Hi-GAL survey \citep{2010PASP..122..314M} have been used to study the properties of cold dust in the molecular cloud. This survey used two instruments of the 3.5-meter telescope of the \textit{Herschel Space Observatory}\footnote{Herschel is an ESA space observatory with science instruments provided by European-led Principal Investigator consortia and with important participation from NASA.} in parallel mode. The instruments used in the survey were the
Photodetector Array Camera and Spectrometer \citep[PACS; ][]{2010A&A...518L...2P} and the Spectral and Photometric Imaging Receiver \citep[SPIRE; ][]{2010A&A...518L...3G} covering the spectral range: 70 - 500 $\mu$m. We used Level-2 PACS images at 70 and 160 $\mu$m, and Level-3 SPIRE images at 250, 350 and 500 $\mu$m images from the archive for our analysis. The beams sizes are $5''.9$, $11''.6$, $18''.5$, $25''.3$ and $36''.9$ at 70, 160, 250, 350 and 500~$\mu$m, respectively. The size of a pixel in the image at the respective wavelengths are $3''.2$, $3''.2$, $6''$, $10''$ and $14''$.  The \textit{Herschel} Interactive Processing Environment (HIPE)\footnote{HIPE is a joint development by the Herschel Science Ground Segment Consortium, consisting of ESA, the NASA Herschel Science Center, and the HIFI, PACS and SPIRE consortia.} has been used to download the images as well as for processing (regridding and convolution) the images .

\subsection{\textit{Spitzer}}
The Infra-Red Array Camera (IRAC) and Multiband Imaging Photometer (MIPS) are two instruments onboard the \textit{Spitzer Space Telescope}. IRAC obtains simultaneous broadband images at 3.6, 4.5, 5.8 and 8 $\mu$m. The FWHMs of the point spread function are $1''.6$, $1''.6$, $1''.8$ and $1''.9$ at 3.6, 4.5, 5.8 and 8 $\mu$m, respectively \citep{2004ApJS..154...10F}. The Level-2 Post-Basic Calibrated Data (PBCD) images from the Galactic Legacy Infrared Mid-Plane Survey Extraordinaire (GLIMPSE) \citep{2003PASP..115..953B} were used, while the point sources for this region were extracted from the Glimpse II Spring'08 Archive. In addition, we examined the MIPS 24~$\mu$m image obtained as part of the MIPSGAL survey \citep{2009PASP..121...76C}.

\section{Results and Discussion}

\subsection{Ionized Gas Emission}

\begin{figure*}[hbt!]
\includegraphics[width=85mm]{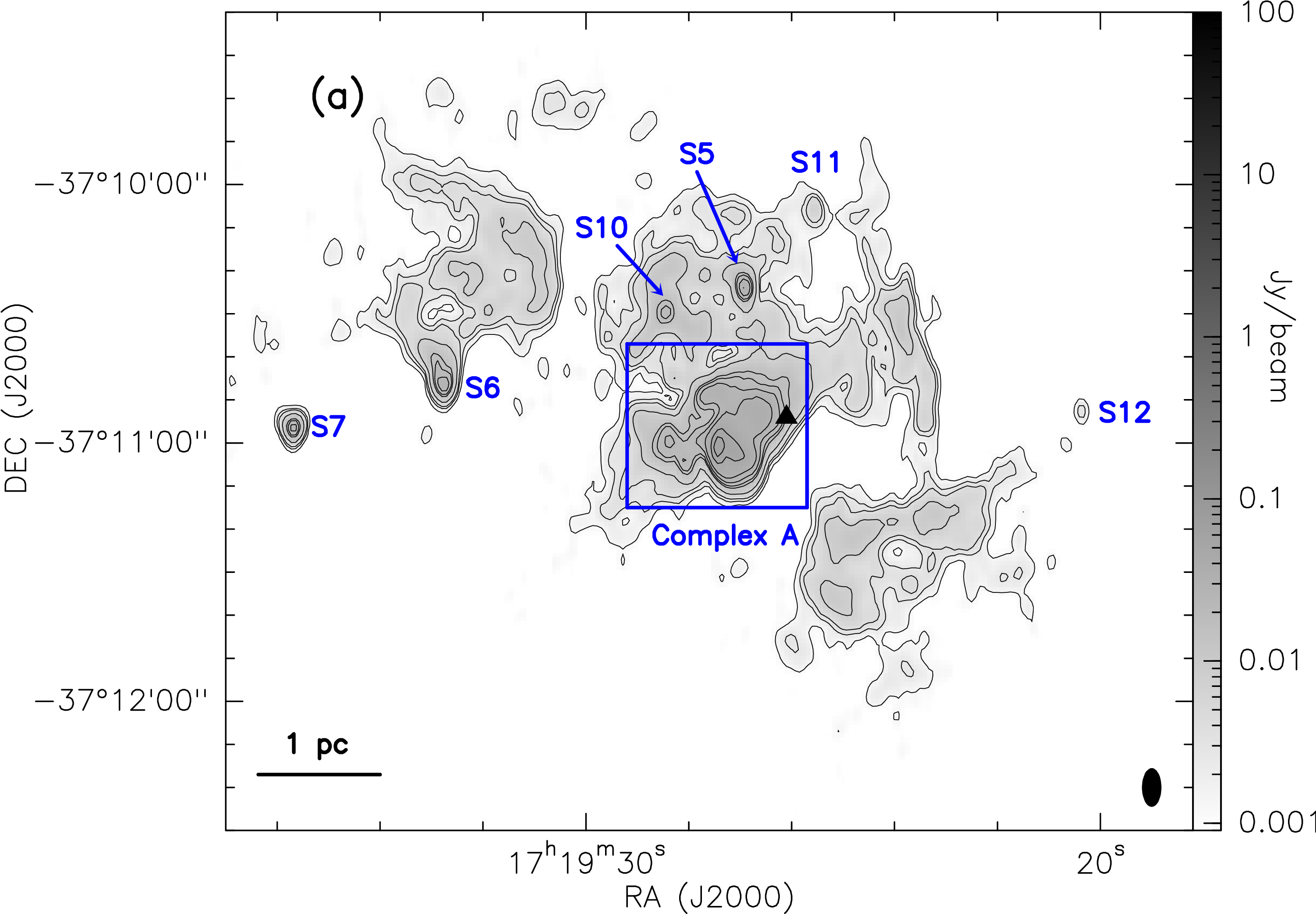}
\quad
\includegraphics[width=80mm]{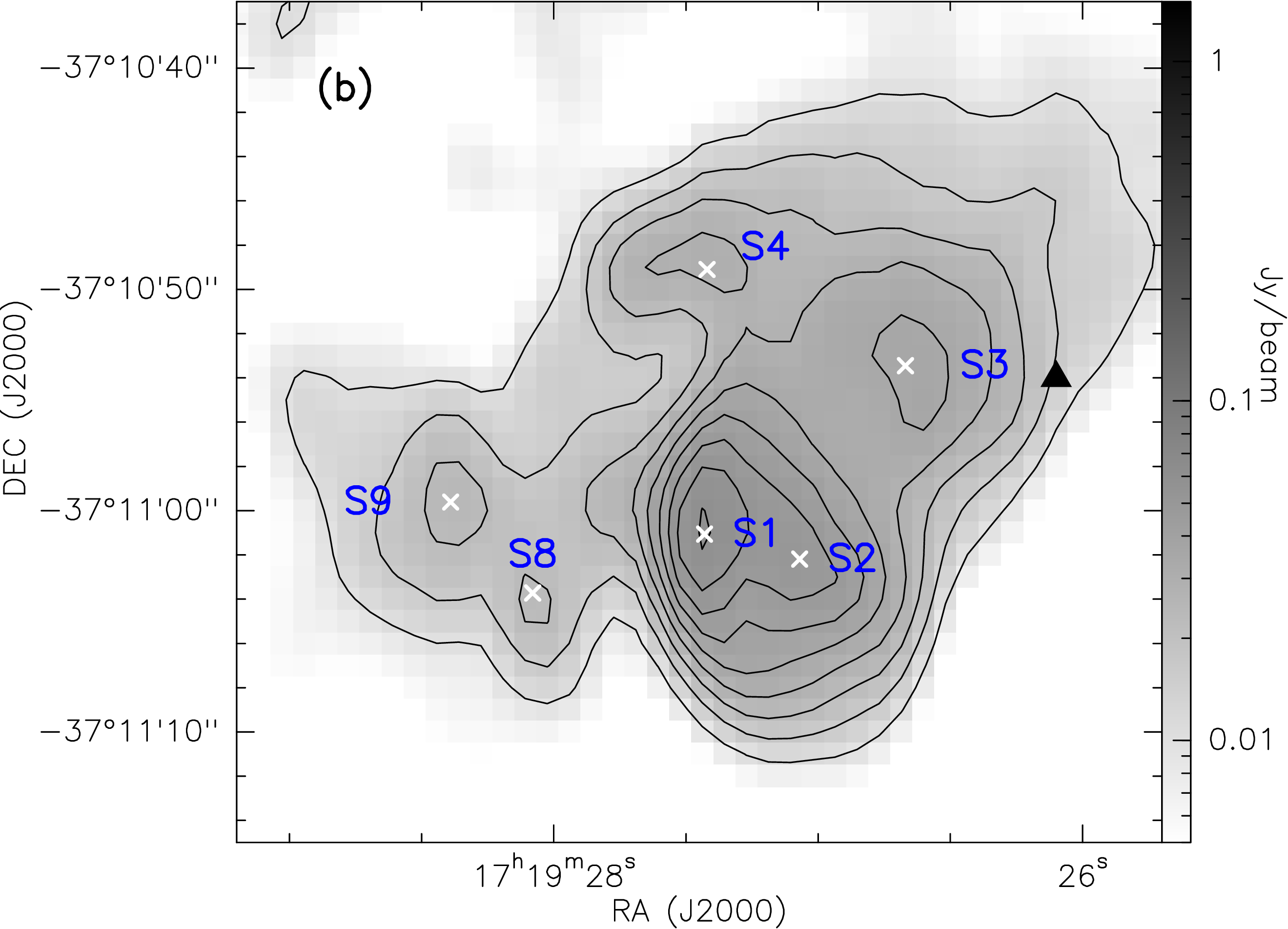}
 \caption{Radio continuum map of IRAS 17160--3707 at 1280 MHz. The entire region is shown in (a) with the rectangle enclosing complex A. The contour levels shown in (a) are at 1.3, 2.5, 3.7, 7.3, 15, 18, 23, 30, 46 and 55 mJy/beam where beam is  $4''.5\times2''.3$ (represented as a filled ellipse towards the bottom right). The region associated with complex A is enlarged in (b) with the compact sources in this region marked as crosses.  The contour levels in (b) are from 10 to 60 mJy/beam in steps of 6 mJy/beam. The solid triangle marks the position of the IRAS source. The radio compact sources S1 - S12 are marked on the images.}
 \label{gmrt_1280}
\end{figure*}

\begin{figure*}[hbt!]
\begin{center}
\includegraphics[width=85mm]{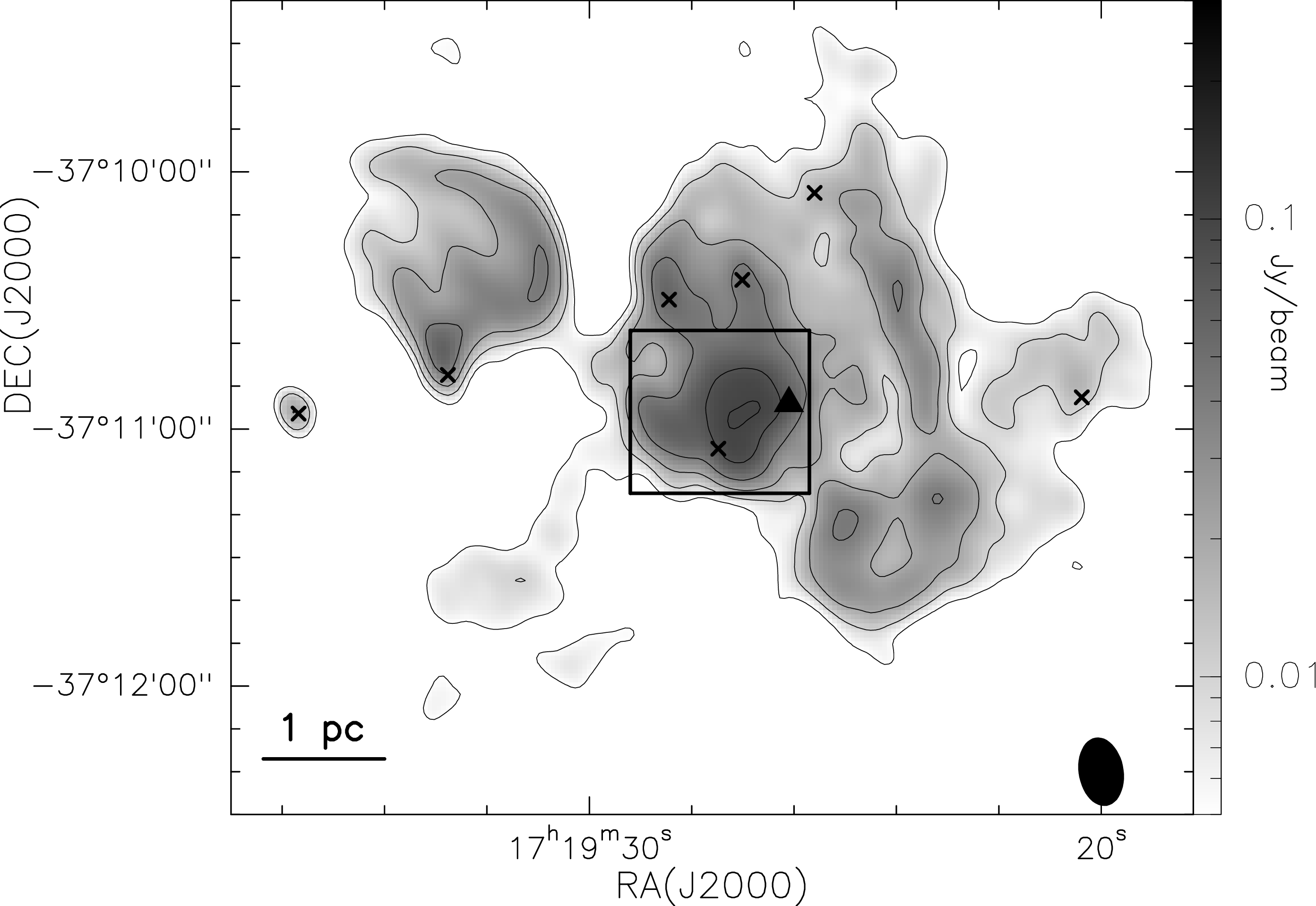}
 \caption{Radio continuum map of IRAS 17160--3707 at 610 MHz. The contour levels are at 4, 10, 20, 40, 50, 80, 90 and 100 mJy/beam, where beam is  $8''.1\times5''.2$ (represented as a filled ellipse towards the bottom right). The box encloses complex A shown in Fig.~\ref{gmrt_1280}~(b) and the crosses represent compact sources. The solid triangle marks the position of the IRAS source. }
 \label{gmrt_610}
 \end{center}
\end{figure*}

The radio continuum emission at 1280 and 610 MHz towards IRAS~17160--3707 is shown in Figs.~\ref{gmrt_1280} and \ref{gmrt_610}. The ionised gas emission  in this region shows a complex morphology with a number of compact sources apart from associated diffuse emission. The resolution of the maps are $4''.5\times2''.3$ and $8''.1\times 5''.2$ at 1280 and 610 MHz, respectively, while the corresponding rms noise levels are 300~$\mu$Jy/beam and 1.4 mJy/beam, respectively. 

\begin{table*}[hbt!]
\scriptsize
\caption{Extracted radio sources and their flux densities at 1280 and 610 MHz.}
\label{radiosource}
\vspace{0.2cm}
\begin{tabular}{l c c c c c c c c c}

\hline \hline
 & & &\multicolumn{3}{c}{1280 MHz} & & \multicolumn{3}{c}{610 MHz} \\
 \cline{4-6}  \cline{8-10}\\
 S.No.  & RA (J2000) & Decl (J2000) & Peak Intensity & Flux Density (\%)$^a$ & Size  & & Peak Intensity & Flux Density (\%)$^a$& Size\\
 & hr:min:sec & $^\circ$:$''$:$''$ & (mJy/beam) & (mJy)& $\theta_{src}$ ($''$) & & (mJy/beam)& (mJy) & $\theta_{src}$ ($''$)\\
 \hline 
 \\
S1$^*$ & 17:19:27.41 & -37:11:00.6 & $59.8$ &$277.8\pm27.8$ & $6.0 $& &... & 1430.0$^b$ & ...\\
S2$^*$ & 17:19:27.07 & -37:11:02.2  &$50.4$ & $271.8\pm27.2$ & $ 6.3$& &... & 1430.0$^b$ & ...\\
S3$^*$ & 17:19:26.60 & -37:10:55.0 & $36.1$ &$302.4\pm30.3$ &$7.2$ & &... & 1430.0$^b$ &...\\
S4$^*$ & 17:19:27.41 & -37:10:48.6 & $30.7$ &$180.1\pm18.0$ &$6.4$ & &... & 1430.0$^b$ &...\\
S5 & 17:19:26.87 & -37:10:23.8 & $30.5$ &$70.9\pm7.1$ (25) & $4.2$ & &... & ... & ...\\
S6 & 17:19:32.76 & -37:10:47.0 & $28.3$ & $77.0\pm7.7$ (40) & $ 4.2$ & & ... & ... & ...\\
S7 & 17:19:35.71 & -37:10:56.6 & $26.6$  & $35.5\pm3.6$ (20) & $ 3.5$ & &$20.6$ & $24.2\pm3.9$ (10) & $6.6$\\ 
S8$^*$ & 17:19:28.08 & -37:11:04.6 & $24.3$ &$39.8\pm4.0$ & $3.2$ & &... & 1430.0$^b$ &...\\
S9$^*$ & 17:19:28.41 & -37:10:59.8 & $24.3$ &$64.9\pm6.5$ & $4.3$ & &... & 1430.0$^b$ &...\\
S10 & 17:19:28.41 & -37:10:29.4 & $19.9$ & $41.7\pm4.2$ (62) & $4.0$ & & ... & ...  &...\\
S11 & 17:19:25.53 & -37:10:05.4 & $7.2$ & $22.6\pm2.3$ (30) & $3.4$ & & ... & ... &...\\
S12 & 17:19:20.38 & -37:10:52.6 & $3.6$ & $4.0\pm0.7$ (20) & $3.3$ & & ... & ... & ...\\
\hline
\end{tabular}
$\,\,\, $ \\
$^a$ Integration of flux density upto \% contour level of peak intensity given in brackets. \\
$^b$ Value of integral flux density of complex A as sources are unresolved. \\
$^*$ Flux estimation using 2D-Clumpfind. \\

\end{table*}


\par Twelve compact radio sources have been identified from the 1280 MHz map. These are listed in Table~\ref{radiosource}. While S7 and S12 are isolated from the main complex, sources such as S5, S6, S10 and S11 appear as compact sources encompassed in diffuse emission. The rest of the sources: S1, S2, S3, S4, S8 and S9 lie close to each other and we call this ``complex A'' after \cite{2003A&A...407..957M}. The identification of compact sources in the presence of  diffuse emission
is non-trivial. We identify compact sources as those having peak emission (local maxima) with  large signal-to-noise ratio, S/N$>50$. This explains the classification of sources S1 - S10 as compact.  In addition to the 10 compact sources, we used the high frequency information (8.6 and 4.8 GHz maps) of this region from \cite{2003A&A...407..957M} and identified the source S11. This is an ultracompact \hii~region at high frequencies but shows weak emission in our maps and hence does not satisfy the S/N criterion described earlier. Further, a seemingly isolated compact source (S12) that showed up as bright source in the mid-infrared was also included. 

To obtain the integrated flux densities of compact sources, we use the following methods. For sources other than those belonging to complex A, we 
obtain the flux densities by using an approximate method wherein the flux densities upto a certain percentage of peak intensity are integrated. The contour level upto which the flux densities are integrated are roughly upto the contour level where the contribution from diffuse emission starts becoming large ($>10''$).  The flux densities and sizes obtained this way can be considered as illustrative at best as it is difficult to disentangle the effects of diffuse emission completely. For the complex A, as the sources are located very close to each other, it is difficult to define contours isolating these sources. Hence we use the flux allocation method adopted in $\it{2{\rm D}-Clumpfind}$, where the flux at each pixel is allocated according to the friends-of-friends algorithm \citep{1994ApJ...428..693W}. The algorithm works by contouring the data at multiples of  the rms noise ($\sigma$) in the map from a given threshold and searches for peaks of emission that are identified as clumps. The advantage of $\it{2{\rm D}-Clumpfind}$ is that it does not \textit{a priori} assume any clump profile, and the total flux is conserved. We considered a threshold of 50$\sigma$ and contour spacing of 3$\sigma$ to identify the compact sources in complex A. We do not prefer to use elliptical gaussian fits as the flux density distribution of most sources, particularly those within radio nebulosity, are not gaussian. Hence, the integrated flux densities determined from the fits are over-estimates compared to the actual integrated fluxes.

\par The peak and integrated flux densities as well as the size estimates of all the sources are listed in Table~\ref{radiosource}. 
The source size ($\theta_{src}$) is taken as the geometric mean of the largest and smallest values of FWHM of the contour level upto which the flux is integrated.  This is used in the estimation of errors in the flux densities as well as in the determination of other radio properties.
The errors in the flux densities listed in Table~\ref{radiosource} are estimated using the expression $\sqrt{(2\sigma\sqrt{\theta_{src}/\theta_{bm}})^2+(2\sigma')^2}$ where $\sigma$ is the rms noise level of the map,  $\sigma'$ is the the error in flux scale calibration, while $\theta_{bm}$ represents the size of the beam \citep{2013ApJ...766..114S}. The uncertainty in the flux calibration is taken to be 5\%  \citep{2007MNRAS.374.1085L}.

The sources in complex A are unresolved at 610 MHz as the resolution of the image is lower. We are able to estimate the flux density solely for S7 based on the contour integration method (described earlier) as it is relatively isolated.  For the other sources, it is difficult to estimate the contribution of compact sources, due to (i) diffuse nebulosity, and (ii) flux density of the thermal compact source is expected to be weaker at such low frequencies due to optical depth effects. 


\begin{table*}[hbt!]
\small

\caption{Radio properties of compact radio sources.} 
\label{radiocomp}
\vspace{0.2cm}
\begin{tabular}{l c c c c l c }
\hline \hline
\multicolumn{4}{c}{Radio (GMRT)} & & \multicolumn{2}{c}{Radio (ATCA)$^a$} \\
 \cline{1-4}  \cline{6-7}
 S.No.  & EM & $n_e$ & $\log( \rm{\dot{N}_{Lyc}})$ & & Name & $\log( \rm{\dot{N}_{Lyc}})$\\ 
            & ($10^6$~pc~cm$^{-6}$) & ($10^3$~cm$^{-3}$) &(${\rm ph\,\,s^{-1}}$) & & & (${\rm ph\,\,s^{-1}}$)\\ \hline
S1 & 7.4 & 6.9 & 47.99 & & A$^b$& 48.63\\
S2 &6.7 & 6.4 & 47.99  & & A$^b$&48.63\\
S3 &5.5 & 5.4 &  48.03 &  & A$^b$&48.63\\
S4 & 4.6 & 5.4 & 47.81 & & A$^b$&48.63\\
S5 & 2.9 & 4.8 & 47.40 &  & B&47.09\\
S6 & 3.5 & 5.4 & 47.44 & & C &47.00\\
S7 & 2.5 & 5.1 & 47.1 & & D &47.61\\ 
S8 & 3.9 & 7.0 & 47.15 & & A$^b$&48.63\\
S9 & 4.0 & 6.3 & 47.36 & & A$^b$&48.63\\
S10 & 3.1 & 5.8 & 47.17 & & - & -\\
S11 & 1.3 & 3.6 & 46.91& & E&46.85\\
S12 & 3.2 & 1.9 & 46.15 & & - & -\\
 \hline
\end{tabular}
$\,\,\,$ \\
$^a$ \citet{2003A&A...407..957M}. \\
$^b$ Part of Complex A. \\
\end{table*}


The diffuse emission in our maps shows a large arc-shaped feature to the north of S6, extending upto $\sim1'$, visible at both frequencies. The emission shows a sharp gradient towards the west suggesting that it is ionisation bounded, while towards the eastern side, it is density bounded as the emission gradient is shallower. Keeping this morphology in view, we searched in literature to establish if this region harbours a bubble, and we find that \citet{2007ApJ...670..428C} have listed this as the infrared bubble CS-112. In addition to CS-112, diffuse emission is seen towards the west and south-west of complex A. Besides, low level nebulous emission is also seen to the south-east of complex A.   The extent of ionised gas in the \hii~region complex is $5.5\times 3.0$ pc$^2$. The total integrated flux densities down to the 5$\sigma$ level at 1280 and 610 MHz are $4.0\pm0.5$~Jy and $5.3 \pm 0.5$~Jy respectively. 

\par Using flux densities at 1280 MHz, we next estimate the radio properties of these compact sources such as the emission measure (EM), electron density ($n_e$) and the Lyman continuum photon rate (${\rm \dot{N}_{Lyc}}$) required to keep the compact sources ionised, under the assumptions that the emission is optically thin and there is negligible absorption by dust \citep{2016A&A...588A.143S}.  

\begin{equation}
 \left( \rm{\frac{EM}{pc\,\,cm^{-6}}}\right) = 3.217\times10^7 \left( \frac{S_\nu}{\rm Jy}\right) \left( \frac{T_e}{\rm K}\right)^{0.175} \left( \frac{\nu}{\rm GHz}\right)^{0.1} \left( \frac{\theta_{src}}{\rm arcsec}\right)^2
\end{equation}

\begin{equation}
\left( {\frac{n_e}{\rm cm^{-3}}}\right) = 2.576\times10^6 \left( \frac{S_\nu}{\rm Jy}\right) ^{0.5}\left( \frac{T_e}{\rm K}\right)^{0.35} \left( \frac{\nu}{\rm GHz}\right)^{0.05} \left( \frac{\theta_{src}}{\rm arcsec}\right)^{1.5} \left( \frac{d}{\rm pc}\right)^{-0.5}
\end{equation}

\begin{equation}
\left( {\frac{\rm \dot{N}_{Lyc}}{\rm s^{-1}}}\right) = 4.1\times10^{-10}\frac{4\pi}{3} \left( \frac{r_o}{\rm cm}\right) \left( \frac{n_e}{\rm cm^{-3}}\right) \left( \frac{T_e}{\rm K}\right)^{0.8}
\end{equation}


Here, $S_{\nu}$ is the flux density at frequency $\nu$, $T_e$ is the electron temperature, $\theta_{src}$ is the angular source size, $r_o$ is the effective radius of the region in physical units, and $d$ is the distance to the source. We use an electron temperature of $\sim$ 6710 K obtained using high-precision radio recombination line and continuum observations by \citet{2006ApJ...653.1226Q}. The radio properties for each source are listed in Table~\ref{radiocomp}. The Lyman continuum values obtained are compared with those obtained at 8.6~GHz by \cite{2003A&A...407..957M}.  Whether each radio source harbours  zero-age main sequence (ZAMS) star(s) or is an externally ionised high density clump is discussed later (Section 4).

A comparison of these low frequency maps with higher frequency images of \citet{2003A&A...407..957M} shows that many of our sources have been detected by these authors.  The resolution of the ATCA maps are better than GMRT ($1''-3''$) and the array configuration is such that compact sources are clearly detected and nebulous emission seen solely near the complex A.

A comparison of the source nomenclature is tabulated in Table~\ref{radiocomp}. The table also lists the six sources comprising complex A as seen in our 1280 MHz map. In addition, there are no ATCA counterparts to the S10 and S12. In case of S12, we find that this source lies outside the field-of-view of the ATCA images. A scrutiny for S10 among the higher frequency images shows diffuse emission at 5 GHz VLA image (described in the next section), and isolated emission at low flux levels (5$\sigma$)  at 8.6 GHz image. It is likely that S10 is an extragalactic source that has stronger emission at lower frequencies due to non-thermal emission.

\subsubsection{Spectral Index Map}

\par In order to investigate the nature of the ionised emission towards this region, we constructed a spectral index map using data at 1280 and 610 MHz. The spectral index $\alpha$ is defined as  $S_\nu\propto$ $\nu^{\alpha}$, where $S_{\nu}$ is the flux density at frequency $\nu$. As we are interested in a spectral index map that includes regions with large scale diffuse emission, we used 1280 and 610 MHz images constructed from visibilities within the same UV range: $ 0.2-50.0~{\rm k}\lambda$. The new 1280 MHz image was convolved and regridded to that of the new 610 MHz map using the AIPS tasks {\tt CONVL} and {\tt LGEOM}. Only pixels with flux densities larger than 5$\sigma$ in both bands were retained for generating the spectral index map. The AIPS task {\tt COMB} was used for this. An error map was also obtained to compare the pixel uncertainties in the spectral index image. We find that the errors 
 in spectral indices are $<0.5$. The envelope exhibits relatively large spectral index errors: $0.3 - 0.4$ as compared to the interior.
However, we would like to mention that these errors do not include uncertainties due to calibration processes.

\par
\begin{figure*}[hbt!]
\includegraphics[width=83mm]{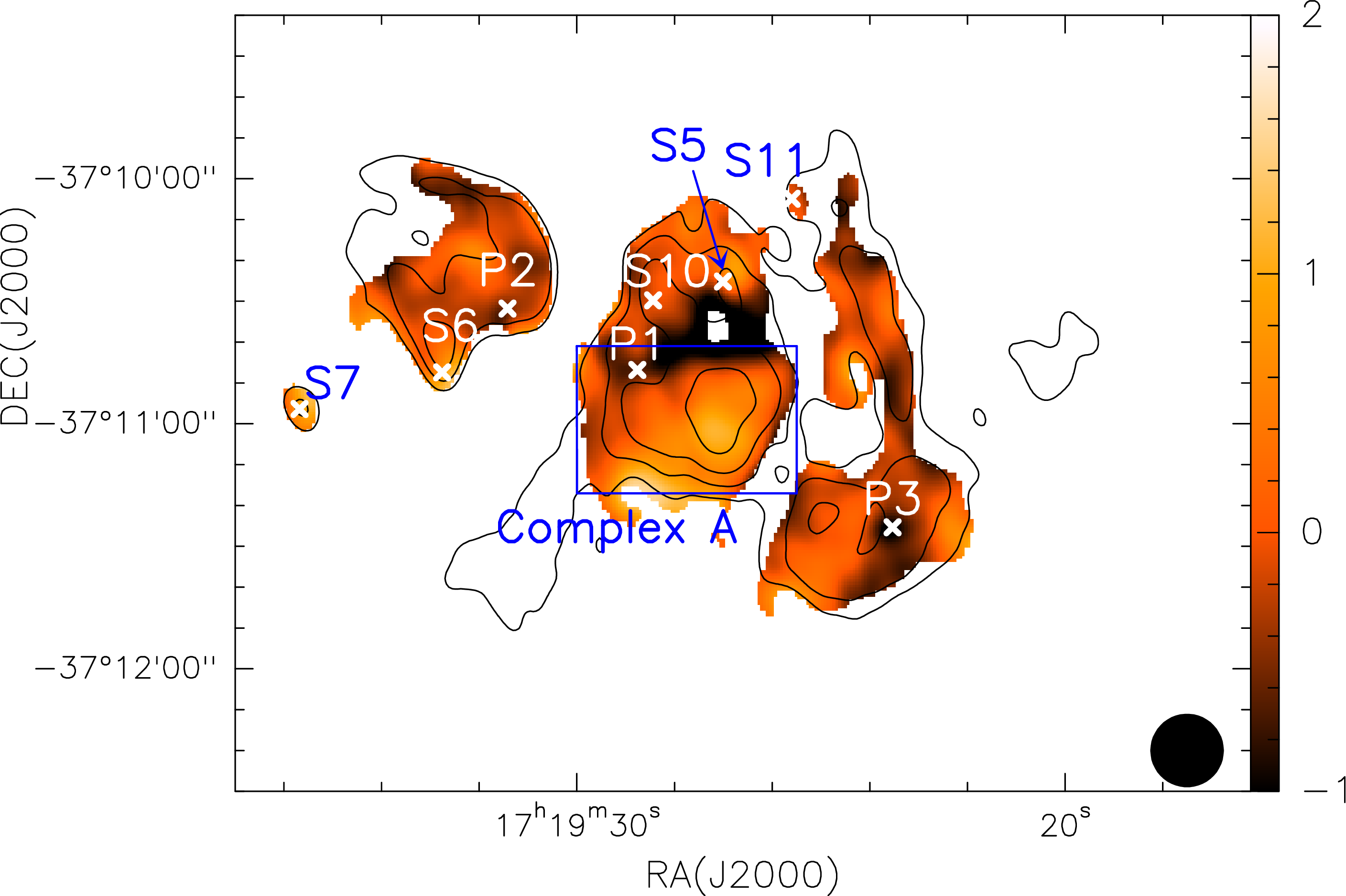}
\quad
\includegraphics[width=82mm]{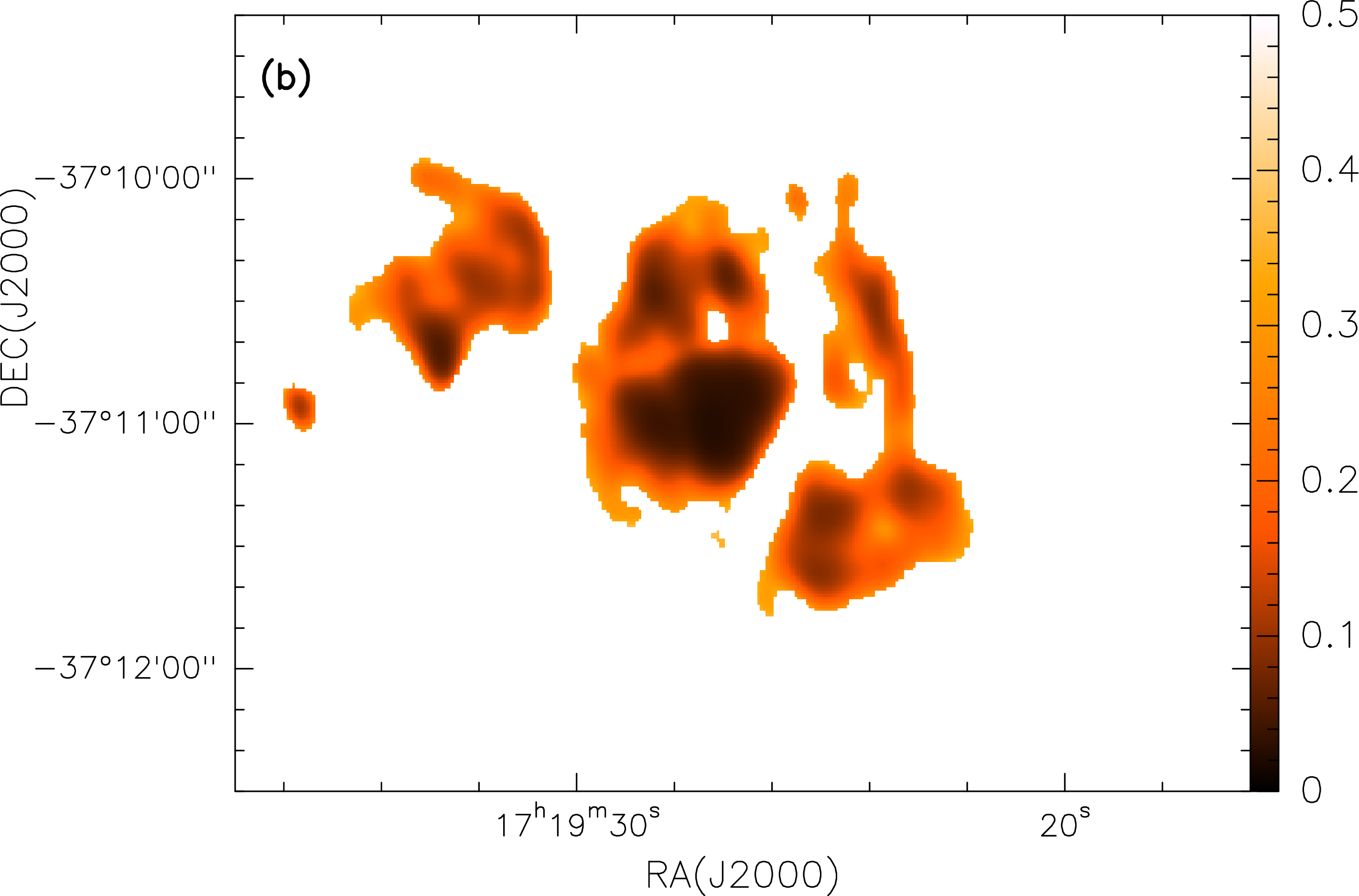}
 \caption{The 1280 - 610 spectral index map overlaid with 610~MHz radio contours is shown in (a) and the corresponding error map in (b). The eight representative positions at which spectral indices are estimated are marked over the images. The beam is shown as a filled ellipse in panel (a).}
 \label{specin}
\end{figure*}

 The 1280 - 610 spectral index map along with the  corresponding error map is shown in Fig.~\ref{specin}. This map shows the variation of the spectral index towards each of the radio sources we have identified, enabling us to infer the mechanism behind the radio emission.   The range of spectral indices in the map lies between -1 and 1.5, indicating the presence of both thermal free-free emission as well as non-thermal synchrotron emission. Thermal emission including those from inhomogeneous \hii~regions with density gradients give rise to spectral index values $-0.1\le \alpha \le 2$ \citep{1975A&A....39..217O}, while spectral indices $\alpha< -0.5$ are believed to be mostly due to non-thermal mechanisms \citep{1999ApJ...527..154K}. 
Towards the radio compact sources: S5, S6, S7 and complex A, the spectral index is positive, lying between 0.7 - 1 suggesting thermal emission. Away from the peak, the spectral index decreases as we move towards the envelopes. On the other hand, the spectral indices towards S10 and S11 are flat signifying optically thin emission. The spectral index decreases to a value of $\sim$ -1 in the envelopes in some pockets, indicating that non-thermal synchrotron emission dominates here. However, we treat this with caution as the errors are also larger. In addition, we see negative values of spectral index towards the rim of CS-112, implying the influence of non-thermal emission. This is discussed further in Section 4.1.3. 


\par
\begin{figure*}[hbt!]
\begin{center}
\hspace*{-1.2cm}
\includegraphics[scale=0.55]{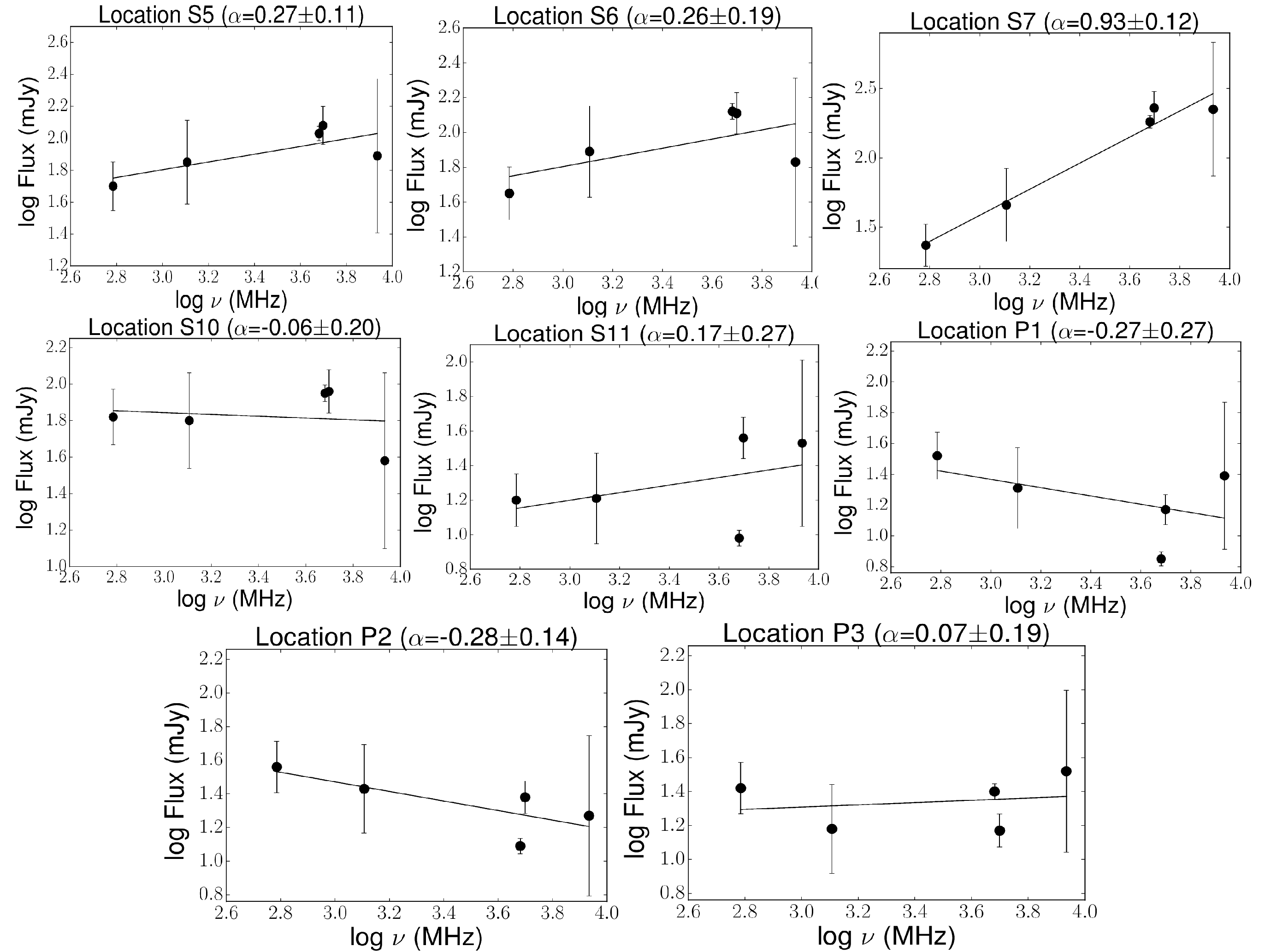} 
\caption{Spectral index plots of the eight representative locations shown in Fig.~\ref{specin}.}
\label{specinpl} 
\end{center}
\end{figure*}

 We have also compared the spectral indices using higher frequency maps: 4.8 and 8.6~GHz ATCA maps  from \citet{2003A&A...407..957M} as well as the 5 GHz VLA map. The 5~GHz VLA image is pipeline-calibrated and processed using AIPS for observations carried out in the BC configuration on 23 June 1989 under the proposal ID AB544. The image has been extracted from the NRAO VLA Archive Survey\footnote{http://archive.nrao.edu/nvas/} (NVAS). The resolution of this image is 9$\arcsec$.3$\times$5$\arcsec$.5 that is the lowest among the five radio images. The resolution of the 4.8 GHz ATCA map is 3$\arcsec$.1$\times$1$\arcsec$.7 while it is 1$\arcsec$.4$\times$0$\arcsec$.8 at 8.6~GHz. All the images are convolved to the lowest resolution to compare the spectral indices. The range of visibilities in the higher frequency images are different from the UV range of the GMRT images and hence the comparison is just illustrative. The spectral indices using five frequencies (610 MHz, 1280 MHz, 4.8, 5 and 8.6~GHz) towards 8 representative positions: five compact sources and three locations in the envelope (P1, P2, P3) are compared with those obtained from the two-frequency GMRT spectral index map. These locations are marked in Fig.~\ref{specin} and the plots of spectral index are shown in  Fig.~\ref{specinpl}. A comparison of these spectral indices is listed in Table~\ref{spectb}. Column 1 lists the locations, Column 2 lists the GMRT 1280 - 610 spectral index values from the map, and Column 3 lists the spectral index estimated using five frequencies: 8.6~GHz, 5~GHz, 4.8~GHz, 1280 and 610 MHz. The spectral indices (from five frequencies) consistently indicate thermal emission towards compact sources while the spectral indices are negative or flat near the locations in the envelope. In Fig.~\ref{specinpl}, we observe that the flux density towards S11 at 4.8~GHz (ATCA) is considerably lower than the flux densities at other wavebands. This is likely to be due to the sensitivity of ATCA and VLA observations to different spatial scales.

It is to be noted that the emission at each frequency could include contributions from both thermal and non-thermal processes and this contribution varies as a function of frequency. Towards P1, P2 and P3, the lower frequency spectral indices (1280 - 610) are more non-thermal in nature (i.e. steeper negative index) when compared to all the five frequencies. The likely reason could be that thermal contribution dominates at higher frequencies. 


\tabcolsep=0.10cm
\begin{table*}[hbt!]
\footnotesize
\caption{Spectral indices of the eight locations shown in Fig.~\ref{specin}. Note that the errors are likely to be larger than the quoted values as they do not include errors due to calibration processes.}
\begin{center}
\hspace*{-0.5cm}
\begin{tabular}{c r r } \hline \hline
Location &$\alpha_{1280-610}$ & $\alpha_{\text{\tiny{GMRT+VLA+ATCA}}}$   \\
\hline\\
S5 &0.60$\pm$0.06 &0.27$\pm$0.11  \\
S6  &0.78$\pm$0.07 &0.26$\pm$0.19 \\
S7&0.86$\pm$0.11& 0.93$\pm$0.12  \\
S10 &-0.08$\pm$0.06 & -0.06$\pm$0.20 \\
S11&-0.05$\pm$0.23 & 0.17$\pm$0.27 \\
P1 &-0.67$\pm$0.17 & -0.27$\pm$0.27\\
P2&-0.52$\pm$0.13 & -0.28$\pm$0.14 \\
P3 &-1.07$\pm$0.26 & 0.07$\pm$0.19 \\
\hline
\end{tabular}
\label{spectb}
\end{center}
\end{table*}

\par Overall, we see that the compact sources are thermal in nature while the co-existence of a non-thermal component is seen towards few pockets in the outer envelope as well as towards the rim of CS-112. Similar variations in the spectral index map (interior versus envelope) are witnessed in other star forming regions \citep{{1993ApJ...415..191C},{1996ApJ...459..193G}} where non-thermal emission is detected towards the outer envelope. \cite{1996ApJ...459..193G} propose that the emission from the diffuse envelope corresponds to synchrotron emission from electrons accelerated in the region of interaction between a central star wind and the ambient cloud material. According to \cite{1993ApJ...415..191C}, this can be expected in a shock wave where most of the electrons having a thermal distribution of velocity will produce the thermal free-free component while a fraction of electrons will be accelerated to relativistic velocities producing the non-thermal component.

\subsection{Dust Emission}

\par
\begin{figure}[hbt!]
\includegraphics[width=150mm]{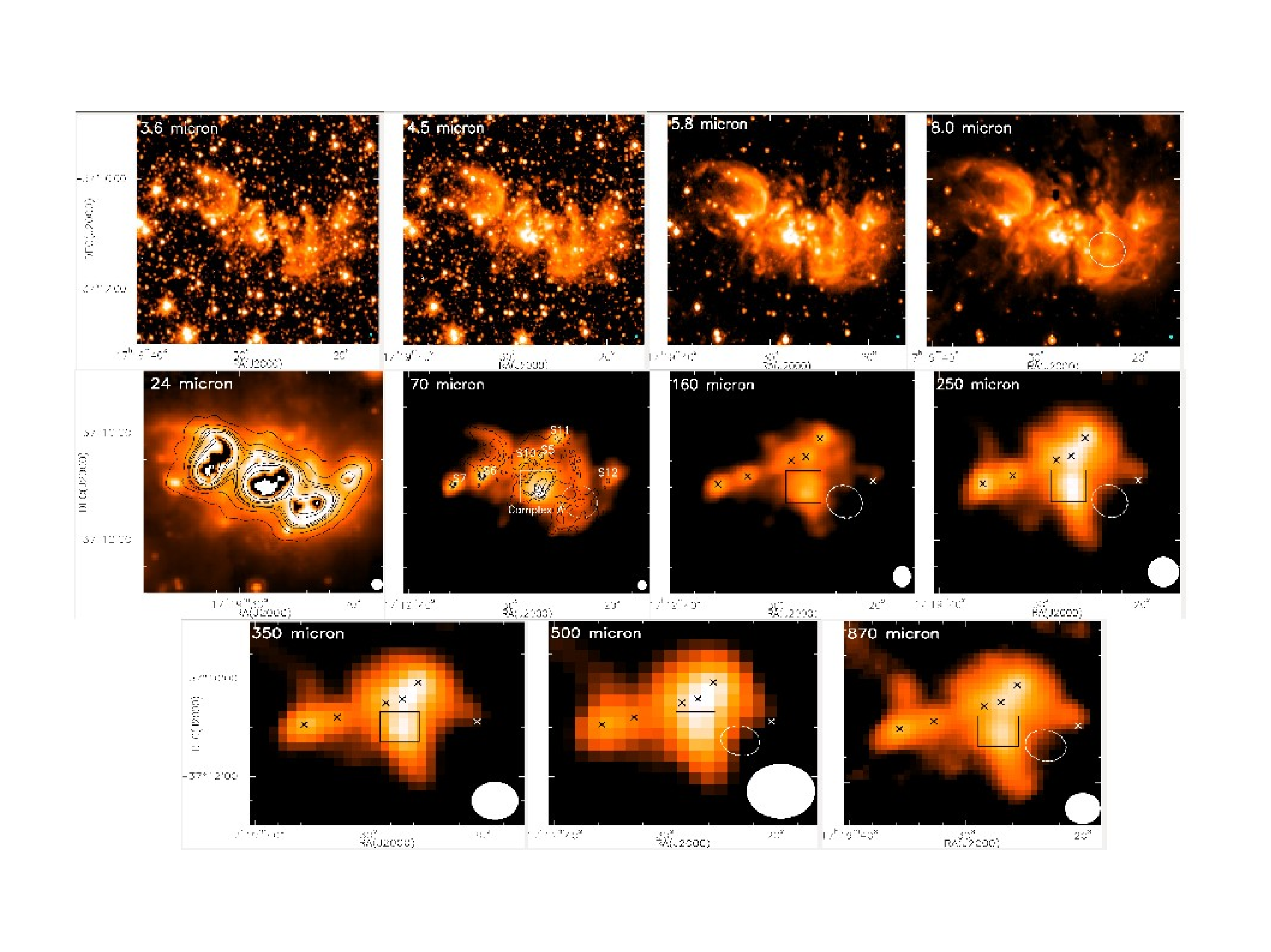}
\caption{Dust emission towards IRAS 17160--3707 observed in the four IRAC bands (3.6, 4.5, 5.8 and 8.0~$\mu$m), MIPSGAL 24~$\mu$m, five HiGAL bands (70, 160, 250, 350 and 500~$\mu$m) and the ATLASGAL 870 $\mu$m. The radio compact sources (crosses) and complex A (enclosed in a box) are marked over the HiGAL and ATLASGAL images. The MSX emission at 21 $\mu$m is shown as black contours overlaid on the MIPSGAL image. The region exhibiting temperature enhancement in the dust temperature map is shown within an ellipse in the 70 and 500 $\mu$m images. The size of the beam is shown in the right hand corner of each image (cyan in the IRAC images to distinguish from point sources).}
\label{ir}
\end{figure}

 The mid-infrared emission in this region shows a nebulosity in the IRAC images displayed in Fig.~\ref{ir}.  The mid-infrared emission morphology matches that of radio emission at 1280 MHz map and we see emission corresponding to the compact sources detected in radio. We also recognise the bubble-like feature corresponding to CS-112. The mid-infrared emission could have contributions from (i) thermal emission from warm dust in the circumstellar envelope
 heated by direct stellar radiation, (ii) heating of dust due to Lyman-$\alpha$ photons resonantly
scattering in the ionised region \citep{1991MNRAS.251..584H}, and  (iii) emission due to excitation of polycyclic aromatic hydrocarbons (PAHs) by UV photons in the Photo Dissociation Regions \citep[PDRs,][]{2011A&A...535A.128B}. Towards complex A in particular, it has been seen that the mid-infrared emission is dominated by PAHs \citep{2004ApJ...613..986P, 2003A&A...407..957M}. At the longer wavelength of 24~$\mu$m, the central bright regions are saturated in the MIPSGAL image. However, a number of point sources can be seen in the image. A comparison with the Midcourse Space eXperiment \citep[MSX,][]{2001AJ....121.2819P} map at 21~$\mu$m  shows regions of bright emission that are saturated in the MIPS image (Fig.~\ref{ir}).


Emission at six wavelength bands have been used to study the physical properties of the cold dust emission. These include five bands from $Herschel$ Hi-GAL (70, 160, 250, 350 and 500~$\mu$m), and the  870~$\mu$m emission from the ATLASGAL survey. These are shown in Fig.~\ref{ir}. The 70 and 160~$\mu$m images show bright clumpy distribution of dust emission. The infrared sources associated with radio compact sources are marked in the figure. The far-infrared images show emission corresponding to S5, S6, S7, S10, S11 and S12. Complex A emerges as a region with bright emission in most of the mid and far-infrared bands. The bubble CS-112 is detected clearly at 70~$\mu$m emission but not at the longer wavebands indicating the lack of cold dust emission.  The longer wavelength emission maps trace high density cold clumps embedded in large scale diffuse emission.  

\subsubsection{Cold Dust Clumps}

\begin{figure}[hbt!]
\hspace*{-0cm}
\includegraphics[height=70mm]{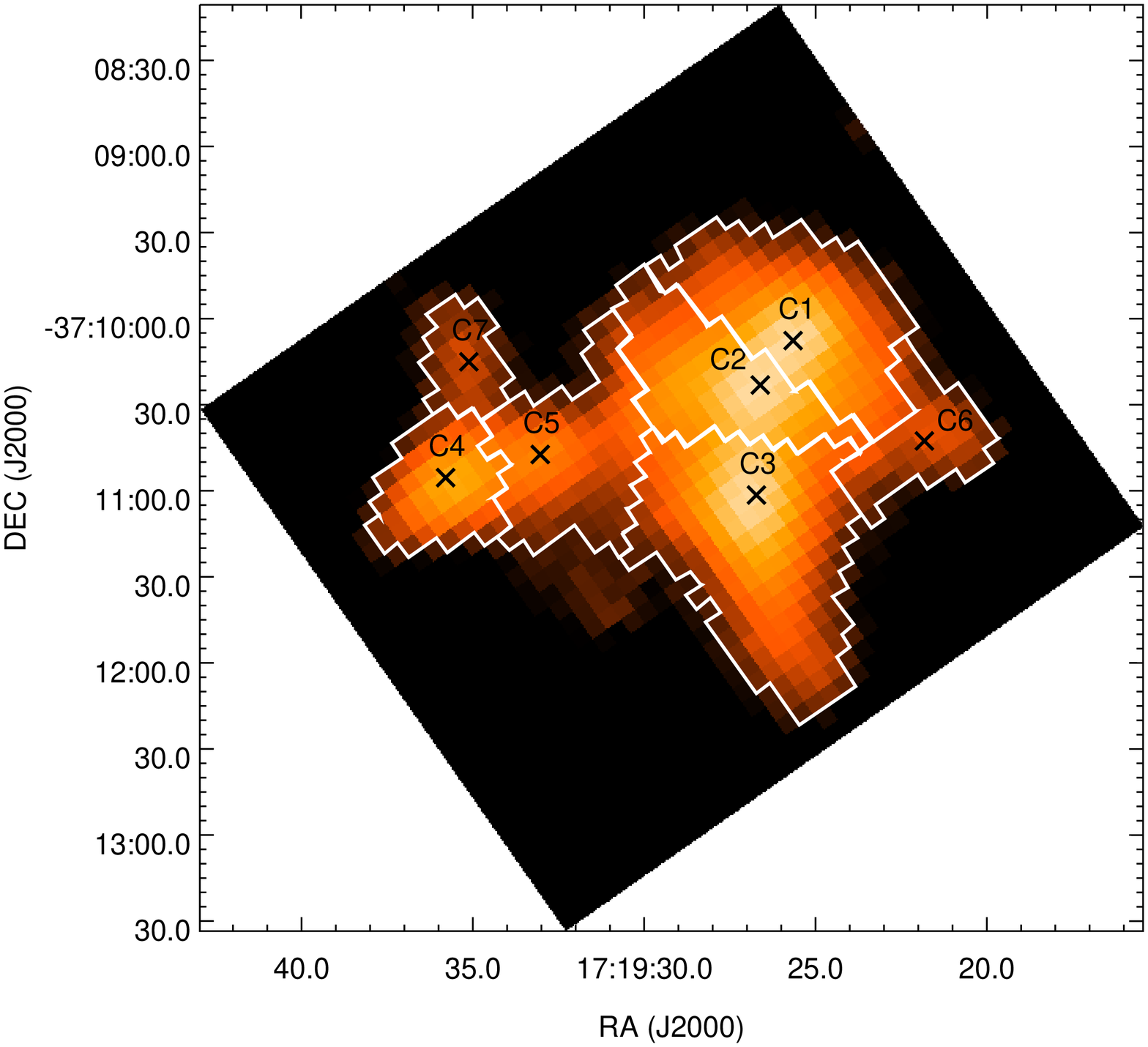}
\hspace*{-1.5cm}
\includegraphics[height=60mm]{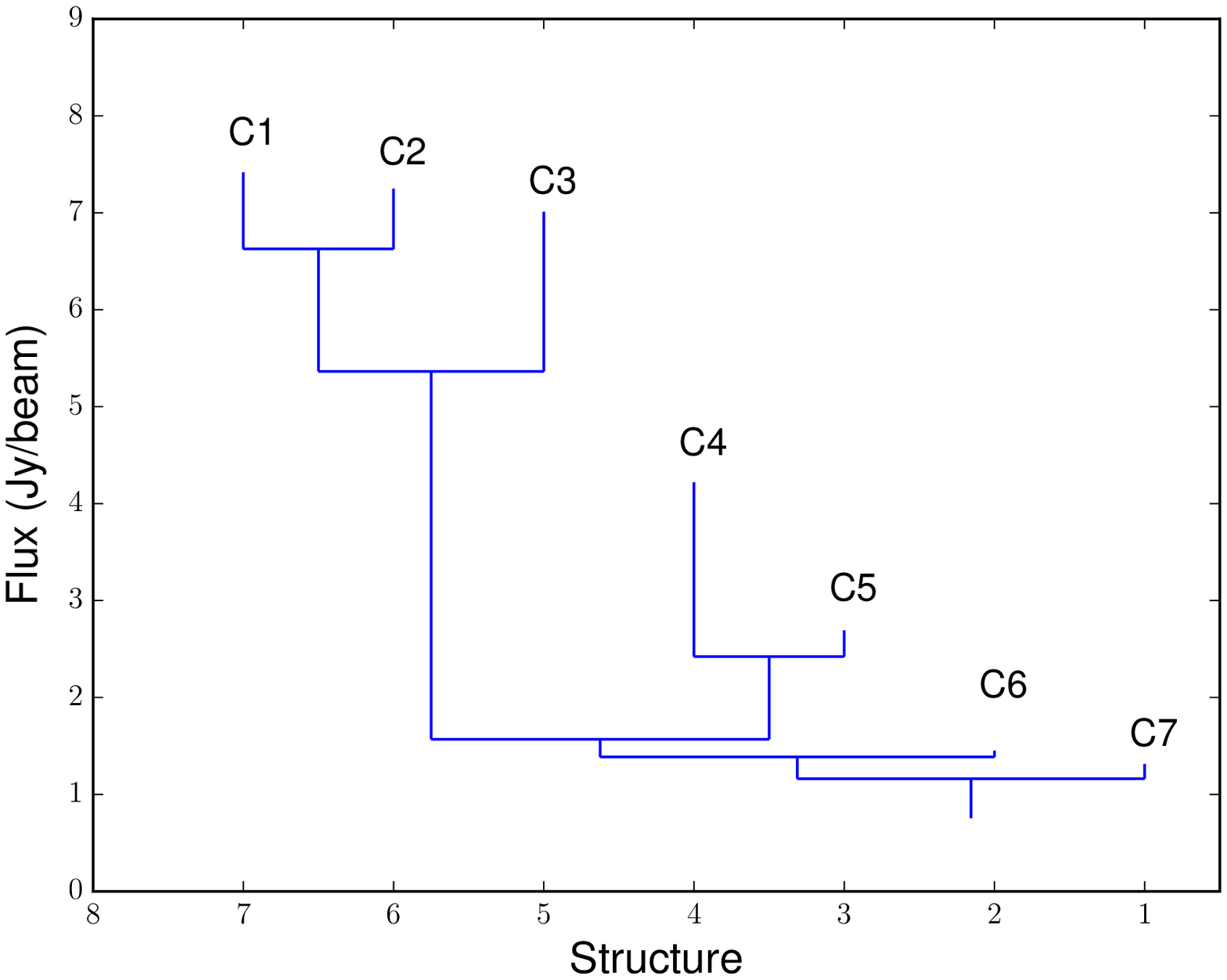}
 \caption{(Left) Clump apertures (obtained from $\it{2{\rm D}-Clumpfind}$ analysis) in white marked over the 870~$\mu$m emission map. The peak positions of clumps are marked with crosses. (Right) Dendrogram structure of the clumps in this region.}
 
\label{cl_aper}
\end{figure}

A scrutiny of cold dust emission at 160~$\mu$m and longer wavelengths reveals the presence of a number of dust clumps. We prefer to use the 870~$\mu$m image to identify the dust clumps for two reasons: (i) this wavelength traces the coldest dust clumps, and (ii) the dust emission is optically thin. 
We used the $\it{2{\rm D}-Clumpfind}$ algorithm discussed in Section 3.1 with a threshold of 15$\sigma$ and contour spacing of 1$\sigma$ to identify seven clumps, where the rms  ($\sigma$) value is $\sim50$~mJy/beam. The clump structure has also been explored through the dendrogram algorithm \citep{2008ApJ...679.1338R} and the resulting tree structure shown in Fig.~\ref{cl_aper} identifies these seven clumps for the same threshold and contour spacing. The seven clumps found in this region are labelled as C1, C2, ..., C7 (in order of decreasing brightness). The dendrogram format distinctly exhibits branches and leaves in the tree structure. The clumps C1, C2 and C3 lie within a large structure that also contains C4 and C5 as leaves. The hierarchial structure is evident \citep{1992ApJ...393..172H} although the statistics are inadequate to comment on the general nature of the molecular cloud. In Fig.~\ref{cl_aper} we show the clumps associated with the 870~$\mu$m emission in this region. The arbitrary apertures corresponding to the area covered by each clump in the 870~$\mu$m image retrieved from the $\it{2{\rm D}-Clumpfind}$  algorithm are overplotted in Fig.~\ref{cl_aper}. As we are interested in the average properties of each clump,  we have constructed a clump spectral energy distribution (SED) using wavelengths 70 - 870~$\mu$m by integrating flux densities in the selected apertures. An average background value is estimated from a nearby field that is $\sim13'$ away (centred at $\alpha_{J2000} =17^h$ 20$^m$14.25$^s$, $\delta_{J2000}=-37^\circ$~19$\arcmin$~23.4$\arcsec$) devoid of bright diffuse emission and this is appropriately subtracted to account for the zero offsets at each wavelength.  We fitted the SEDs with a modified blackbody model of the form \citep[][]{{2013A&A...551A.111O},{2011A&A...535A.128B}, {1990MNRAS.244..458W}}:

\begin{equation}
F_{\nu} = \Omega B_{\nu}(T_{D})(1-e^{-\tau_{\nu}})
\end{equation}
\begin{equation}
\tau_{\nu} = m_{H}\mu N(H_{2})k_{\nu}
\end{equation}

Here $F_{\nu}$ is the flux density at frequency $\nu$, $\Omega$ is the solid angle subtended by the clump, $B_{\nu}(T_{D}$) is the blackbody function at temperature $T_{D}$, $m_{H}\mu$ is the mean particle mass, $N(H_{2})$ is the column number density of molecular hydrogen, and $k_{\nu}$ is the dust mass opacity.  $m_H$ represents the mass of a hydrogen atom and $\mu$ is taken to be 2.86 under the assumption that the gas is 70$\%$ molecular hydrogen by mass \citep{2010A&A...518L..92W}. The dust mass opacity is estimated using the following expression \citep{{1990AJ.....99..924B},{1983QJRAS..24..267H}},
\begin{equation}
k_{\nu} = 0.1\left(\frac{\nu}{1000 \mathrm{GHz}}\right)^{\beta} \mathrm{cm^{2}g^{-1}}
\end{equation}

\par
\begin{figure*}[hbt!]
\centering
\includegraphics[width=5.1cm]{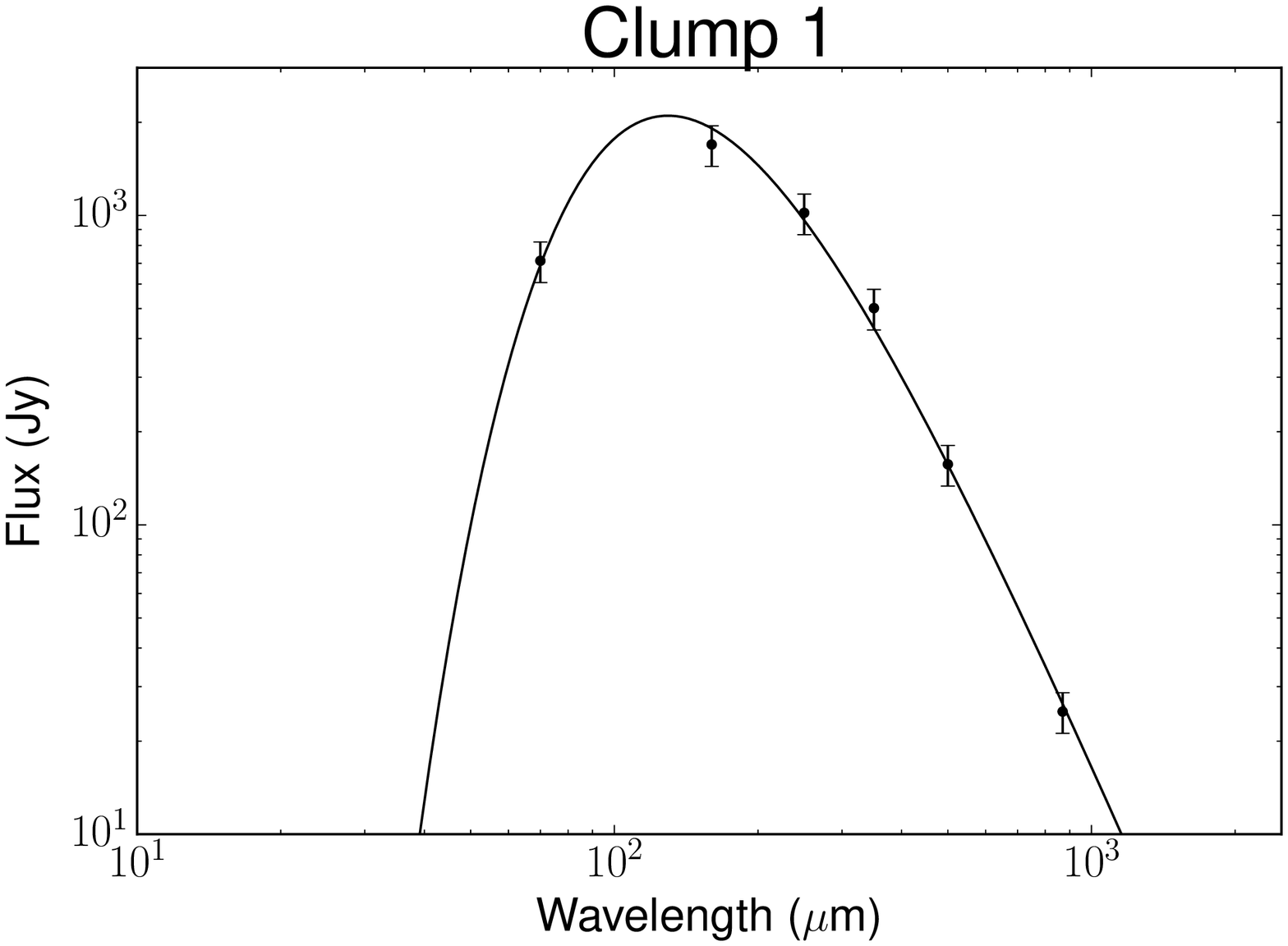}\quad \includegraphics[width=5.1cm]{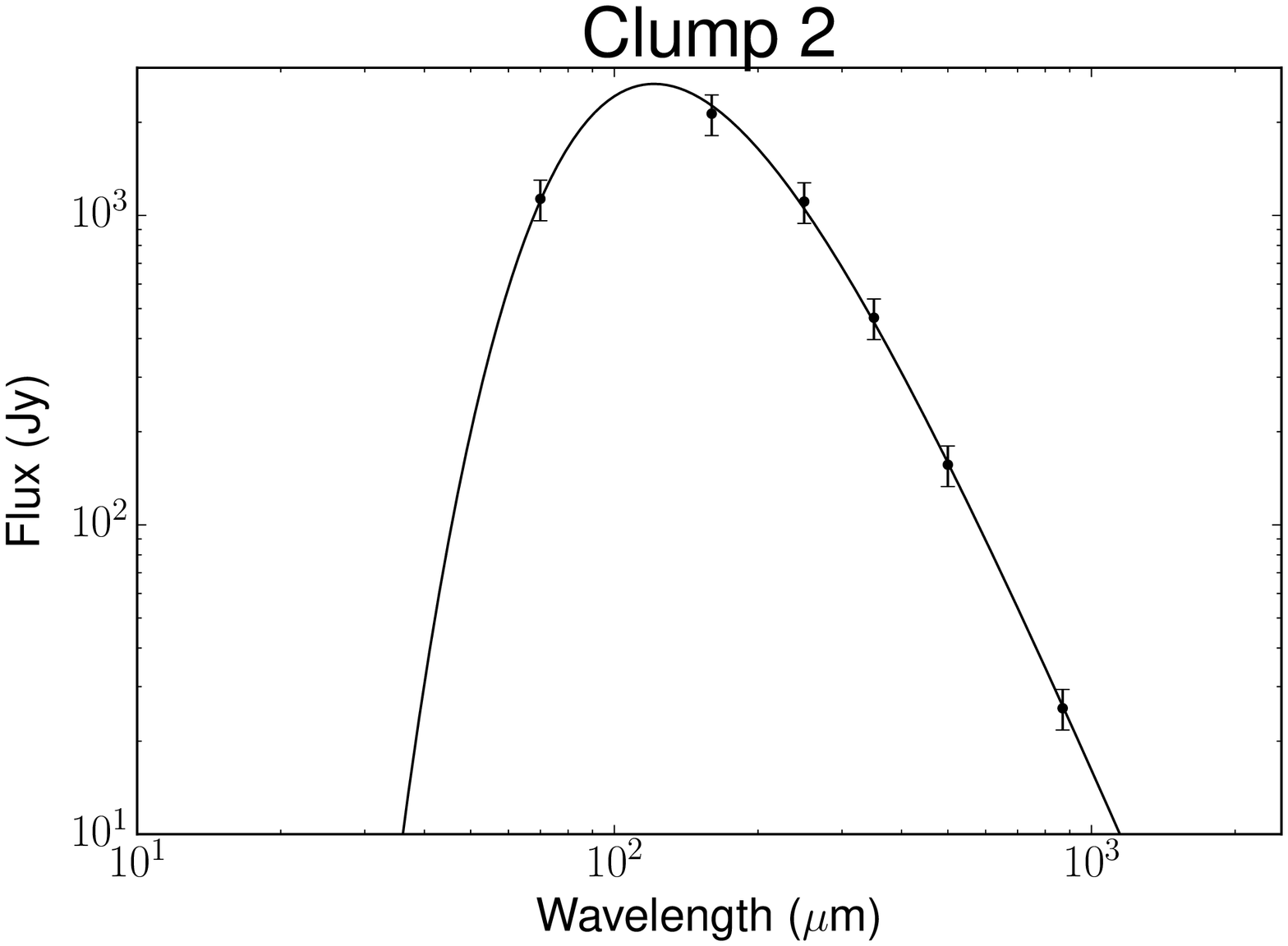}\quad\includegraphics[width=5.1cm]{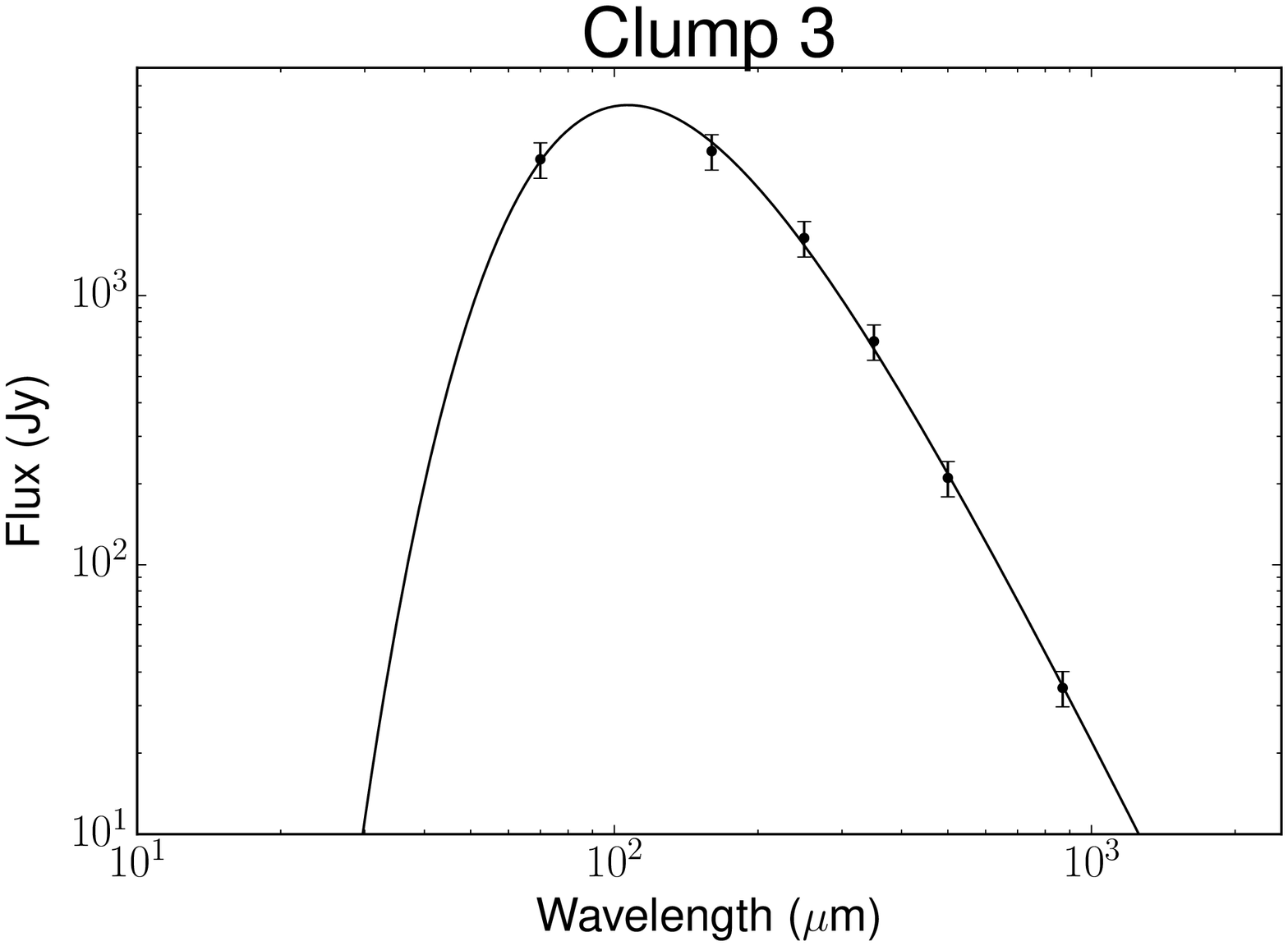}\quad\includegraphics[width=5.1cm]{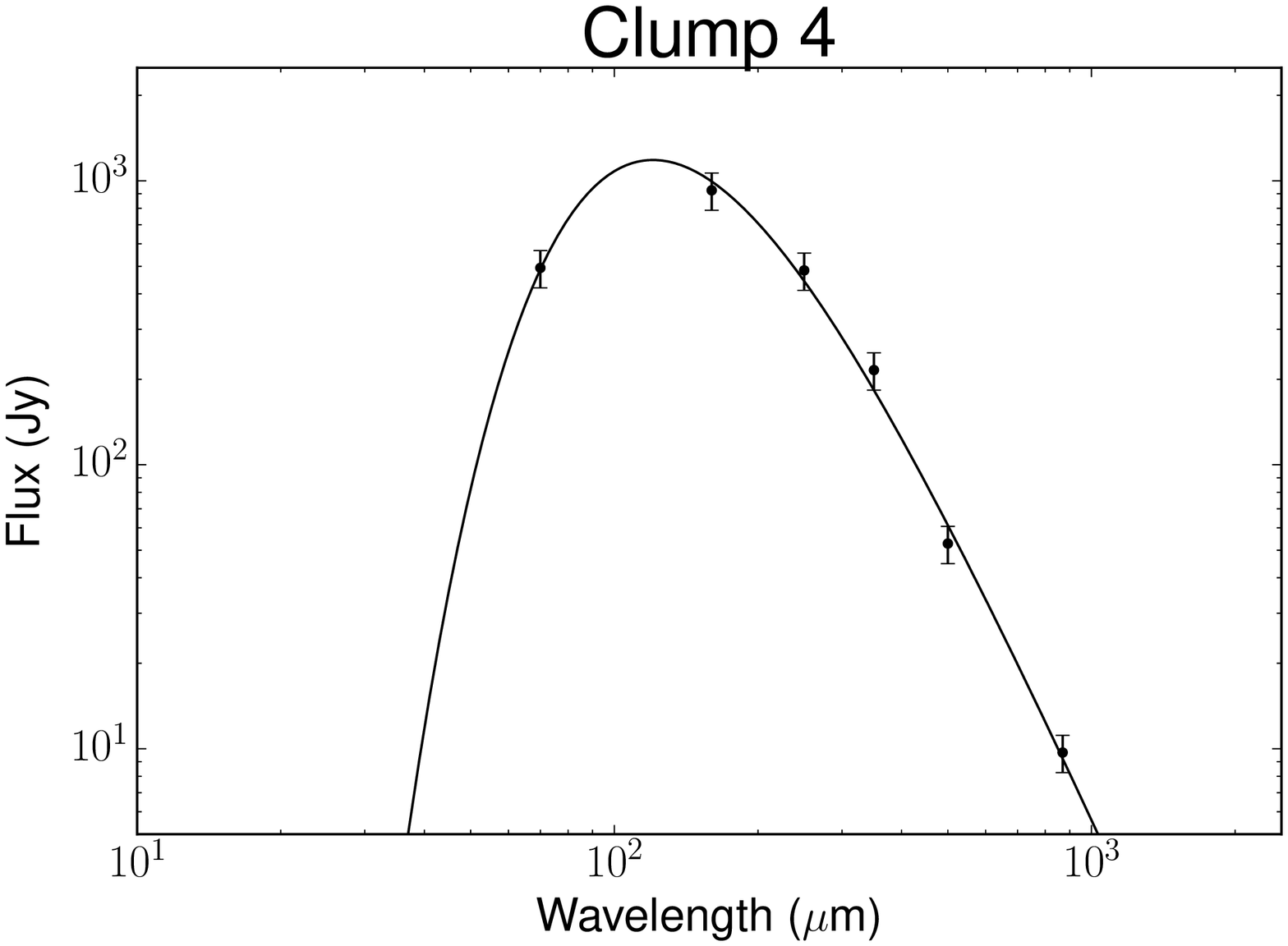}\quad\includegraphics[width=5.1cm]{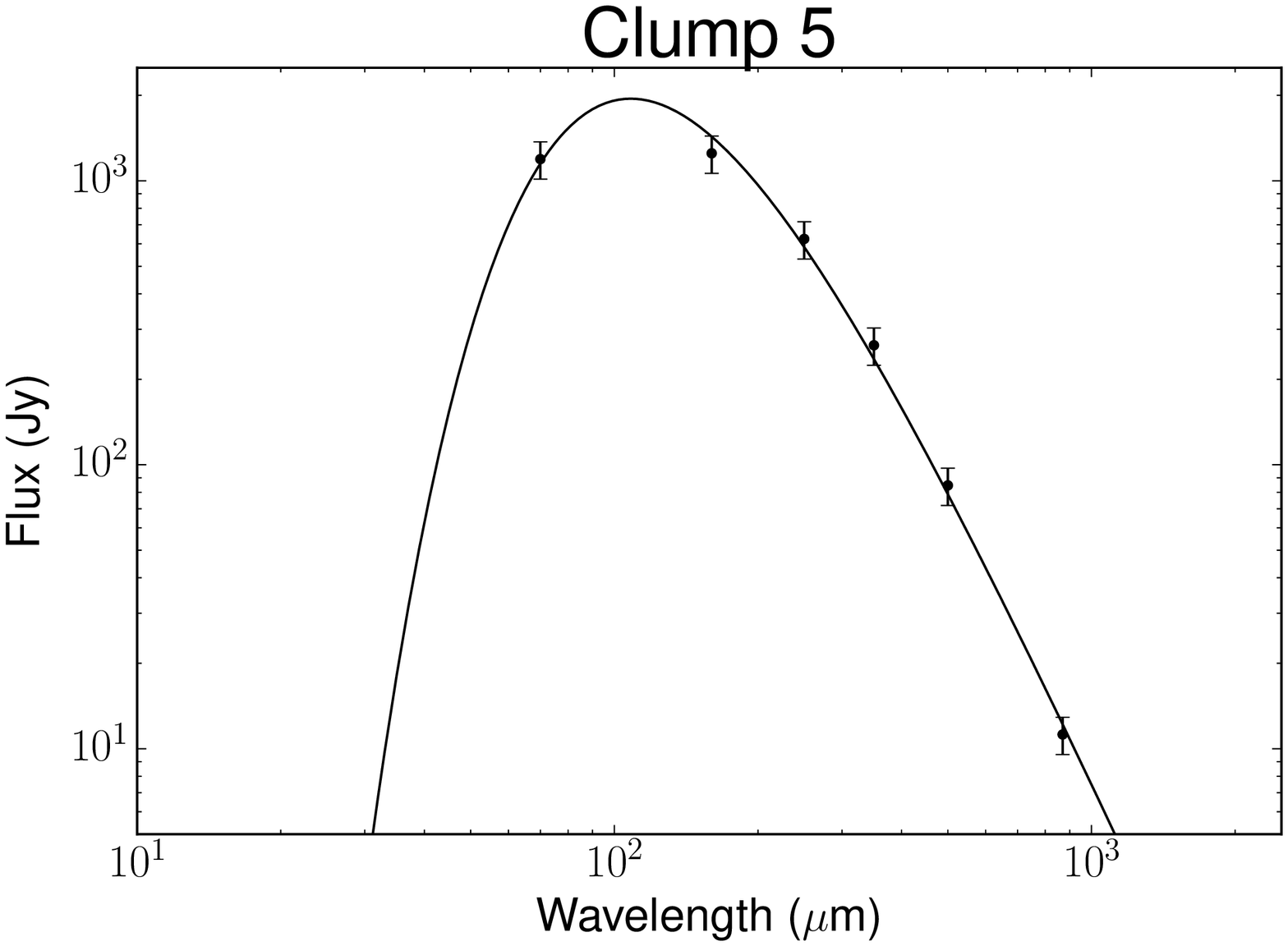}\quad \includegraphics[width=5.1cm]{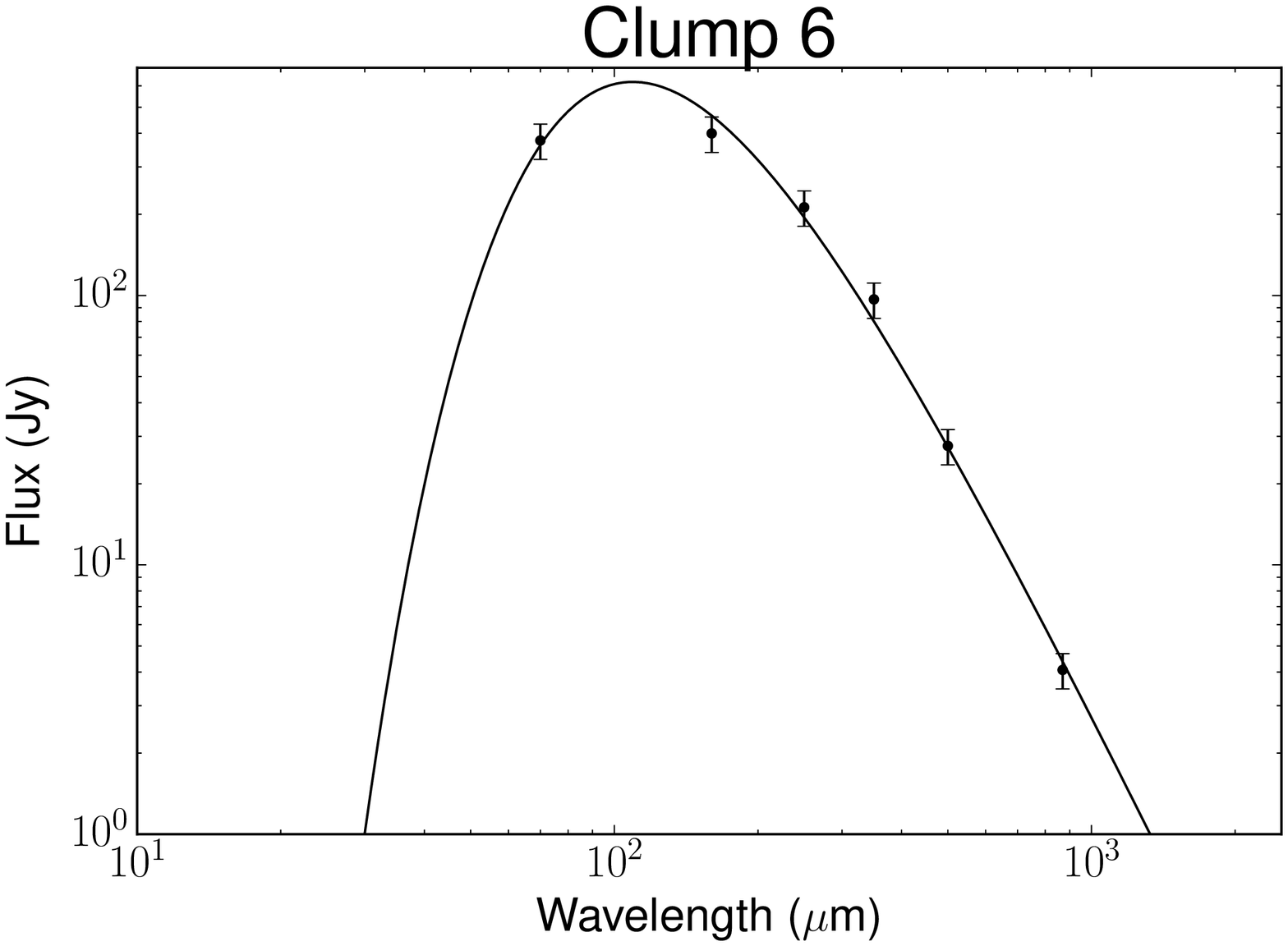}\quad\includegraphics[width=5.1cm]{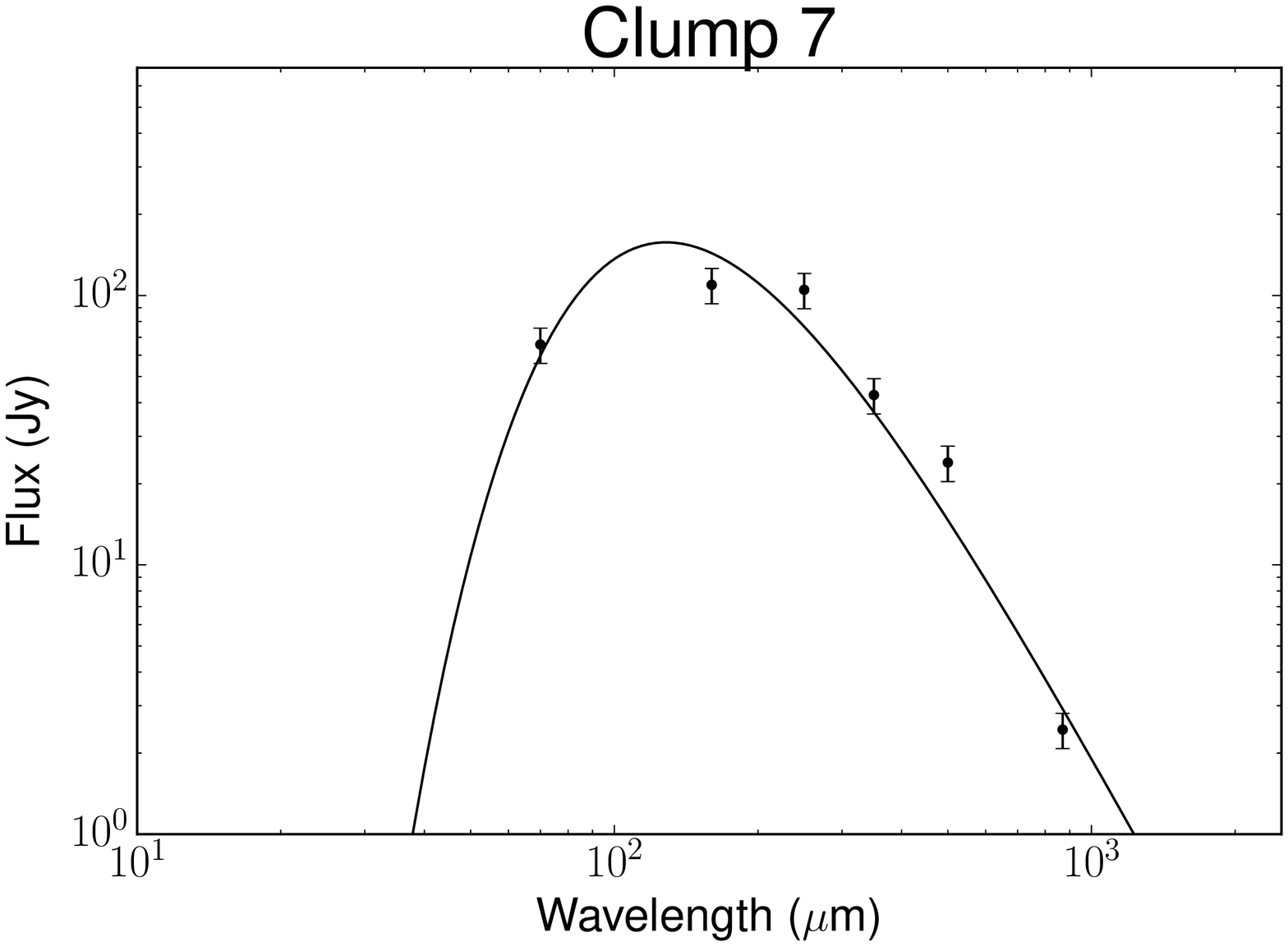}

\caption{Spectral energy distributions of the cold dust clumps in IRAS~17160-3707. Data from \textit{Herschel} and ATLASGAL images are represented as solid circles. The best-fit modified blackbody function for each SED is shown as a solid line curve. The error bars correspond to 15\% uncertainties in flux densities. }
\label{clsed}
\end{figure*}

 In this equation, $\beta$ represents the dust emissivity index.  The best fits were obtained using nonlinear least squares Levenberg-Marquardt algorithm considering $N(H_2)$, $T_D$, and $\beta$ as free parameters. Based on other studies, we assumed flux uncertainties of the order $\sim15$\% in all bands \citep{{2013A&A...551A..98L}, {2009A&A...504..415S}}. The resolution of the 350 and 500~$\mu$m images are lower than the 870~$\mu$m image. As a result, clumps located close to each other, such as C1 - C2 and C4 - C5 are not resolved in these images. Nevertheless, we have integrated flux densities at 350 and 500~$\mu$m in the apertures and find that the best-fit parameters change negligibly ($<10$\%) when we exclude fluxes at these two wavelength bands for these clumps. The best-fit SEDs are shown in Fig.~\ref{clsed} and the values of the derived parameters for all the clumps are listed in Table~\ref{cltb}.

\tabcolsep=0.08cm
\begin{table*}[hbt!]
\footnotesize
\caption{Properties of the cold dust clumps identified in this region.}
\begin{center}
\hspace*{-0.5cm}
\begin{tabular}{ c c c c c c c c c c c}\hline \hline
\setlength{\tabcolsep}{1pt}
Clump & $\rm{\alpha_{J2000}}$ & $\rm{\delta_{J2000}}$ &Area &$\beta$ &Temperature &Column density& $\chi_{red}^2$&Mass &L &$\Sigma$\\
& $(^{h~m~s})$ &$(^{\degr~\arcmin~\arcsec})$&(pc$^{2}$) &&(K)&(10$^{22}$ cm$^{-2}$)&&($\times$10$^2$~M$_\odot$) &($\times$10$^2$~L$_\odot$) &(g cm$^{-2}$)\\
\hline\\
C1 &17:19:25.65 &-37:10:07.9 &3.3 &1.7 $\pm$ 0.1 & 24.2 $\pm$ 0.9 & 5.9 $\pm$ 0.8 &0.6&48&106&0.3\\
C2 &17:19:26.60 &-37:10:23.3 &2.3 &1.8 $\pm$ 0.1 & 26.0 $\pm$ 0.5 & 7.6 $\pm$ 0.5 &0.1&44&143&0.4\\
C3 &17:19:26.81 &-37:11:01.9 &4.7 &1.7 $\pm$ 0.1 & 29.6 $\pm$ 0.9 & 4.1 $\pm$ 0.4 &0.2
&48&317&0.2\\
C4 &17:19:35.73 &-37:10:55.2 &1.5 &1.9 $\pm$ 0.2 & 25.1 $\pm$ 1.2 & 5.1 $\pm$ 0.9 &0.9&19&63&0.3\\
C5 &17:19:33.12 &-37:10:48.2 &2.5 &1.8 $\pm$ 0.1 & 28.4 $\pm$ 1.5 & 3.0 $\pm$ 0.5 &0.7&19&118&0.2\\
C6 &17:19:21.87 &-37:10:41.9 &1.2 &1.8 $\pm$ 0.2 & 28.3 $\pm$ 1.9 & 2.2 $\pm$ 0.5 &1.0&7&38&0.1\\
C7 &17:19:35.12 &-37:10:14.0 &0.9 &1.4 $\pm$ 0.4 & 25.9 $\pm$ 3.9 & 1.7 $\pm$ 0.8 &5.7&3&9&0.1\\
\hline
\end{tabular}
\label{cltb}
\end{center}
\end{table*}


\par The dust temperature in the clumps lie between $24 - 30$~K, with C3 showing the highest temperature and C1 having the lowest dust temperature. The molecular hydrogen column densities range within $1.7-7.6\times~10^{22}$~cm$^{-2}$ with the clump C2 showing the largest column density value among the clumps. The range of dust emissivity index $\beta$ values is $1.4 - 1.9$. These values represent average values over the entire clump region and the distribution of column density and dust temperature are discussed in more detail in Section 3.2.3. We have also estimated the mass of each clump from the column density value using the expression of the form

\begin{equation}
M_c= {N(H_2)}\mu\ m_H\,A
\end{equation}
Here $M_c$ is the clump mass and $A$ is the physical area of the clump. The area and mass of each clump is listed in Table~\ref{cltb}. The clumps C1 and C3 are the most massive clumps with a mass of 4800~M$_\odot$ each, while C7 is the least massive clump with a mass of 300~M$_\odot$.  The total mass of the cloud is 18,800~M$_\odot$. 
We estimated the effective clump diameter as the geometric mean of the largest and smallest size of the irregular clump aperture from the $\it{2{\rm D}-Clumpfind}$  algorithm. The diameters of clumps in this region range between $1.0 - 2.4$~pc. 
\citet{2013MNRAS.435..400U} investigated the properties of a large number of clumps associated with compact \hii~regions and they find that the clump masses range between $10^2-10^5$~M$_\odot$ for clumps of size 0.2 to 6 pc. Our clumps fall well within this range. In addition, \cite{2015arXiv151108810S} studied the properties of starless and massive star forming clumps from the Bolocam Galactic Plane survey and find that the mass of clumps harbouring ultracompact \hii~regions lie in the range 30 - 30000 M$_\odot$ with radius lying between 0.1 to 4~pc. 
\citet{2013ApJ...773..102F} for instance adopt a lower limit of 40~M$_{\odot}$ for identifying potential molecular cores that can form a high mass star. The sizes of cores in their sample range between $0.4-1.1$~pc. In our case,  all the clumps have masses larger than 40~M$_\odot$. This is not surprising as the clumps other than C6 and C7 are associated with compact radio sources lying close to the peak emission (within $5''$). If we presume that each of these compact radio source is excited by ZAMS star(s), this would necessitate the presence of massive star(s) in these clumps. Thus, these clump parameters characterise those of massive star forming clumps and lie within the mass range mentioned in above studies. We have also estimated the luminosity of each clump by integrating under the best-fit SED. The clump C3 has the largest luminosity of $3.17\times10^4$~L$_\odot$ while C7 is the least luminous clump with 900~L$_\odot$. 

\par Another parameter used to probe massive star formation in molecular clouds is the surface density ($\Sigma$) defined as $M_c/A$. According to \citet{2008Natur.451.1082K}, clouds with a minimum surface density of $\Sigma \sim$1~g~cm$^{-2}$ would be able to suppress fragmentation to form massive stars. The surface density values for clumps in this region are listed in Table~\ref{cltb}. The values of $\Sigma$ for the seven clumps lie in the range $0.1-0.4$~g~cm$^{-2}$. The maximum $\Sigma$ is observed towards clump C2 that also displays the largest column density among clumps.  None of our clumps satisfy the massive star forming criteria of \citet{2008Natur.451.1082K}. As the clumps C1 - C5 have massive embedded (proto)stellar objects, the surface density of these clumps would signify values where massive stars are forming. These are surface density values post massive star formation in a sense. Therefore, such a comparison of $\Sigma$ makes sense under the assumption that the density of the clump has not changed notably. It is worth noting though, that C6 and C7 have the lowest surface densities of 0.1~g~cm$^{-2}$ while the values of clumps with massive embedded stars are in the range $0.2-0.4$~g~cm$^{-2}$. These values are consistent with the values obtained towards a large sample of massive star forming cores by \cite{2010A&A...517A..66L}, \cite{2010A&A...520A.102M} and \cite{2013A&A...556A..16G}. These authors have shown that it is likely that massive star forming cores possess lower surface densities of the order of $\sim$0.2~g~cm$^{-2}$. 
However, it must be borne in mind that these values are sort of average values given the large size of clumps and the actual surface density could be higher near the peak emission or dense core. This is also emphasized by \citet{2009ApJS..181..360C} who carried out a survey of massive star forming clumps and attribute lower surface densities to massive protostars in cores of lower densities. 

\subsubsection{Bolometric Luminosity}

\begin{figure}[hbt!]
\begin{center}
\includegraphics[scale=0.5]{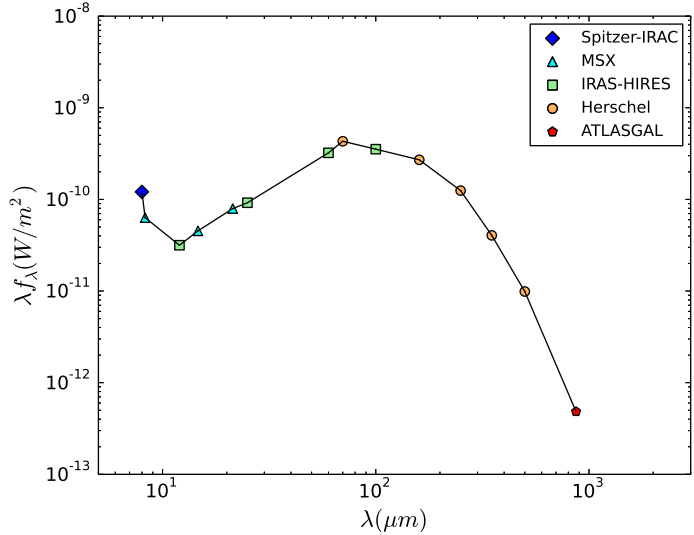}
\caption{Spectral energy distribution of the IRAS~17160--3707 region obtained by integrating fluxes within a circular region of diameter $3\arcmin.2$ centred on $\alpha_{J2000}=17^h19^m30.3^s$, $\delta_{J2000}=-37^\circ10'25.7''$.}
\label{lum}
\end{center}
\end{figure}

We decided to estimate the bolometric luminosity of this region and compare it with the IRAS luminosity obtained earlier by others considering that the wavelength coverage is larger now with various scientific surveys and missions. The global SED of this region is constructed using flux densities from \textit{Spitzer}-IRAC, IRAS-HIRES \citep{1990AJ.....99.1674A}, MSX, \textit{Herschel}-HiGal, and ATLASGAL within a circular region of radius $3\arcmin.2$ centred on $\alpha_{J2000}=17^h19^m30.3^s$, $\delta_{J2000}=-37^\circ10'25.7''$. This region is selected keeping in view the extent of diffuse emission as well as the low resolution of the IRAS-HIRES 100~$\mu$m image ($\sim2\arcmin.1$). We interpolate the flux densities of the observed SED using power laws and numerically integrate under the SED to obtain the bolometric luminosity. The total luminosity is  $8.3\times10^5$~L$_\odot$ that is nearly a factor of two higher than the previous estimate of $\sim3.6\times$10$^{5}$ L$\odot$ \citep{2002A&A...381..571P} where the latter value assumes a distance of 5.7~kpc. The larger flux density at 8~$\mu$m as compared to the 12 $\mu$m band is explicable on the basis of PAH emission in the former and silicate absorption in the latter. The full SED of this region is shown in Fig.~\ref{lum}. A comparison of the bolometric luminosity with the total estimated luminosity of the clumps (Table~\ref{cltb}) shows that the former is an order of magnitude larger than the latter. This is attributed to (i) non-negligible contribution of near and mid-infrared fluxes to the bolometric luminosity unlike the modified blackbody fits to the clumps, and (ii) contribution of diffuse emission from the larger area used to estimate the bolometric luminosity. Although the bolometric luminosity corresponds to a single ZAMS star of spectral type O5 \citep{1973AJ.....78..929P}, the presence of seemingly isolated compact radio sources and young stellar objects coincident with few compact sources ascertains that this complex is excited by a cluster of massive stars.

\subsubsection{Column density, Temperature and Dust Emissivity Index Maps}

\par In addition to clump SEDs, we have also constructed the line-of-sight averaged molecular hydrogen column density, dust temperature and $\beta$ maps by carrying out pixel-to-pixel greybody fits in the wavelength regime (70 - 870~$\mu$m). While in the previous section we obtained the clump averaged properties, here we obtain the column density and dust temperature on the pixel scale.  Although few clumps are not resolved because of poor resolution ($36''.9$), we can investigate the morphological distribution of dust properties. The 70, 160, 250, 350 and 870~$\mu$m images are convolved and regridded to the lowest resolution ($36''.9$) and pixel size ($14''$) of the 500~$\mu$m image. The kernels used for the convolution of the \textit{Herschel} images are taken from \citet{2011PASP..123.1218A}, while a Gaussian kernel is used for  the 870~$\mu$m image.  As earlier, the average background value estimated from a nearby field is subtracted. We fitted the flux densities at each pixel with the modified blackbody model defined earlier (Eqns. $2-4$). 

\begin{figure*}[hbt!]
\includegraphics[width=150mm]{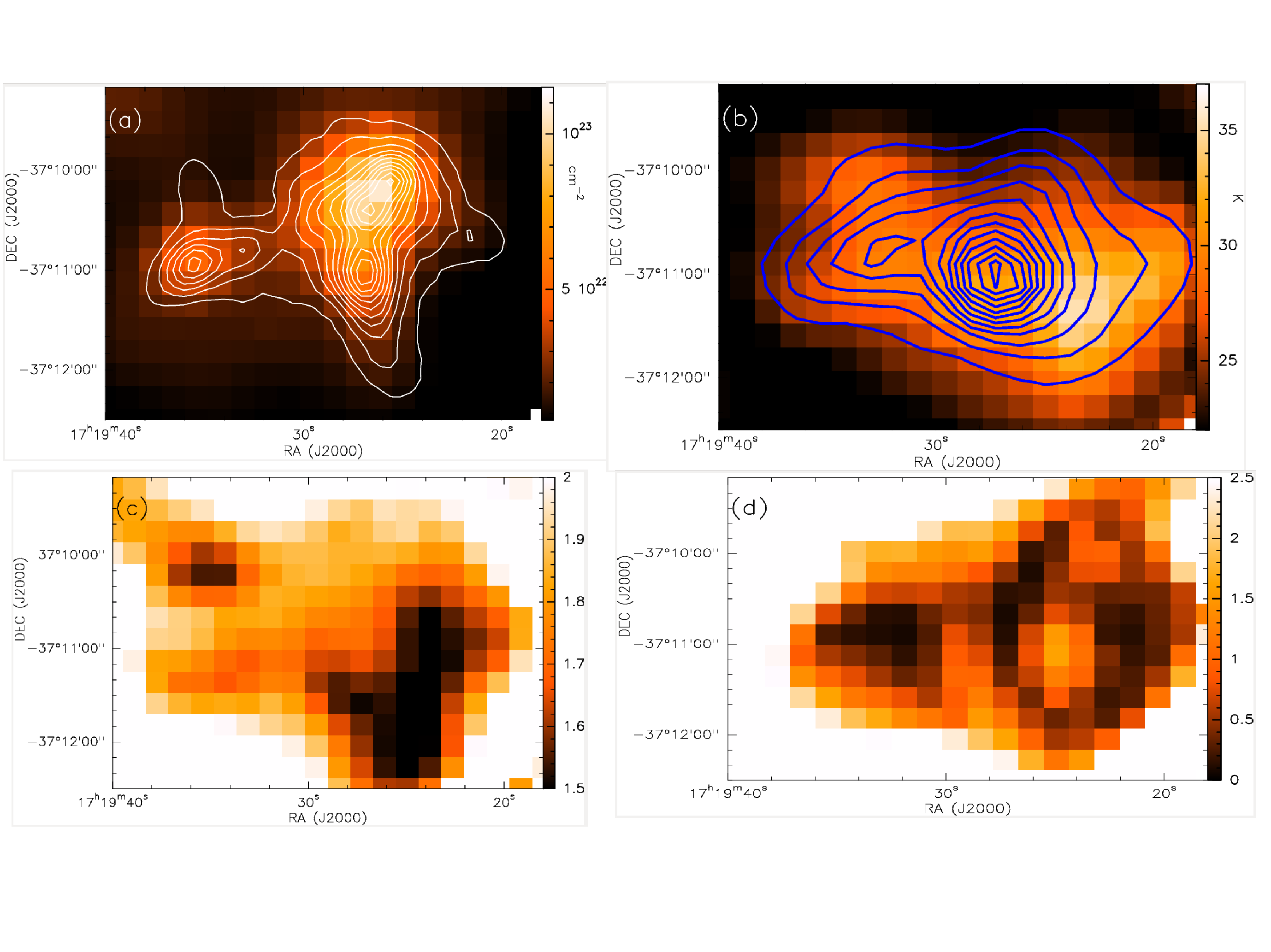}
 \caption{(a) Column density, (b) dust temperature, (c) dust emissivity index $\beta$, and (d) $\chi_{red}^{2}$ maps constructed using the pixel-to-pixel modified blackbody fits. The 870~$\mu$m emission contours are overlaid on the column density map with the contour levels same as that in Fig.~\ref{cl_aper}. The dust temperature map is overlaid with contours of 70~$\mu$m emission convolved to the resolution of 500~$\mu$m emission map. The contour levels of the 70~$\mu$m emission are 20.0, 38.5, 57.0, 75.5, 94.0, 112.5, 131.0, 149.5, 168.0, 186.5, 205.0, 223.5 and 242.0~Jy/pixel. }
 \label{colden}
\end{figure*}


\begin{figure*}[hbt!]
\begin{center}
\includegraphics[height=65mm]{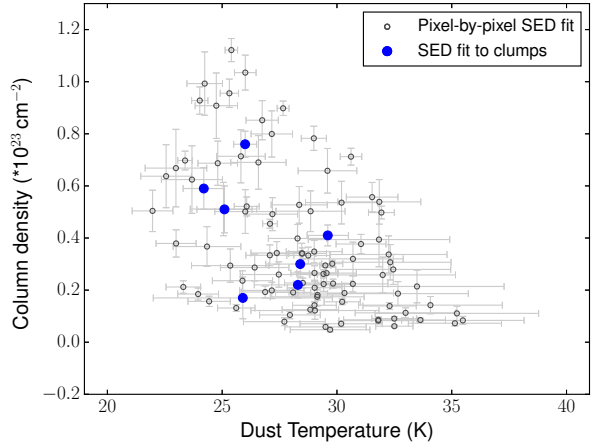}
 \caption{The dust temperature (T$_d$) versus column density plot for pixels within the 15$\sigma$ contour of the 870~$\mu$m map, shown as open circles. The corresponding values for the 7 cold dust clumps are also shown in the image as filled circles.}
 \label{T_cd}
 \end{center}
\end{figure*}


\par The column density, temperature and dust emissivity index ($\beta$) maps alongwith the reduced $\chi^2$  ($\chi^2_{red}$) image are shown in Fig.~\ref{colden}. For further analysis and comparison with average clump values, we consider pixels within the 15$\sigma$ contour of the 870~$\mu$m emission, same as the threshold used for clump identification. The fits are relatively good as $\chi^2_{red}<2$ for these pixels. The peak column density is $1.1\times10^{23}$~cm$^{-2}$ towards clumps C1 - C2, while the median column density in our region of interest is $3.0\times10^{22}$~cm$^{-2}$. The column density values agree well with that of other massive star forming regions \citep[e.g.][]{{garay2007multiwavelength},{2010ApJ...717.1157D}}. The column density distribution shows two peaks: towards C1 - C2 and C4, separated by 3.2~pc. The clumps are not resolved owing to the poor resolution of the map. A comparison of the column density and temperature values obtained within our region of interest is compared with the average values obtained from clumps in Fig.~\ref{T_cd}. Clearly, the average clump estimates fall within the values occupied by the pixels.

\par The maximum temperature in this region is 35.5~K and peaks $\sim70''$ to the west of the column density peak. The mean value of dust temperature is 28.9~K. We would like to emphasize that the $\chi^2_{red}$ for pixels near the highest temperature are larger than the surrounding pixels. This is evident from Fig.~\ref{colden}~(d). The dust temperature peak can be understood when we take into consideration the 70~$\mu$m emission. The 70~$\mu$m flux density is the shortest wavelength used in the SED construction. Consequently, pixels with significant emission at 70~$\mu$m and negligible cold dust emission are weighed by the 70~$\mu$m emission leading to higher dust temperature.  This explains the large temperature to the west of complex A. This region is enclosed within a dashed line ellipse in Fig.~\ref{ir} over the 70 and 500~$\mu$m images, emphasizing the contrast in emission. This is also conspicuous from the 870~$\mu$m and 70~$\mu$m emission contours overlaid over images in Fig.~\ref{colden}. In addition to thermal emission from warm dust that also contributes to the SED, the 70 $\mu$m emission traces the contribution of very small dust grains \citep{2013A&A...554A..42R} and many studies have excluded this wavelength in their analyses \citep{{2010A&A...518L..99A},{2010A&A...518L..98P}}. 
Our attempts to exclude the 70~$\mu$m data point from the pixel-to-pixel SED fitting leads to unreasonably high values of parameters towards many pixels as only the longer wavelength edge of the greybody function is sampled (considering that the maximum mostly occurs between 70 and 160~$\mu$m). The temperature enhancement seen to the south-west of complex A is also associated with warm-hot dust emitting at 8, 24 and 70~$\mu$m and with diffuse ionised gas (see radio contours overlaid on the 70~$\mu$m image in Fig.~\ref{ir}). This prompts us to speculate that this region could be more evolved compared to the rest of the complex associated with cold dust emission. This is based on the premise that the molecular cloud (hence cold dust) here has dispersed leading to relatively low column density values. Alternately, it is possible that this region is not affiliated to the molecular cloud of IRAS 17160-3707 but lies along the same line-of-sight. Considering that IRAS  17160-3707 lies in the inner Galactic plane, this possibility cannot be ruled out. 

The dust emissivity index $\beta$ gives information about the properties of dust grains and hence the $\beta$ map of a star forming region is very useful in gauging the dust properties of the cloud. The variation of $\beta$ in this region is within a relatively narrow range $1.4-2.0$ unlike few other star forming regions where the range is larger, extending from 0.8 to 3.1 \citep[e.g.][]{{2010A&A...518L..99A},{tabatabaei2014variation},{2016MNRAS.456.2425V}}.  The mean value of  $\beta$ is 1.7, that is lower than the value of $\sim2.0$ typically quoted in literature \citep{{2002MNRAS.329..257W},{2013ApJ...767..126S}}. The lower values of $\beta$ have been attributed to the presence of fractal dust grains, which are created from the coagulation of smaller sub-units \citep{1987ApJ...320..818W}. These fractal dust grains are of low mass and are porous, electrically conducting structures. 
Previous studies of dust temperature ($T_D$) and $\beta$ towards molecular clouds and interstellar medium in general have shown that there exists an anti-correlation between $T_D$ and $\beta$ \citep[e.g.][]{{2003A&A...404L..11D},{2008A&A...481..411D},{2010A&A...520L...8P},{2012A&A...542A..10A}}. An anti-correlation is also apparent from the $\beta$ and $T_D$ maps shown in Fig.~\ref{colden} where the region displaying high temperature shows lower dust emissivity values and vice-versa. While this effect may have implications on the composition and size of dust grains, there could be pseudo anti-correlation effects introduced either due to the fitting procedure itself as shown by \cite{2013ApJ...767..126S} or due to noise effects and dust temperature variations in the line of sight \citep{{2009ApJ...696..676S},{2009ApJ...696.2234S}}. Hence we treat this anti-correlation effect between $T_D$ and $\beta$ with caution as the temperature map could have contribution from very small grains at 70~$\mu$m.

\begin{figure}[hbt!]
\begin{center}
\hspace*{-2.0cm}
\includegraphics[width=70mm,angle=-90]{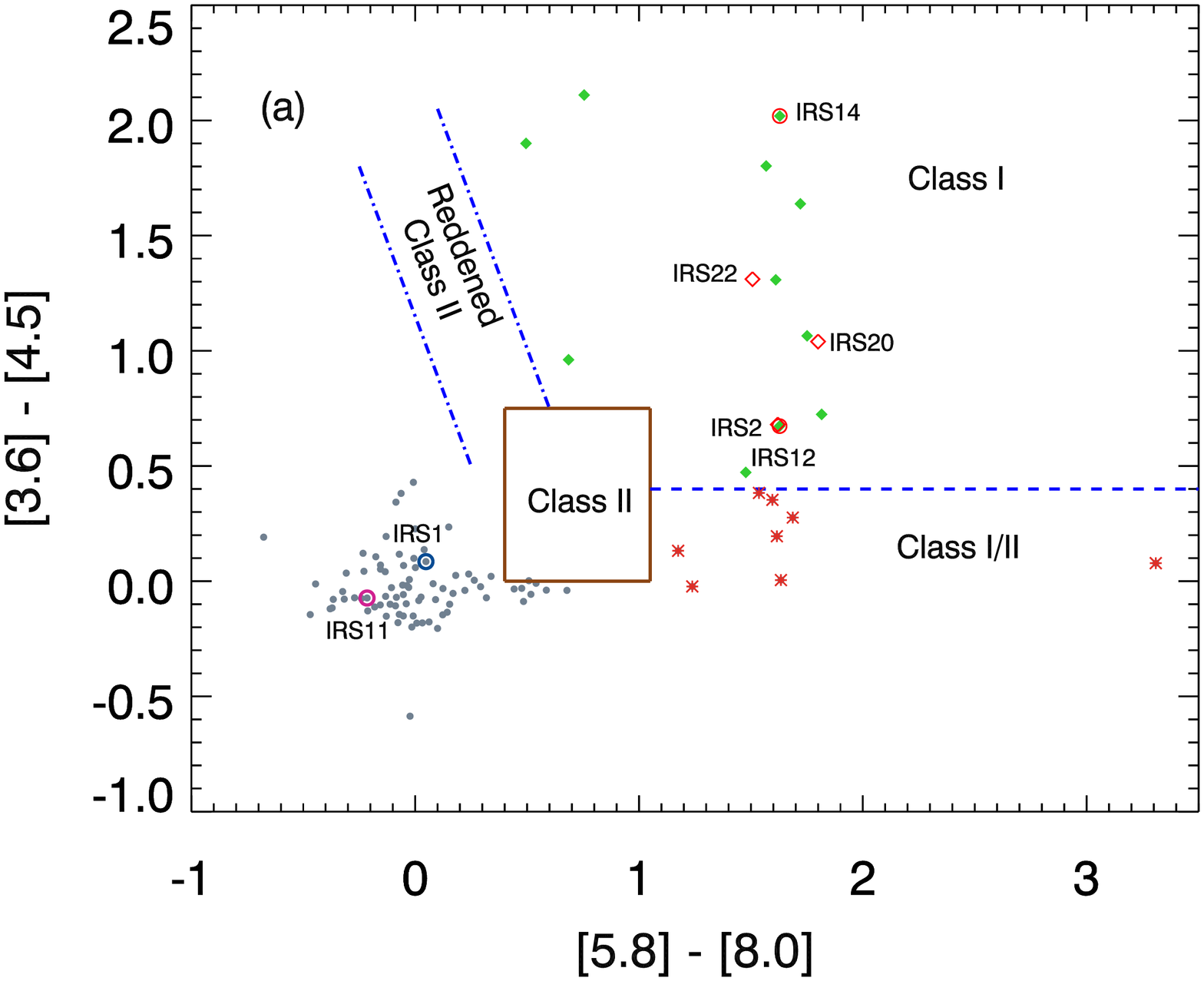}
\hspace*{-0.2cm}
\includegraphics[width=70mm,angle=-90]{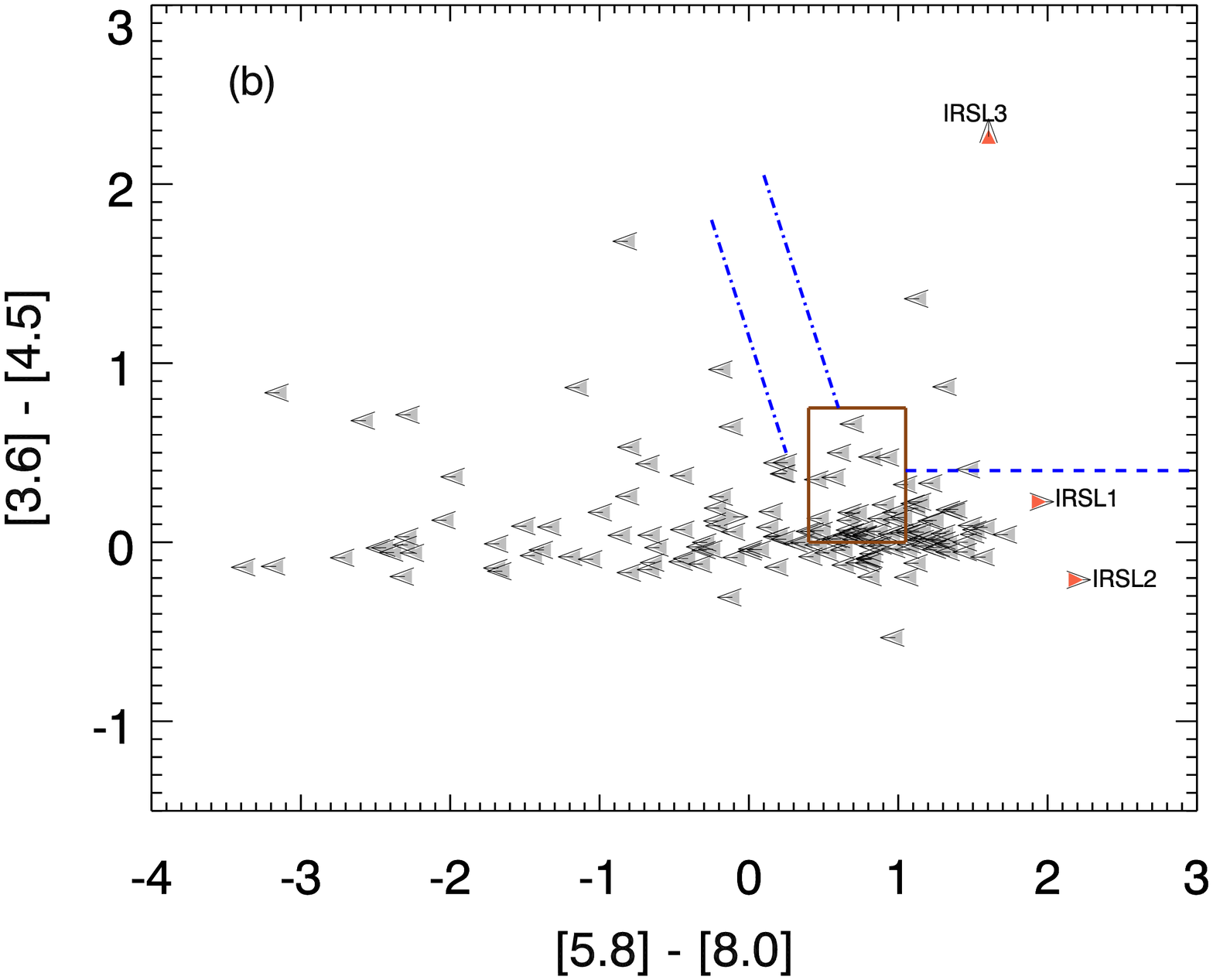}
 \caption{Color-color diagram using \textit{Spitzer}-IRAC colors [3.6] - [4.5] versus [5.8] - [8.0]. The solid line box indicates the position occupied by Class II sources while the parallel dash-dotted lines represent the region of reddened Class II sources. The region to the top right is occupied by Class I sources and the region to the right of Class II sources could be occupied by Class I or Class  II sources. (a) The filled diamonds are sources lying in Class I region while asterisks represent Class I/II sources. The dots represent other late type objects such as Class III objects or field stars. The sources encircled have 24~$\mu$m counterparts and open diamonds represent objects whose photometry was carried out by us. (b) Color-color diagram of IRAC sources not detected in one of the bands. The arrowheads pointed 
up, down, right and left represent sources not detected in 3.6, 4.5, 5.8 and 8.0 $\mu$m bands, respectively (details in text). The red arrowheads mark the sources likely to be young stellar objects based on their position in the color-color diagram.}
 \label{cc}
 \end{center}
\end{figure}


\subsection{Identification of Young Stellar Objects}

\par In order to examine the exciting sources of this region and their nature, we search the  Glimpse II Spring'08 Archive for mid-infrared point sources in a circular region of radius $2'$ centred on the location  $\alpha_{J2000}$=17$^h$19$^m$28.45$^s$, $\delta_{J2000}$=-37$^\circ$10$'$55.4$''$. This circular region encompasses the radio as well as the warm and cold dust emission, and is shown in Fig.~\ref{ir}.
 We find 916 sources of which 98 are detected in all the four IRAC bands. To identify the YSOs, we use the color scheme of
\cite{2004ApJS..154..363A} and \cite{2004ApJS..154..367M}. The colors  [5.8] - [8.0]  versus [3.6] - [4.5] are used to locate the YSOs based on their position in the color-color diagram. This color-color diagram is shown in Fig.~\ref{cc}(a). Also shown are the regions occupied by Class I, Class II sources, and reddened Class II sources based on predictions of existing models for disks and envelopes \citep{2004ApJS..154..363A}. We find 11 sources having colors associated with Class I protostars. Eight sources lie to the right of the region occupied  by Class II and these could be either Class I or Class II sources and are denoted as Class I/II. The details of these sources are listed in Table~\ref{yso}. We also searched for YSOs among the other sources that are not detected in one of the bands. We used the sensitivity limit as an upper limit to the flux density to identify their probable locations in the color-color diagram. We would like to add a note of caution by commenting that although the sensitivity limits have been used, the actual upper limits could depend on a number of factors that require an inspection of the sources in the image. These sources  are represented as arrowheads in the color-color diagram in Fig.~\ref{cc}(b) with the direction of arrow indicating the direction in which the actual source position would be located. 
Among these sources, we find only three that are of interest to us as their positions suggest that they could be YSOs. These sources are shown as red triangles in the figure and they are listed in Table~\ref{yso} as IRSL1, IRSL2 and IRSL3. 

\begin{figure*}[hbt!]
\begin{center}
\includegraphics[width=70mm,angle=-90]{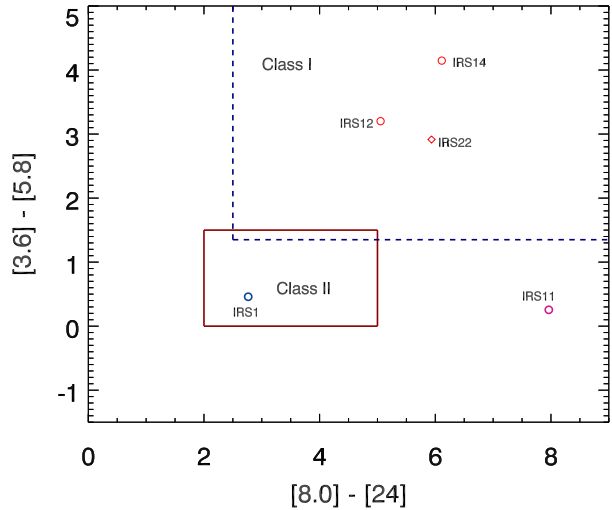}
 \caption{Color-color diagram of [3.6] - [5.8] versus [8.0] - [24] using mid-infrared fluxes from \textit{Spitzer}-IRAC.
  The open circles represent sources that have 24~$\mu$m counterparts and diamonds represent objects whose aperture photometry was carried out. The solid line box indicates location of Class II sources, while the dashed lines suggest the location occupied by Class I sources.}
 \label{24cc}
 \end{center}
\end{figure*}



\begin{table*}[hbt!]
\scriptsize
\caption{Young stellar objects in the IRAS 17160-3707 region identified from the mid-infrared color-color diagrams.}
\begin{center}
 \hspace*{-1.2cm}
\begin{tabular}{r c c c c c c c c c c} \hline \hline
YSO & $\rm{\alpha_{J2000}}$ & $\rm{\delta_{J2000}}$ & J & H & K & 3.6~$\mu$m & 4.5~$\mu$m & 5.8~$\mu$m & 8.0~$\mu$m & 24~$\mu$m \\
& $(^{h~m~s})$ &$(^{\degr~\arcmin~\arcsec})$& (mag) & (mag) & (mag) & (mag) & (mag) & (mag) & (mag) & (mag) \\ \hline
\\

IRS1 & 17:19:19.52 &	-37:11:46.2 & $10.32\pm0.04$ & $8.37\pm	0.03$ & $7.44\pm0.02$ & $7.11\pm0.04$ & $7.03\pm0.03$ & $6.65\pm0.02$ & $6.61\pm0.03$ & $3.84\pm 0.25$ \\
IRS2 & 17:19:20.39 &	-37:10:52.9 & ... & ... & ... & $9.28\pm0.02$ & $8.60\pm0.01$ & $6.18\pm0.02$ & $4.56\pm0.02$ & ... \\

IRS3 & 17:19:20.75 & -37:11:42.9 &  ... & ... & ... & $12.59 \pm 0.06$ & $12.58 \pm 0.10$ & $10.91 \pm 0.11$ & $ 9.28 \pm 0.23$ & ... \\
IRS4 & 17:19:22.88 & -37:09:55.6 &  ... & ... & ... & $13.57 \pm 0.08$ & $13.29 \pm 0.15$ & $11.14 \pm 0.31$ & $ 9.45 \pm 0.11$ & ... \\
IRS5 & 17:19:23.04 & -37:09:27.2 & ... & ... & ... & $11.61 \pm 0.04$ & $10.65 \pm 0.04$ & $ 9.85 \pm 0.04$ &  $9.16 \pm 0.04$ & ... \\
IRS6 & 17:19:23.05 & -37:10:09.2 &  ... & ... & ... & $12.34 \pm 0.05$ & $12.21 \pm 0.11$ & $10.33 \pm 0.14$ & $ 9.16 \pm 0.27$ & ... \\
IRS7 & 17:19:23.64 & -37:10:16.7 &   ... & ... & ... & $14.58 \pm 0.22$ & $13.52 \pm 0.35$ & $11.00 \pm 0.10$ & $ 9.25 \pm 0.12$ & ... \\
IRS8 & 17:19:23.65 & -37:10:24.0 &  ... & ... & ... & $12.02 \pm 0.08$ & $ 9.91 \pm 0.04$ & $ 8.86 \pm 0.14$ & $ 8.10 \pm 0.10$ & ... \\
IRS9 & 17:19:23.88 & -37:09:30.8 & ... & ... & ... & $13.27 \pm 0.08$ & $13.07 \pm 0.16$ & $11.31 \pm 0.19$ & $ 9.69\pm  0.24$ & ... \\
IRS10 & 17:19:24.38 & -37:10:21.6 & ... & ... & ... & $13.76 \pm 0.29$ & $12.45 \pm 0.12$ & $ 9.83 \pm 0.13$ & $ 8.22 \pm 0.20$ & ... \\
IRS11 & 17:19:25.48 &	-37:10:05.4 & $14.02\pm0.05$ & $12.00\pm0.03$ & $11.17\pm0.03$ & $10.56\pm0.04$ & $10.64\pm0.06$ & $	10.31\pm0.06$ & $10.52\pm0.18$ & $2.56\pm0.13$ \\
IRS12 & 17:19:25.80 & -37:11:49.6 & ... & ... & ... &  $10.81 \pm 0.04$ & $10.14 \pm 0.05$ & $ 7.61 \pm 0.03$ & $ 5.98 \pm 0.03$ & $0.93\pm0.04$ \\
IRS13 & 17:19:25.95 & -37:10:06.6 & ... & ... & ... &  $13.79 \pm 0.13$ & $11.99 \pm 0.09$ & $10.02 \pm 0.05$ & $8.45 \pm 0.06$ & ... \\
IRS14 & 17:19:27.03 & -37:10:24.8 &   ... & ... & ... & $12.49 \pm 0.10$ & $10.47 \pm 0.14$ & $ 8.34 \pm 0.06$ & $ 6.71 \pm 0.06$ &  $0.60\pm0.19$  \\
IRS15 & 17:19:27.49 & -37:09:42.3 & ... & ... & ... & $12.30\pm0.05$ & $10.40 \pm 0.04$ &  $9.26 \pm 0.03$ & $ 8.77 \pm0.03$ & ...\\
IRS16 & 17:19:29.46 & -37:11:27.9 &  ... & ... & ... & $13.04\pm  0.10$ & $12.69 \pm 0.13$ & $11.18 \pm 0.10$ & $ 9.59 \pm 0.08$ &... \\
IRS17 & 17:19:29.98 & -37:11:27.5 &  $13.84 \pm 0.06$ & $12.51 \pm 0.05$ & $11.93 \pm 0.04$ & $11.42 \pm 0.05$ & $11.44 \pm 0.05$ & $10.73 \pm 0.09$ &  
$9.49  \pm0.13$ & ... \\
IRS18 & 17:19:30.11 & -37:11:37.2 & ... & ... & ... &  $13.47 \pm 0.13$ & $13.09 \pm 0.16$ & $10.32 \pm 0.07$ & $ 8.79 \pm 0.14$ & ...\\
IRS19 & 17:19:30.38 & -37:11:50.7 & $13.35 \pm 0.03$ & $12.72 \pm 0.03$ & $12.43 \pm 0.03$ & $12.04 \pm 0.05$ & $11.96 \pm 0.07$ & $11.87 \pm 0.14$ &  $8.56\pm  0.14 $ & ... \\
IRS20 & 17:19:32.76 & -37:10:47.0 & ... & $13.45\pm0.04$ & $11.43\pm0.04$ & $8.25\pm 0.01$ & $7.21  \pm0.01$ &  $5.05 \pm 0.01$ & $ 3.25 \pm 0.01$ & ...\\
IRS21 & 17:19:34.72 & -37:10:49.9 & ... & ... & ... & $12.71\pm 0.07$ & $12.24  \pm0.15$ &  $9.33 \pm 0.07$ & $ 7.85 \pm 0.08$ & ...\\
IRS22 & 17:19:35.63 & -37:10:56.7 &   ... & ... & ... & $11.23\pm 0.04$ &	$9.92 \pm 0.02$ & $ 8.31 \pm 0.04$ & $ 6.81 \pm 0.04$ & $0.87\pm0.03$\\ 
IRS23 & 17:19:36.00 & -37:10:54.4 & ... & ... & ... & $13.57 \pm 0.15$ & $11.93 \pm 0.28$ & $9.47 \pm 0.10$ & $ 7.75 \pm 0.16$ & ...\\
IRS24 & 17:19:36.27 & -37:10:52.3 & ... & ... & ... & $13.59 \pm 0.17$ & $12.86 \pm 0.14$ & $10.33 \pm 0.10$ & $ 8.51 \pm 0.20$ & ...\\ \hline
IRSL1 & 17:19:22.53 & -37:10:17.8 &   ... & ... & ... & $13.90\pm 0.07$ &	$13.68 \pm 0.15$ & ... & $ 9.56 \pm 0.35$ & ...\\ 
IRSL2 & 17:19:26.38 &-37:12:25.3 & ... & ... & ... & $9.35 \pm 0.14$ & $9.56 \pm 0.07$ & ... & $ 9.31 \pm 0.03$ & ...\\
IRSL3 & 17:19:35.51 & -37:10:56.4 & ... & ... & ... & ... & $10.74 \pm 0.34$ & $9.13 \pm 0.09$ & $ 7.52 \pm 0.22$ & ...\\
\hline
\end{tabular}
\label{yso}
\end{center}
\end{table*}


\par In addition, we searched for  24~$\mu$m point sources in this region from the catalog of \cite{2015AJ....149...64G}. This catalog also identifies IRAC-GLIMPSE and 2MASS counterparts to the 24~$\mu$m point sources. 24~$\mu$m point sources towards young star forming regions represent active star formation sites where the emission is mostly the result of dust heated due to accretion from the core on to the central protostar \citep{2010ApJ...715..310R}. We find nine sources in this region, of which seven have IRAC counterparts. However, only four of these: IRS1, IRS11, IRS12 and IRS14, are detected in all the four bands. To understand the nature of these sources, we have also constructed a color-color diagram based on colors from both IRAC and MIPS bands: [8.0] - [24] versus [3.6] - [5.8]. These colors have been used to locate young protostars such as Class I/0 and Class II \citep{2006ApJ...643..965R}. The four sources detected in IRAC and 24~$\mu$m bands are represented as open circles in the color-color diagram in Fig.~\ref{24cc}. Two sources: IRS12 and IRS14 are classified as Class I protostars while IRS1 belongs to the Class II category. IRS1 and IRS11 lie in the region occupied by later class / field stars according to the IRAC color-color diagram, see Fig.~\ref{cc}(b). While the classification and color space occupied by a given class is not stringent as there is a variation in properties leading from one class of objects to another, one cannot rule out the possibility of spurious association considering that this region is overpopulated (being located towards the inner Galactic midplane). 


\begin{table}[hbt!]
\footnotesize
\caption{The association of radio sources with YSOs.}
\begin{center}
\hspace*{-0.5cm}
\begin{tabular}{c c c } \hline \hline
Radio &IRAC &MIPS \\
source& YSO&24~$\mu$m source\\
\hline
S1&\text{\sffamily X} &\text{\sffamily X}  \\
S2& \text{\sffamily X}&\text{\sffamily X}   \\
S3& \text{\sffamily X}& \text{\sffamily X} \\
S4 & \text{\sffamily X}&\text{\sffamily X}  \\
S5 &IRS14  &\checkmark   \\
S6 &IRS20  & \text{\sffamily X}\\
S7 &IRS22, IRS23, IRSL3  &\checkmark \\
S8 &\text{\sffamily X} &\text{\sffamily X}  \\
S9 &\text{\sffamily X} &\text{\sffamily X}  \\
S10 &\text{\sffamily X} &\text{\sffamily X}\\
S11 &IRS11  &\checkmark   \\
S12 &IRS2 &\text{\sffamily X} \\
 
\hline 
\end{tabular}
\\ 
\label{rads}
\checkmark - Detection; \text{\sffamily X} - Non detection\\
\end{center}
\end{table}


\par
\begin{figure*}[hbt!]
\begin{center}
\includegraphics[width=10.5cm]{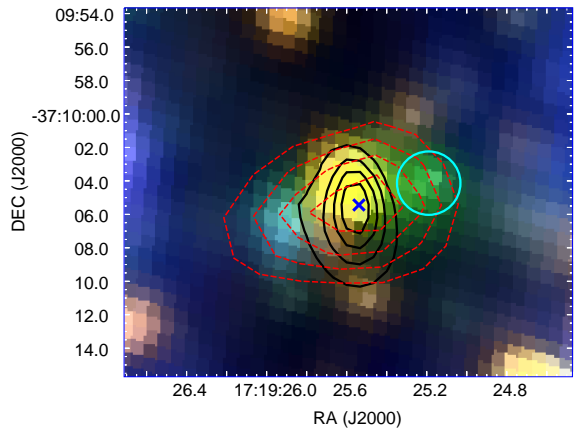}
 \caption{IRAC color-composite image including 8~$\mu$m (red), 4.5~$\mu$m (green) and 3.6~$\mu$m (blue), associated with the young stellar object IRS11 (indicated with a cross). Overlaid on the image are 1280~MHz radio emission depicted in black solid line contours and MIPS 24~$\mu$m emission in red dashed line contours. The contour levels are at 2.3, 3.5, 4.7, 6.0 and 7.2~mJy/beam for 1280~MHz and 350, 412.5, 475, 537.5 and 600~MJy/sr for 24~$\mu$m respectively. The location of green knot is encircled in cyan.}
 \label{ego}
 \end{center}
\end{figure*}

 IRS11 displays colors characteristic of later class / field stars in the IRAC color-color diagram and the fluxes decrease steadily from 3.6 to 8.0~$\mu$m. However, it has a large 24~$\mu$m excess in the color [8.0] - [24] = 7.9. This is unusually large and one cannot rule out a false association in this region due to population overdensity. However, sources with such extreme colors have been observed in young clusters of Elephant Trunk Nebula \citep{2004ApJS..154..385R}, Trifid Nebula \citep{2006ApJ...643..965R} and NGC~2467 \citep{2009ApJ...700..506S}. These colors have been attributed to 
debris disk, or massive disk-like emission with inner holes \citep{2004ApJS..154..385R}.  In some cases, extended emission associated with the 24~$\mu$m source is likely to give such extreme colors. A visual inspection of the IRAC color-composite image shows the presence of excess 4.5~$\mu$m emission towards the west of IRS11 in the color-composite image. This is reminiscent of extended green objects or green-fuzzies detected in other star forming regions \citep{{2009ApJS..181..360C},{2012ApJ...748....8T}}. Such objects are characterized by their extended green emission in the IRAC three-color composite images. Their angular sizes vary from a few to 30$\arcsec$ and are believed to be associated with massive YSOs \citep{2009ApJ...702.1615C}. A search in the Glimpse II Spring'08 Archive towards this knot yields a point-like source with emission detected solely at 4.5~$\mu$m. We also carefully checked for this knot of emission in the 3.6, 5.8 and 8.0~$\mu$m images but we do not find any emission here. The location of this green knot is encircled in Fig.~\ref{ego}. There is a likelihood that IRS11 is a young massive protostellar source with outflow signatures. A high resolution image at 24~$\mu$m is required to pin-point the exact location of the YSO.

\par Considering that the 24~$\mu$m emission indicates an early evolutionary stage of YSOs, we searched for 24~$\mu$m point source counterparts to the radio sources. For this, we considered a search radius of $\mathbf{5''}$ from the radio peak positions. This search radius is selected keeping in view the beam size of the 1280 MHz radio image. We find three 24~$\mu$m sources corresponding to S5, S7 and S11. The 24~$\mu$m counterparts are IRS14, IRS22 and IRS11, respectively.  Analogous to 24~$\mu$m, we searched for GLIMPSE counterparts to the radio compact sources. Only S5 and S11 have infrared counterparts detected in all the four IRAC bands.  
S7 has 3 associated mid-infrared GLIMPSE sources at distances $0.9''$, $2.3''$ and $4.1''$ from the peak emission. The mid-infrared source located closest to S7 is detected solely at 3.6 and 5.8~$\mu$m bands. The other two sources correspond to IRSL3 and IRS23, respectively.
In addition, we noticed that there is a bright mid-infrared source associated with S6 in the images that has no IRAC detection but has a 2MASS counterpart. To obtain photometric fluxes of  the mid-infrared counterparts to S6 and the source closest to S7 in all the IRAC bands, we carried out aperture photometry with a diameter of $6''$ (using the task {\tt qphot} in IRAF) centered on the source and subtracted the background emission estimated from an annulus within diameters $8.5''$ and $12''$. The magnitudes of the sources corresponding to S6 and S7 are listed in Table~\ref{yso} as IRS20 and IRS22, respectively. Both these sources classify as Class I in Fig.~\ref{cc}~(a). While IRS22 is also classifed as Class I in Fig.~\ref{cc}~(b), IRS20 is saturated at 24~$\mu$m. Further, we observe that  S12 has a bright mid-infrared counterpart that is not listed as a source in the GLIMPSE Archive. This could be due to the extended nature of this object. We extracted the flux densities for this source in the IRAC bands using an aperture of diameter  $10''$ after subtraction of background from an annulus between diameters $16''$ and $26''$. Those magnitudes are listed under the label IRS2 in Table~\ref{yso}. Table~\ref{rads} lists the radio sources and the associated YSO(s).

\par The YSOs are all located away from the central complex A where the gas and dust emission is nebulous leading to a difficulty in identification and extraction of point-like sources. Besides, we have only classified sources based on detection in IRAC bands as well as the 24~$\mu$m band. This would imply that the census of YSOs towards this region is not complete and consequently, should be considered as a subsample of YSOs in this region. 

\subsubsection{SED Models}

In order to learn about the evolutionary stages of the YSOs, we have constructed their near and mid-infrared SEDs and fitted them with the radiative transfer models of \citet{2007ApJS..169..328R}.
The models assume an accretion scenario for the star formation process where a central star is surrounded by an accretion disk, an infalling flattened envelope and presence of bipolar cavities. We have used the command-line version of the SED fitting tool where a large number of precomputed models are available. The SEDs are constructed using wavelengths ranging from $1.25-24$~$\mu$m (see Table~\ref{yso} for fluxes upto 24~$\mu$m). Where the 24~$\mu$m point source counterpart is not detected, we have assumed an upper limit to the 24~$\mu$m flux density by carrying out aperture photometry for the parameters given by \citet{2015AJ....149...64G}.  For sources saturated in the 24~$\mu$m image, we used the flux density corresponding to the saturation magnitude of 0.5 \citep{2015AJ....149...64G} as the lower limit. Few sources appear point-like at 70~$\mu$m, such as those associated with IRS14, IRS20 and IRS22. For these, we have estimated fluxes within apertures of size 12$\arcsec$ after subtracting the background from a nearby emission free region. 

Unlike the 24~$\mu$m image, the 70~$\mu$m image shows point sources as elongated, possibly due to different scan speeds in two directions 
\citep{2015MNRAS.449.2784O}. However, the aperture size is selected such that the elongated beam is included within the aperture. For the other YSOs, we assume the 70~$\mu$m clump flux density as the upper limit as it is difficult to isolate flux density corresponding to the source.

For wavelength bands at 160~$\mu$m and longer, we used the associated clump fluxes as upper limits for constraining the models as the resolution is poor.

\begin{figure*}[hbt!]
\begin{center}

\hspace*{-1.6cm}
\includegraphics[width=60mm]{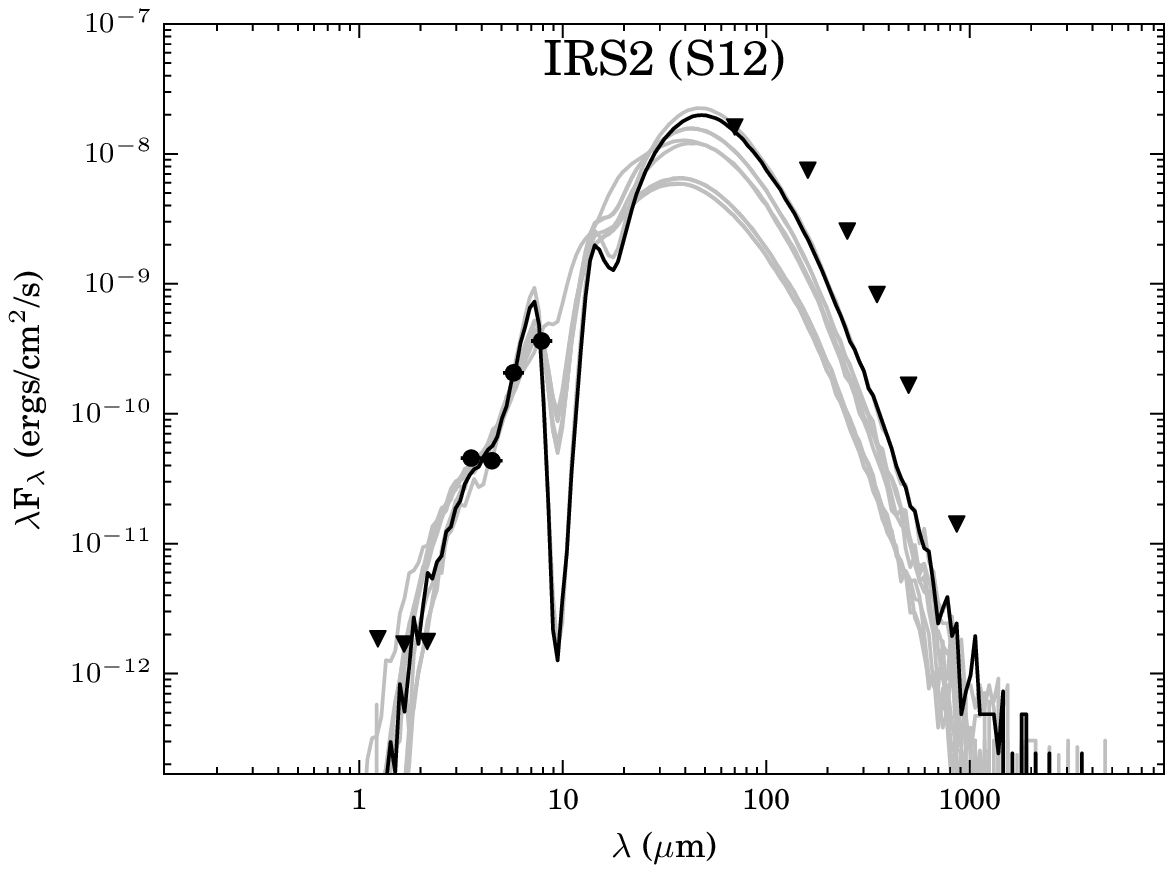} 
\hspace*{-0.2cm}\includegraphics[width=60mm]{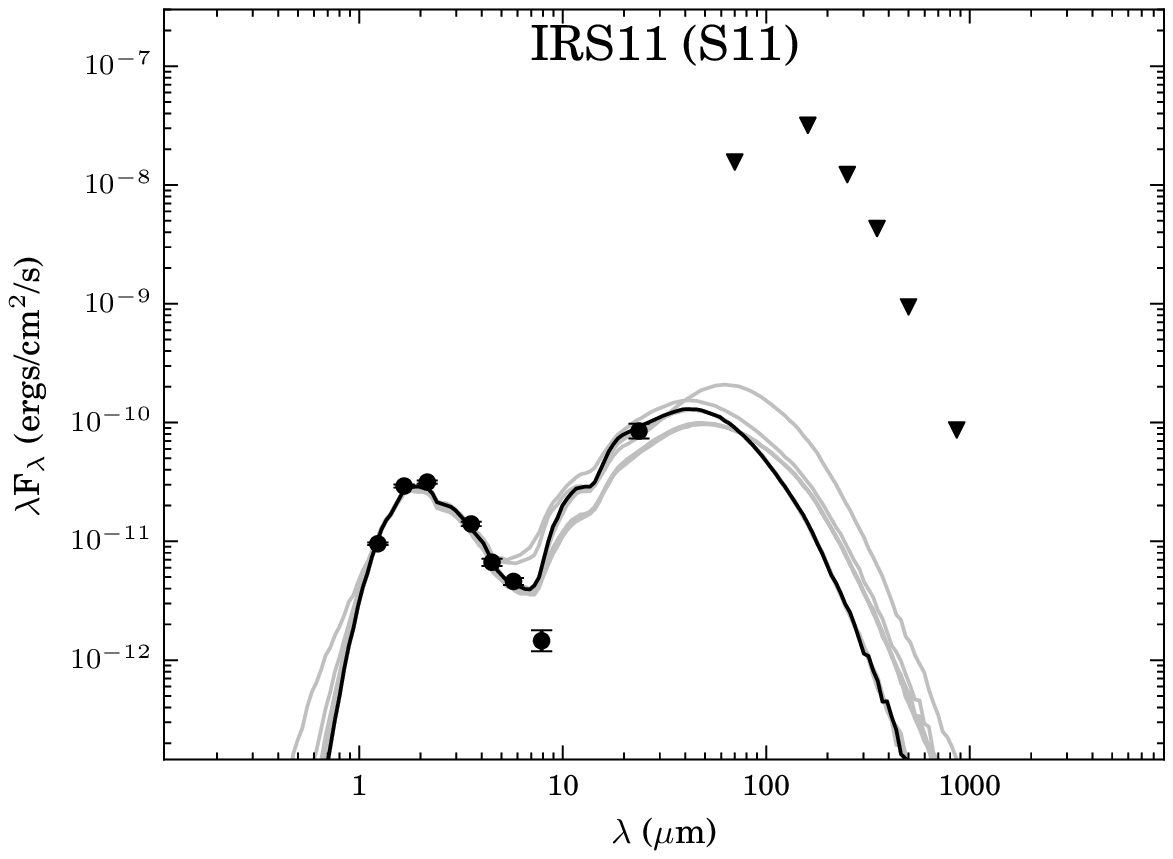}
\hspace*{-0.2cm}\includegraphics[width=60mm]{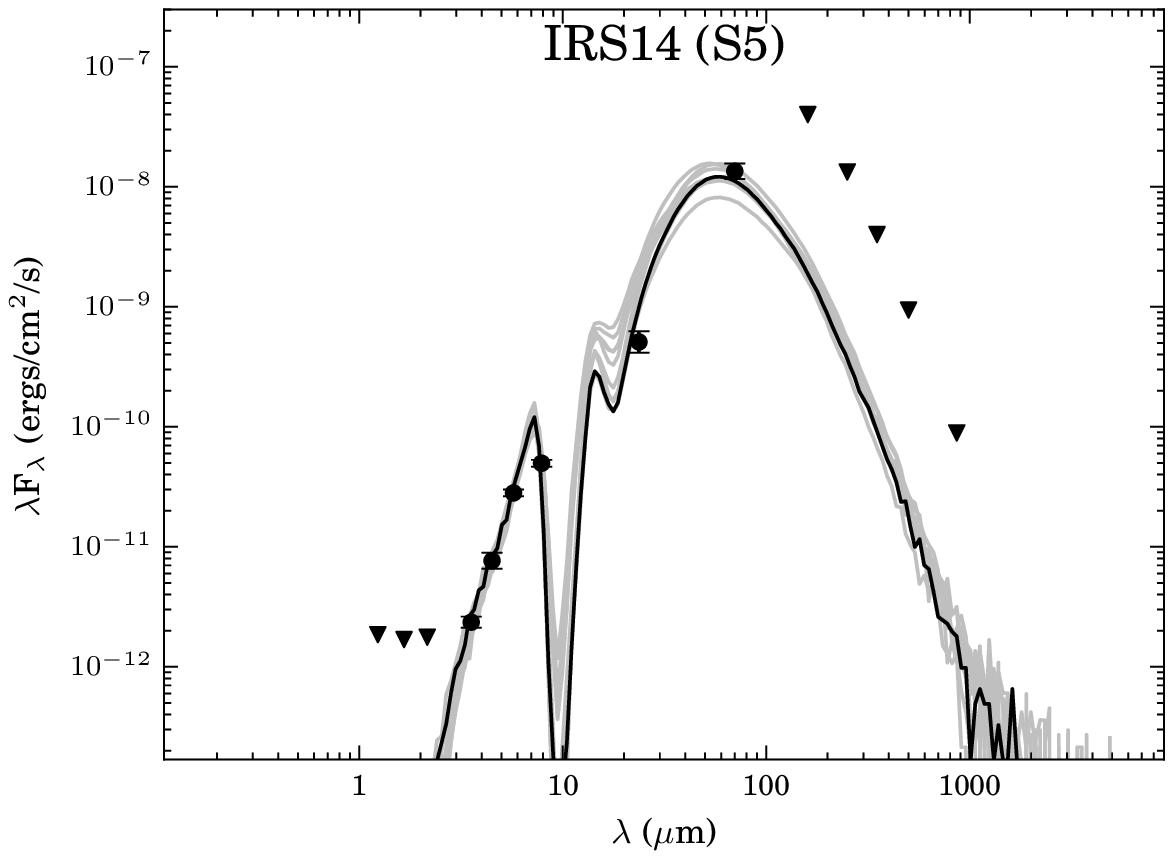} 
\hspace*{-0.2cm}\includegraphics[width=60mm]{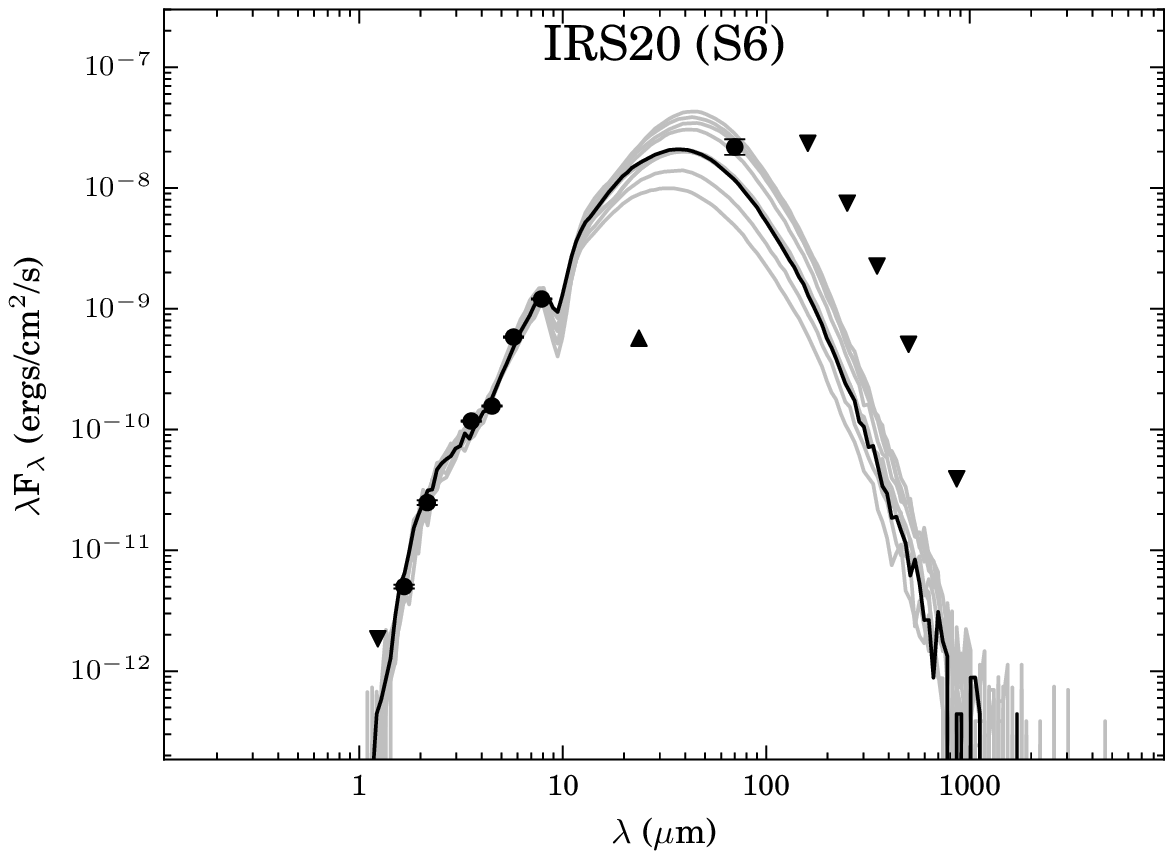}
 \hspace*{-0.2cm}\includegraphics[width=60mm]{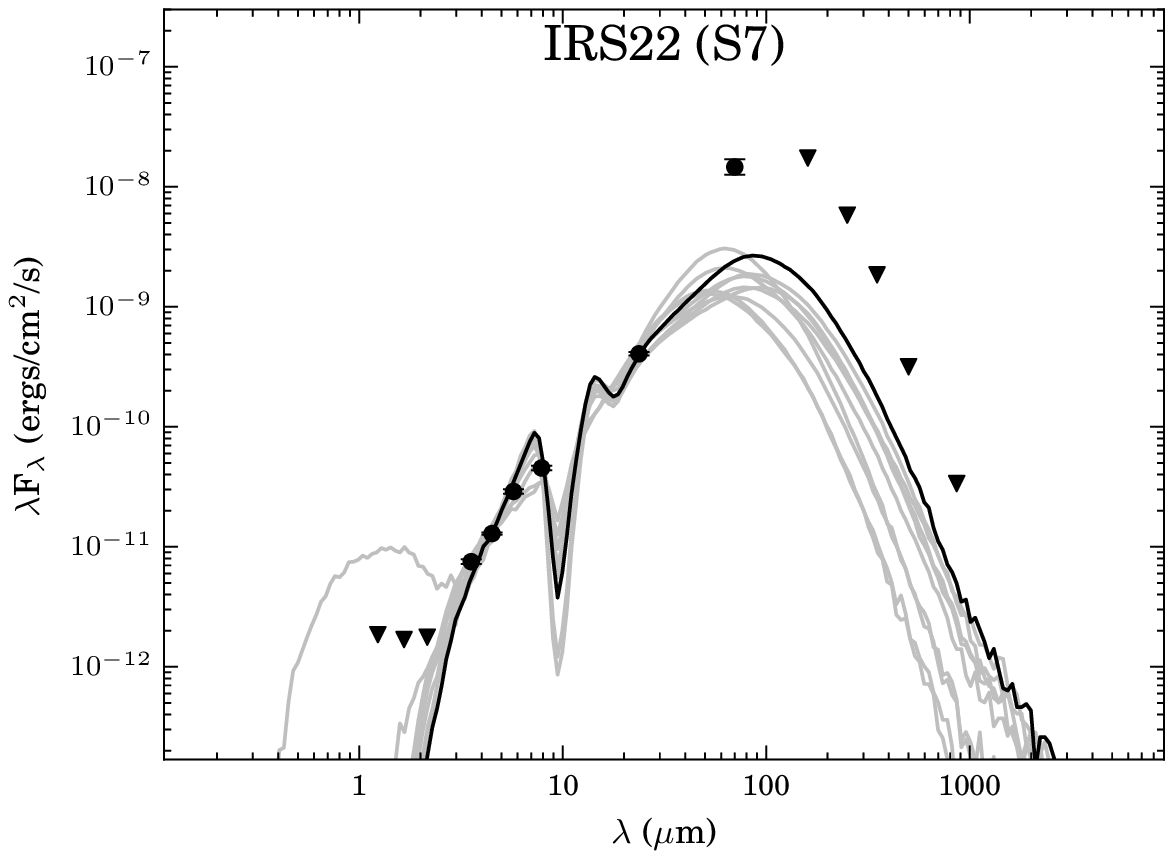}

\caption{The infrared SEDs of IRS2, IRS11, IRS14, IRS20 and IRS22 fitted with the models of Robitaille et al. (2007). The filled circles represent the input fluxes while triangles represent upper limits from the respective clumps. The black line shows the best fitting model, and grey lines show the subsequent nine good fits.}
\label{rwiw}
\end{center}
\end{figure*}


\par The inputs to the model include distance to the source, approximate visual extinction  (A$_V$) and the magnitude in each waveband. The approximate  A$_V$ is estimated by using the column density values at the locations of these sources in the column density map. For converting column density to A$_V$ in magnitudes, we apply the following relation: $\mathrm{N(H_2)} = \mathrm{A}_V\times1.1\times10^{21}$~cm$^{-2}$ \citep{2009MNRAS.400.2050G}. Here, we present the results for IRS2, IRS11, IRS14, IRS20 and IRS22 as they are positionally coincident with the radio compact sources S12, S11, S5, S6 and S7, respectively (see Table~\ref{rads}).  We consider ten best fit models for each source and these  are presented in Fig.~\ref{rwiw}. The range of model parameters are listed in Table~\ref{rwiw_tb}.  The best fit models estimate the mass range as $6.8-29.4$~M$_{\odot}$, with age $\le 0.2$~Myr emphasizing the youth of these sources. This highlights the fact that these are massive objects. The best fit model obtained for IRS20 has a large effective temperature of the host object, $36200$~K. For sources (other than IRS11), the models indicate high effective temperatures of $\sim30000$~K. This is vindicated by the existence of compact radio emission seen towards these sources. IRS11 has a lower effective temperature of few thousand kelvin, suggesting a protostellar evolutionary phase for S7. S7 is also relatively isolated with lack of diffuse emission in its vicinity implying an earlier evolutionary phase. 

\par We have also fit models to all the other YSOs using 2MASS, IRAC and 24~$\mu$m flux densities, or upper limits. For longer wavlength bands, we used the clump flux densities as upper limits. The models are shown and listed in Appendix A. According to the best-fit models, IRS1, IRS7 and IRS13 are evolved (age $\ge1.9$ Myr) compared to the other YSOs. It is important to realise that although the models are consistent with observed flux densities, the results must be treated with caution. An important caveat is the absence of PAH or small-grain continuum emission, which forms an important source of emission to mid-infrared fluxes  in YSOs with hot stellar sources \citep{2006ApJS..167..256R}. This is significant in our case as the SEDs of all the YSOs are constructed mainly using IRAC and 24~$\mu$m fluxes. Another
primary assumption of these models is that the SEDs of the massive YSOs are scaled-up versions of their lower mass counterparts and the massive star formation scenarios have not been incorporated in them \citep{2008ASPC..387..290R}.  This also explains why YSOs associated with radio sources  (other than IRS20) show low temperatures for best fit models in spite of the presence of ionised gas emission.

\begin{table*}[hbt!]
\caption{Parameters of the models shown in Fig.~\ref{rwiw}. Col. 4 -- 9 give  Mass, Effective Temperature, Luminosity, Inclination angle, Envelope accretion rate, Disk Mass, Extinction and Age, respectively. The parameters listed are for the best fit model and the range of all the ten best fits.}
\scriptsize
\begin{center}
\vskip 0.4cm
\label{spec}
\hspace*{-1.4cm}
\setlength{\tabcolsep}{2pt}
\begin{tabular}{c c c c c c c c c c c} \hline \hline \\

Source &&$\chi^2$ &Mass  &T$_{eff}$  &Luminosity  &Inc. angle & Env. accretion rate  &Disk mass  & A$_V$   & Age\\
 & &&(M$_{\odot}$) &(K) &(L$_{\odot}$) & (Deg.) &(M$_{\odot}$/yr) &(M$_{\odot}$)  & (mag) &(Myr)\\
\hline \hline \\
\multirow{ 2}{*}{IRS2 (S12)} & Best fit &1091 & 29.4 & 4288 & 30040 & 31.8 &8.9$\times$10$^{-4}$  & 0 & 2.3 &0.001\\ 
&Range &1091-1467& 16.1-29.4 &4288-33320&15800-30040 &31.8-87.1 &  $1.3$-$8.9\times10^{-4}$ & 0-0.1 & 0-20.0 &0.001-0.3\\ 
\hline  \\

\multirow{ 2}{*}{IRS11 (S11)} & Best fit &123.4 & 6.8 & 4649 & 321.6 & 18.2 &4.9$\times$10$^{-5}$  & 1.2$\times$10$^{-3}$ & 10.9 &0.08\\ 
&Range &123.4-231.4& 6.4-6.8 &4462-4720&260.5-342.6 &18.2-49.5 &  $0.4$-$5.6\times10^{-4}$ & $0.9$-$3.9\times10^{-3}$ & 3.1-10.9 &0.03-0.1\\ 
\hline \\

\multirow{ 2}{*}{IRS14 (S5)} & Best fit &49.7 & 25.4 & 4252 & 20200 & 31.8 &1.3$\times$10$^{-3}$  & 0 & 41.8 &0.002 \\ 
&Range &49.7-57.8& 17.6-28.7 &4240-34260 & 14890-36740 &31.8-87.1 &  $0.7$-$1.3\times10^{-3}$ &0-0.1 &22.8-77.1 &0.001-0.09\\ 
\hline \\

\multirow{ 2}{*}{IRS20 (S6)} & Best fit &911.8 & 21.4 & 36200 & 54160 & 87.1 &2.9$\times$10$^{-4}$ & 7.6$\times$10$^{-4}$ & 10.2 &0.2 \\ 
&Range & 911.8-1471 & 15.9-26.3 & 16320-38940 & 30060-91980 & 75.5-87.1 & $1.1$-$8.7\times10^{-4}$ & $7.6\times10^{-4}$-$ 1.3$ & 0-20.9 & 0.02-0.3 \\
\hline \\

\multirow{ 2}{*}{IRS22 (S7)} & Best fit &248.6 & 12.5 & 4493 & 3563 & 18.2 &3.0$\times$10$^{-3}$ & 0.3 & 52.2 &0.006 \\ 
&Range &248.6-381.4 & 8.1-12.5 &4151-25050 &1608-5248 &18.2-81.4 & $0$-$3.9\times10^{-3}$ &$0.003$-$0.8$ &2.4-55.0 &0.002-0.7\\
\hline \hline \\

\end{tabular}
\label{rwiw_tb}
\end{center}
\end{table*}



\section{Star Formation Activity}

\begin{figure}[hbt!]
\begin{center}
\includegraphics[width=85mm]{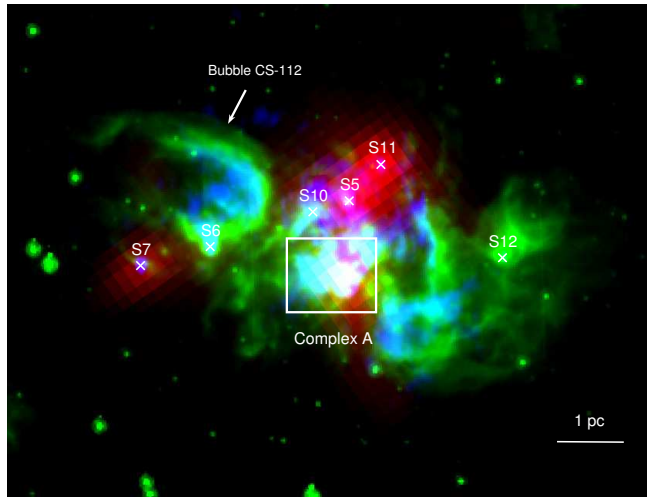}
\caption{Color composite image of the IRAS~17160--3707 region. The ionised gas emission at 1280~MHz is shown in blue, the IRAC~8~$\mu$m emission in green and the ATLASGAL 870~$\mu$m cold dust emission in red. The locations of radio sources are also marked in the image. North is upwards while East is to the left. }
\label{colcom}
\end{center}
\end{figure}

The ionised gas distribution as well as the warm and cold dust emission towards the entire \hii~region complex is shown as a color composite image in Fig.\ref{colcom}. The ionised gas distribution is shown in blue, the warm dust emission at 8~$\mu$m in green and cold dust emission at 870~$\mu$m in red. The radio emission traces the warm dust morphology to a large extent. However, we also see regions of ionised gas that are devoid of 8~$\mu$m emission, for instance towards the interior of CS-112 and towards the west and south-west of complex A. Prominent cold dust emission is seen towards the north and east, enveloping S5, S11 and S7. Based on higher radio frequency information, S11 is categorised as an ultracompact \hii~region, while S5, S6 and S7 are \hii~regions that are compact and resolved \citep{2003A&A...407..957M}. Warm dust emission with little contribution from cold dust emission is seen towards S6 and S12.

\subsection{Morphology of gas and dust}

The overall morphology of the \hii~region in the evolved phase depends on factors such as density inhomogeneities in the molecular cloud,  age as well as dynamics of the ionised and molecular gas. The density gradient structure can be explained using the champagne flow models, wherein the shape of  a \hii~region is
 due to the density gradient of the molecular material in the vicinity of the newly formed massive star \citep{1979A&A....71...59T}. \cite{2001ApJ...549..979K} have explained the extended emission around ultracompact \hii~regions by combining the champagne flow
model with the hierarchical structure of molecular clouds. In the earliest stages of ionisation by massive star(s), such as those associated with hyper and ultracompact \hii~regions, simulations have shown that other dominant factors such as outflow dynamics,  rate of accretion and gravitational instabilities could also shape the ionised gas distribution and evolution \citep{{2007ApJ...666..976K}, {peters2010understanding}, {2011IAUS..270..107K}}. Towards the IRAS~17160--3707 region, we see a number of compact radio sources within extended emission that is seen prominently at 610 MHz. The compact source S7 is relatively isolated and we have not detected any extended emission associated with it. There are other arc and filamentary structures observed in this region, e.g. towards the north-east and west of complex A. The proximity and possible interaction of these compact sources with each other give rise to the overall morphology of IRAS~17160--3707.

\subsubsection{Complex A}

Complex A spanning $\sim1.2$~pc is the brightest region in IRAS~17160--3707 at mid and far-infrared wavelengths extending from 8 to 250~$\mu$m. The infrared and radio nebulosity towards this region indicates the presence of a massive embedded cluster based on morphological evidences. In Fig.\ref{colcom}, we see that the 8~$\mu$m mid-infrared emission is seen away from the cold dust emission peak suggesting a region of high extinction. In the absence of higher resolution mid-infrared images, it is difficult to ascertain if the six radio compact sources in complex A possess an intrinsic ionising source or are externally ionised. Such externally ionised shells on ultracompact \hii~region scales may be produced around turbulent self-gravitating cores when expanding \hii~regions encounter such cores \citep{2007ApJ...668..980M}.  Unlike our radio maps where complex A comprises of compact sources, at higher frequencies, only fragmented emission is seen prompting \cite{2003A&A...407..957M} to label it as a classical \hii~region. We differ with these authors as it is possible that the emission from the compact sources is resolved out at high frequencies due to the interferometric filtering. In addition, this region is near the peak of clump C4 implying that it suffers large extinction as the associated nebulosity is neither seen in the optical~DSS nor in the 2MASS near-infrared images. Our view is that complex~A comprises multiple compact \hii~regions that are prior to the classical expansion phase. 

\subsubsection{Radio Compact Sources}

The radio compact sources S5, S6, S7, and S11 have associated warm and cold dust counterparts implying the presence of  embedded YSO(s) that have not yet shedded their dust cocoons. This is also evident from the SED models discussed in Section 3.4, where the age obtained from the best fit model of the massive embedded objects is 0.2~Myr or less. The radio luminosities indicate  ZAMS spectral types of B0 - O9.5 for S5 and S6, and B0.5 - B0 for S7 and S11. S12 has weak radio flux density indicating a later ZAMS spectral type of B1 - B0.5 of the intrinsic stellar object \citep{1973AJ.....78..929P}. However, it is relatively bright in mid-infrared (unlike radio) when compared to the other sources in this region. The 8~$\mu$m emission shows a bright emission but at shorter wavelengths, only diffuse emission is discernible implying relatively high extinction towards the embedded object.  An arc-shaped emission is seen to the south-east of S10 in mid-infrared as well as ionised gas emission. Such mid-infrared arc-shaped features have been observed toward clusters of massive stars and could be attributed to (i) expansion of an \hii~region, (ii) bow-shocks due to high velocity stars \citep{2012A&A...538A.108P} or (iii) dust or bow wave \citep{2014A&A...563A..65O}.

\subsubsection{Bubble CS-112}

\begin{figure}[hbt!]
\begin{center}
\vspace{0.5cm}
\hspace*{-1.5cm}
\includegraphics[width = 65mm,bb=-30 -30 590 590]{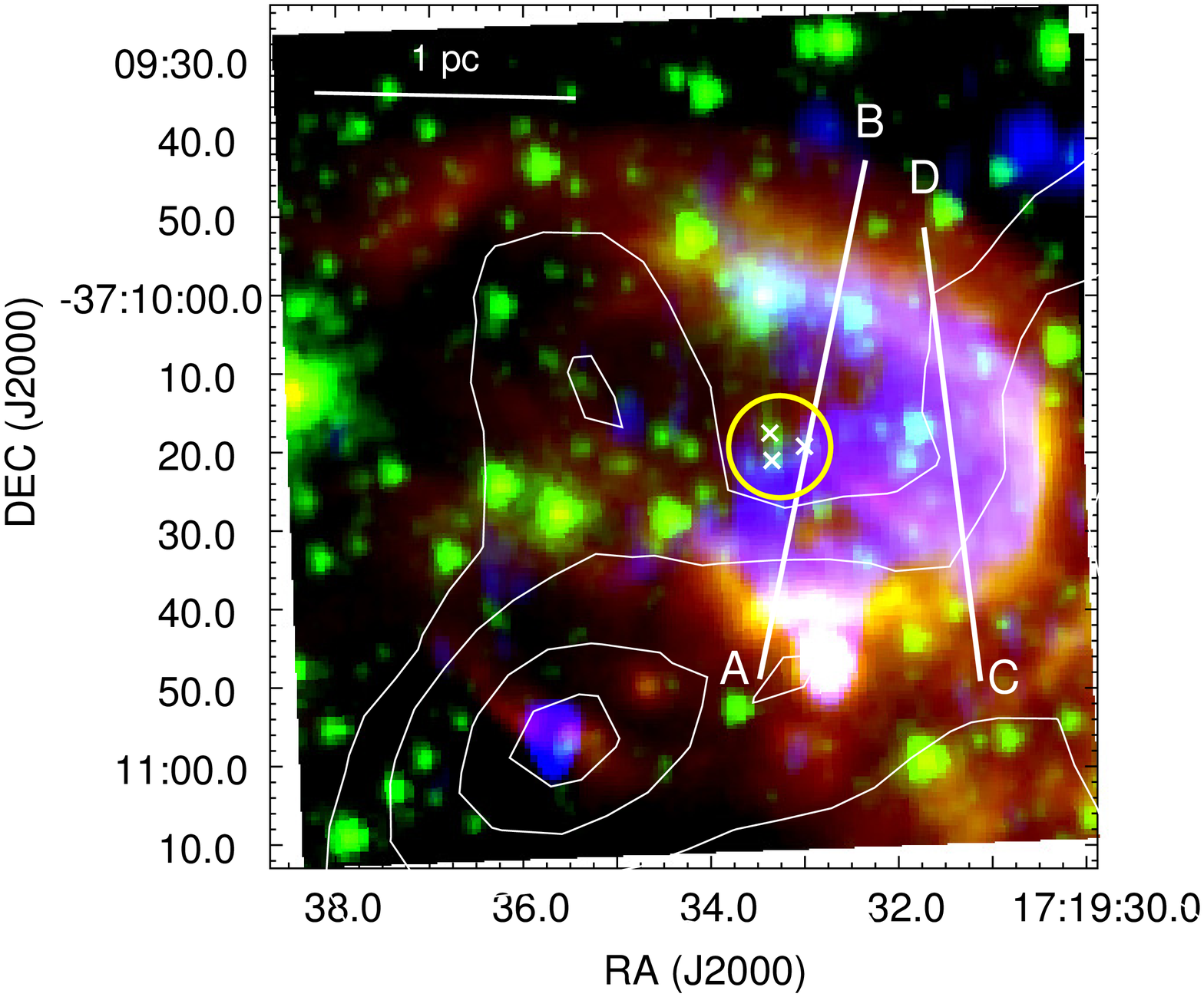}
\hspace*{0.5cm}
\includegraphics[width=85mm]{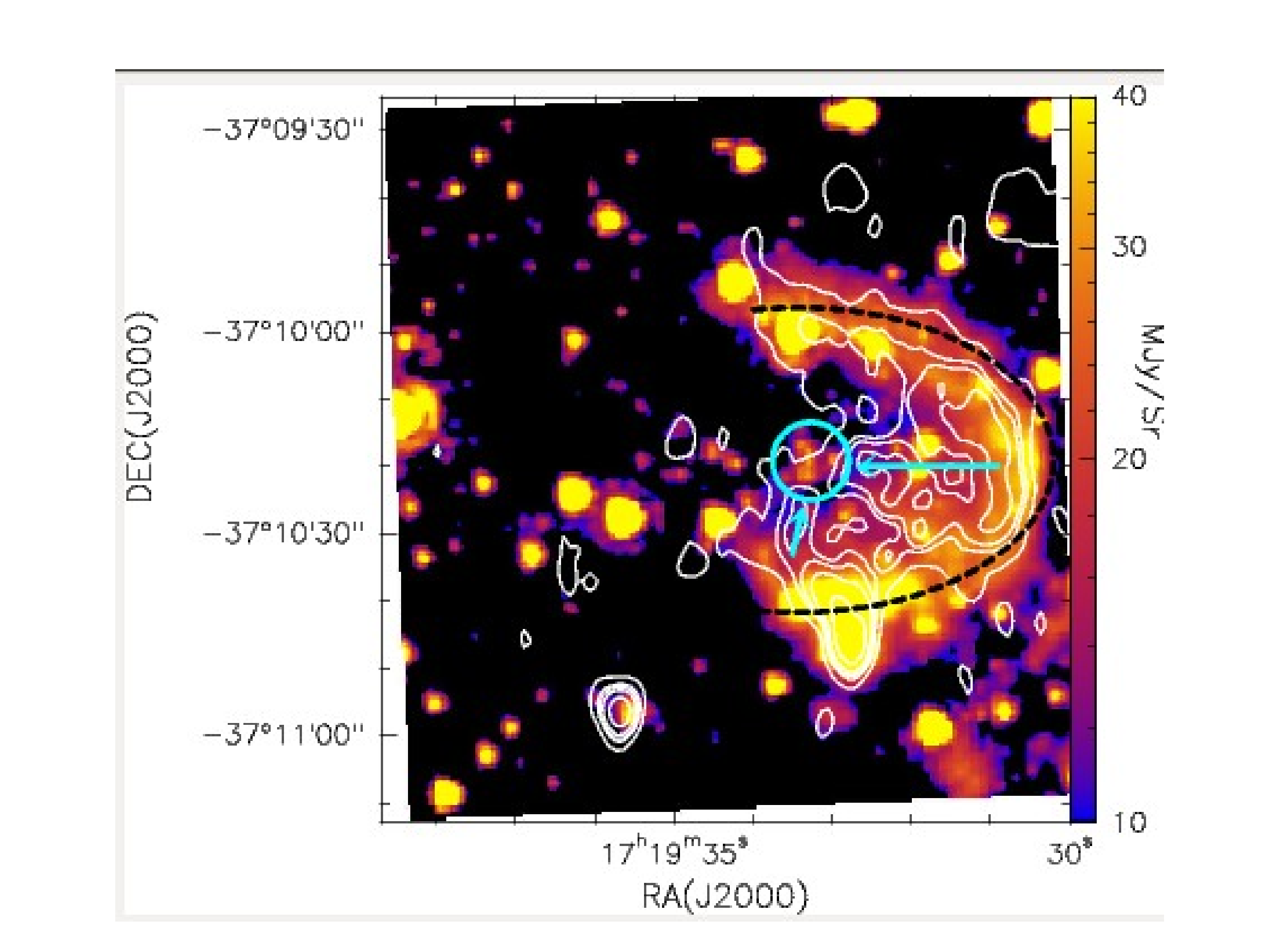}
\caption{(Left) Color composite image of the bubble CS-112, with blue representing the ionised gas emission at 1280~MHz, green showing the IRAC~3.6~$\mu$m emission and red depicting the 8~$\mu$m emission. The contours represent 870~$\mu$m cold dust emission, with contour levels at 0.70, 1.25, 1.92, 2.60, and 3.70~Jy/beam. The likely exciting group of stars is encircled in yellow. AB and CD represent the lines along which intensity profiles are constructed and presented later. (Right) 4.5~$\mu$m image of CS-112 with 1280 MHz radio contours overlaid. The dashed partial ellipse traces the bubble structure. The arrows depict the protruding ionising filaments pointing towards the central group of encircled stars. }
\label{bubble}
\end{center}
\end{figure}

We next attempt to understand the bubble-like morphology of CS-112 seen in ionised gas and warm dust emission.  This bubble has a partially
 broken morphology and is oriented in the east-west direction, with the radio source S6 lying to the south of the bubble. We estimate the extent of the bubble by considering its size along the N-S direction close to the location of S6. The size of the bubble is approximately 2 pc with a  thickness 0.3~pc in the 8~$\mu$m image. A color composite image of CS-112 showing the distribution of mid-infrared, radio and submillimeter emission is presented in Fig.~\ref{bubble}. The region associated with CS-112 is saturated at 24~$\mu$m. 

\begin{figure}[hbt!]
\begin{center}
\hspace{-1.0cm}
\includegraphics[scale=0.40]{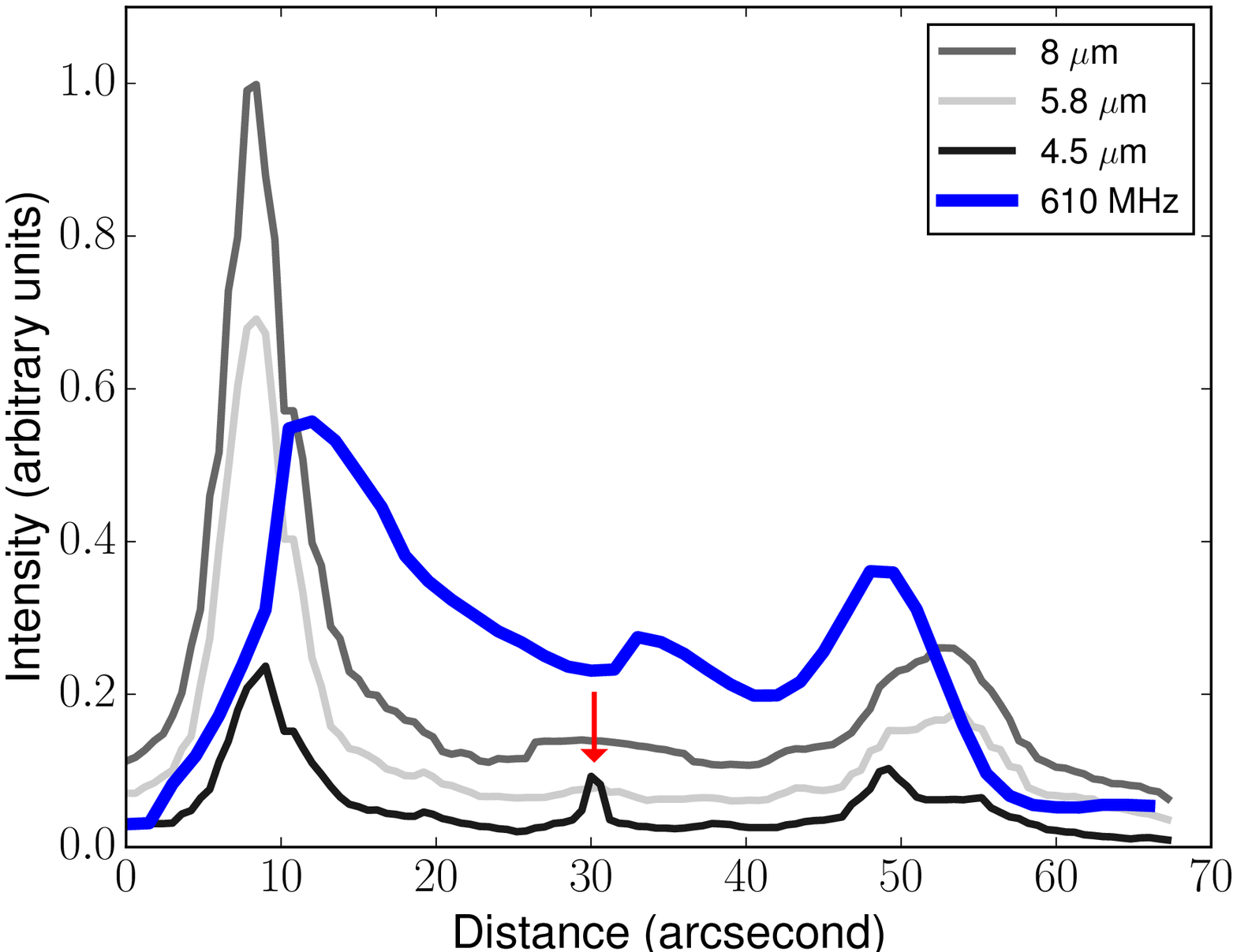}
\hspace*{-0.5cm}
\includegraphics[scale=0.40]{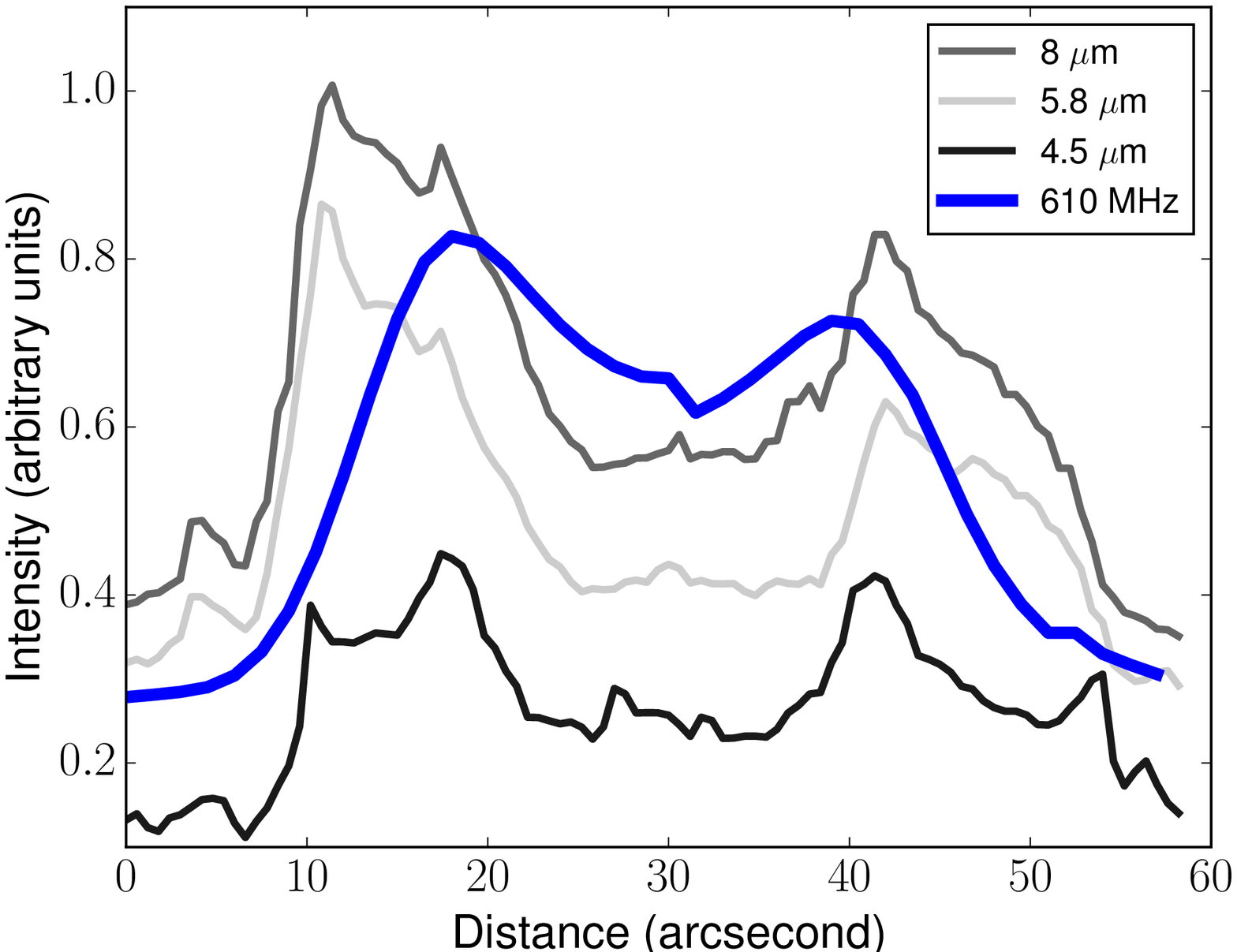}
\caption{Intensity profiles of the bubbles CS-112 along the lines AB (left) and CD (right) shown in Fig.~\ref{bubble}. The thin grey and black lines represent mid-infrared intensity profiles while the thick (blue) line represents the radio intensity profile at 610 MHz. The profiles have been shifted and scaled by arbitrary factors for better viewing. The arrow in the intensity profile along AB indicates the position of the central cluster at 4.5~$\mu$m.}
\label{bubprof}
\end{center}
\end{figure}

The bubble morphology is demonstrated by using intensity profiles along the lines shown in Fig.~\ref{bubble}. We see an emission enhancement in the bubble in the vicinity of S6. To establish that this is the case, we consider two lines through the bubble: (i) AB passing through the emission enhancement near S6, and (ii) CD passing through another portion of the bubble. Both the lines are selected such that the contribution from stellar sources near the bubble edges is negligible. The intensity profiles at 4.5, 5.8 and 8.0~$\mu$m are shown in Fig.~\ref{bubprof}. Also shown are the emission profiles from ionised gas emission at 610 MHz. As we are interested solely in the relative intensities along the edges of the bubble, the profiles are scaled and shifted by arbitrary factors for clarity. From this figure, it is evident that the edge near S6 (i.e near point A in the line AB) is brighter than the bubble rim near B. This is unlike the intensity profile through the line CD where the emission contrast across the bubble edges is lower. We believe that the bubble rim near S6 is brighter in both radio and mid-infrared due to the gas from S6 filling the bubble region bereft of molecular gas leading to higher intensity in the rim. In other words, the ionised gas expansion and leakage of UV photons from S6 would occur more easily in a region where ionised gas (due to the bubble) is already present. The excitation of PAHs towards the bubble rim near S6 could be because of leakage of UV photons from the latter.   This would inherently imply that the bubble lies at the same heliocentric distance as S6.

A comparison of radio and mid-infrared emission reveals that unlike the mid-infrared emission seen at the periphery of the bubble, the ionised gas occupies the interior and inner rim of the bubble. 
This can be understood in terms of excitation of PAHs in the PDR, beyond the ionisation front of the expanding \hii~region of the bubble \citep{2008ApJ...681.1341W}. Similar morphologies have been observed towards numerous other bubbles \citep[e.g.][]{{2010A&A...523A...6D},{2012A&A...544A..39J},{2015ApJ...798...30L}}
It has been postulated that such infrared bubbles are excited by massive star(s) / cluster of stars where the expanding \hii~region excites the ambient medium giving rise to a ring of PAH emission \citep{2006ApJ...649..759C}. An alternate theory proposes the interaction of strong stellar winds with the surrounding interstellar medium to produce a spherical ring of shocked gas \citep{1977ApJ...218..377W}. 

The radio emission within the region denoted by the partial dashed ellipse in Fig. ~\ref{bubble} corresponds to  $N_{Lyc} = 3\times10^{48}$~s$^{-1}$ for 1280~MHz. Assuming this to be optically thin, this is equivalent to a ZAMS star of spectral type O8 - O7.5 \citep{1973AJ.....78..929P}.  However, this is an underestimate as some UV photons are absorbed by dust (e.g. dust emitting at 24~$\mu$m).  The identification of the exciting star(s) is difficult considering the population overdensity in near-infrared. Nevertheless, we have attempted to search for the bubble progenitor(s) towards the centre based on (i) its hemi-spherical shape and (ii) direction of 
pillar-like structures protruding within the bubble towards the central region both in ionised gas and mid-infrared dust emission, shown by arrows in Fig.~\ref{bubble} \citep{2009A&A...494..987P}.  The latter structures are believed to be shaped by the UV radiation field of the exciting star(s).  We visually scrutinised the central region of the bubble in the IRAC~3.6~$\mu$m image and located a favourable candidate: a faint group of stars. This cluster is encircled in Fig.~\ref{bubble}. The emission from the cluster at 4.5~$\mu$m is evident in the intensity profile through AB in Fig.~\ref{bubprof} (left).
The GLIMPSE Archive contains three sources and these are marked in Fig.~\ref{bubble}. 
However, due to lack of flux density information in all the four bands, it is difficult to classify them through color-color diagrams. Towards this region, we observe flux enhancements corresponding to nearly six sources within a small region (circular region of radius $5\arcsec$) unlike its surroundings. Unlike the IRAC images, 2MASS detects only a single source at the location of the cluster. This is likely to be due to the lower resolution of 2MASS ($4''$) as compared to IRAC ($2''$). We speculate that these could be the exciting sources and further studies (high resolution imaging and spectroscopy) are required to confirm this view. 

\begin{figure*}[hbt!]
\begin{center}
\includegraphics[scale=0.35]{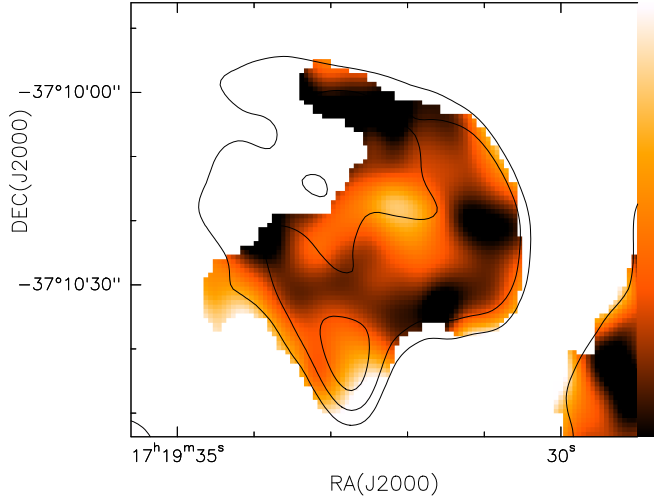}
\caption{Enlarged view of spectral index map towards the bubble CS-112. The black contours represent 610 MHz emission. }
\label{bubspecin}
\end{center}
\end{figure*}

We see a void in the cold dust emission towards the bubble, particularly near the cluster region. This could be explained if we consider dispersal of the ambient molecular cloud by stellar winds from the exciting sources. We note that although there is a clump towards the west of the bubble (C2, see Figs.~\ref{cl_aper} and \ref{bubble}), it is larger and more massive than the one towards the east (C7). This probably explains the shell feature seen distinctly to the west in the broken bubble morphology.  
We also observe that the radio emission towards the periphery of the bubble  shows contribution of non-thermal component. This is evident from the spectral index map of CS-112, shown in Fig.~\ref{bubspecin}. The non-thermal radiation can be attributed to the presence of relativistic electrons produced in shocked regions of gas along the bubble periphery and accelerated by the magnetic field \citep{2013A&A...559A..31B}.

We can estimate the dynamical age, $t_{dyn}$, of the ionised gas emission associated with the bubble based on a simple model of expanding photoionised nebula, in a homogeneous medium \citep{1980pim..book.....D}. If the cluster is indeed the exciting source, it is possible to estimate the age of the bubble based on the time elapsed for the ionisation front to move from the location of the cluster to the present site through the ambient medium using the following formulation. 

\begin{equation}
t_{dyn}=\left(\left[\frac{4R_s}{7c_i}\right]\right)\left[\left(\frac{R}{R_s}\right)^{7/4}-1\right]
\end{equation}

\noindent Here $c_i$ is the isothermal sound velocity in the ionized gas and $R$ is the radius of the bubble.  $R_s$ is the radius of the Str{\"o}mgren sphere, given by 
\begin{equation}
R_s =\left(\frac{3N_{Lyc}}{4 \pi n_o^2 \alpha_B}\right)^{1/3} 
\end{equation}
$\alpha_B$ is the radiative recombination coefficient. We assume $\alpha_B=2.6 \times$ 10$^{-13}$ cm$^3$~s$^{-1}$ and $c_i=10$~km~s$^{-1}$ considered for typical \hii~regions \citep{2005fost.book.....S}. $n_o$ denotes the mean number density of atomic hydrogen and we estimate its value using the expression $3N(H_2)$/2L where L is the molecular core radius. The three clumps surrounding CS-112 are C2 (west), C5 (south) and C7 (east). For estimating the dynamical time, we consider the mean number density of clump C2 as it is in the direction of the shell. Using the clump parameters, we get  $n_o = 4.3\times10^4$~cm$^{-3}$. 

Taking 1.0~pc as the bubble radius, we estimate the Str\"{o}mgren radius to be $R_S=0.04$~pc. This corresponds to an expansion timescale of 0.7~Myr for the bubble CS-112.
 
A simple model such as this has also been used to estimate the age of other bubbles such as N22 \citep{2012A&A...544A..39J} and N115 \citep{2014A&A...569A..36X}. 

In summary, the presence of compact and ultracompact \hii~regions indicates the youth of this region that is consistent with SED modelling of massive YSOs as well as with the approximate age of the bubble. This star forming flurry is reflected in numerous studies of masers carried out for this region. 

\subsection{Masers detected in the region}
This region is rich in masers and has been extensively investigated in various species by a number of researchers. In high mass star forming regions, maser species are believed to trace shocks, infall motions, outflows and circumstellar disks \citep{{2005ApJS..156..179D},{2007IAUS..242...71F}}. Of the different masing species, water (H$_2$O) masers are frequently observed towards outflows from high and low mass YSOs \citep{{2005PASJ...57..595H},{2006A&A...446..985M}}. Hydroxyl (OH) masers are mostly associated with UC\hii~regions \citep{2005ApJS..160..220F} although they are also detected near less evolved massive protostellar objects \citep{2007A&A...465..865E}. Class II 6.7-GHz methanol (CH$_3$OH) maser emission is closely correlated with high-mass star forming regions \citep{1997MNRAS.291..261W}. 

\begin{figure}[hbt!]
\begin{center}
\includegraphics[width=75mm]{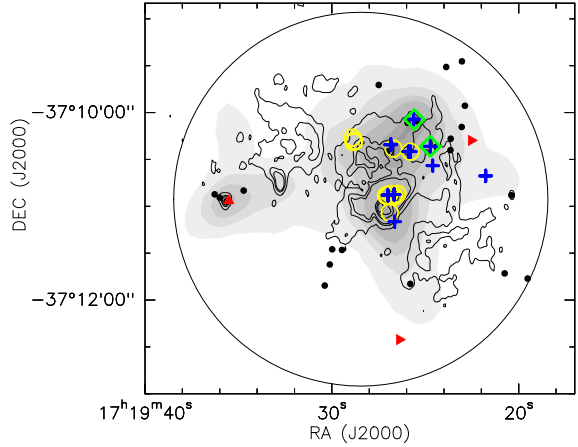}
\caption{The 870~$\mu$m emission map overlaid with 1280~MHz radio contours. The contour levels are at 1.3, 6.3, 11.3, 21.3, 31.3, 41.3 and 51.3~mJy/beam. The positions of water (plus symbols), methanol (open circles), and hydroxyl (diamonds) masers are marked. Also plotted are the location of  YSOs  as filled circles. The red triangles represent YSOs from Fig.~\ref{cc} (b). The large circle encompassing ionised and dust emission represents the region within which sources were extracted for YSO classification.}
\label{maser}
\end{center}
\end{figure}


Towards the IRAS~17160--3707 region, previous studies have detected H$_2$O \citep{{1989A&A...213..339F},{1996ApJ...463..205A},{1999A&AS..137...43F},{2011MNRAS.416.1764W},{2014A&A...563A..68C},{2014MNRAS.442.2240W}}, CH$_3$OH \citep{{1997MNRAS.291..261W},{2003ChJAA...3...49X},{2005A&A...432..737P},{2009PASA...26..454C},{2010MNRAS.404.1029C},{2013MNRAS.431.1752U},{2014A&A...563A..68C},{2014A&A...569A.125H}} as well as OH
 \citep{{1989A&A...213..339F},{2001MNRAS.326..805C}} masers. We could not find studies investigating Class I methanol masers towards this region in the literature. The locations of the masing spots with respect to radio and submillimeter emission are shown in Fig.~\ref{maser}. While in few cases, interferometric observations have been carried out localising the masing region, in most
 cases the positions indicate the direction in which single dish pointing observations were carried out. Hence, there is uncertainty in the locations of these masers.
 In the present case, we associate masers with the clumps in order to fathom the evolutionary stage of the clump (see next section). We consider a maser as associated with a clump if its location is within $10''$ of the clump peak position. This search radius is based on the resolution of the dust emission map used for clump identification. Maser 
emission is observed towards clumps C1, C2, C3 and C6 (see Table~\ref{tab_maser}). Several of these maser spots appear to be associated with the radio peaks such as S5, S11 and S12. Towards S5, we observe water and methanol masers, while water and OH masers are detected towards S11. These locations are also associated with Class I/II YSOs (see Section 3.3.1). Hence, it is likely that the masers in this region trace the outflow and disk activity of young-high mass stars. Water and methanol masers are also detected towards complex A. These masers emphasize the early stages of massive star forming activity towards this region.
  
\begin{table}[hbt!]
\footnotesize
\caption{Location of masers in this region based on other studies. See references in text. }
\begin{center}
\hspace*{-0.5cm}
\begin{tabular}{c c c c } \hline \hline
Clump &Water (H$_2$O)&Class II methanol (CH$_3$OH)&OH \\
No.&maser&maser&maser\\
\hline
C1 &\checkmark &\text{\sffamily X}&\checkmark \\
C2 &\checkmark &\checkmark &\text{\sffamily X} \\
C3 &\checkmark &\checkmark &\text{\sffamily X} \\
C4 &... &\text{\sffamily X}&\text{\sffamily X}\\
C5 &\text{\sffamily X} &\text{\sffamily X}&\text{\sffamily X}  \\
C6 &\checkmark &\text{\sffamily X} &\text{\sffamily X} \\
C7 &... &\text{\sffamily X} &\text{\sffamily X} \\
\hline
\end{tabular}
\label{tab_maser}
\\
\checkmark - Detection; \text{\sffamily X} - Non detection; ... - No data available \\
\end{center}
\end{table}


\subsection{Clump Evolutionary Stages}
Seven clumps are detected in the IRAS 17160--3707 region and we would like to gather leads to the evolutionary stages of these clumps. Various studies have surveyed numerous clumps in star forming regions in order to statistically analyze and sequence them according to their evolutionary stage, based on 
multiwavelength tracers such as the presence of  ionised gas, masers, shocks, outflows, and nebulosity in the near and mid-infrared images \citep{{2009ApJS..181..360C},{2010ApJ...721..222B},{2013A&A...550A..21S},{2013A&A...556A..16G}}. However, although differing in nomenclature, most of these are consistent with each other in terms of the broad evolutionary classification and tracers used. In this paper, we examine the relative evolutionary stages of clumps detected in this region based on their physical properties. For this, we used the broad evolutionary sequence proposed by \citet{2013A&A...556A..16G}. They have classified clumps into two categories, according to the presence and absence of signposts of active star formation. The presence of water masers, green fuzzies, 24~$\mu$m point sources and radio continuum emission (within $10''$ of the clump peak position) were considered and if any of these signposts were observed, then the clump is classified as belonging to the star forming stage (SFS) and in quiescent stage (QS) otherwise. The SFS clumps were further categorized into SFS-1: those typically having embedded high / low mass protostellar objects, and SFS-2: a later stage where there are hypercompact or ultracompact \hii~regions associated with the clumps. The presence of an \hii~region in SFS-2 is indicated by the radio continuum and/or strong mid-infrared emission whereas QS and SFS-1 do not show these properties. 

\par We have extensively scrutinised this region for green fuzzies and could find only one object near clump C1 (discussed earlier in Section 3.2.4). We have searched for 24~$\mu$m point sources associated with the clumps within a search radius of 10$\arcsec$ from the peak position in the 24~$\mu$m catalog and based on the association with water masers, mid-infrared and radio emission, we have categorized the clumps. The final results of the clump classification are presented in Table~\ref{clump_ev}. The columns in the table list the clump name, association with radio source, 8~$\mu$m peaks, 24~$\mu$m point source and presence of H$_2$O masers. Column 6 gives the clump evolutionary stage. 
Five clumps: C1, C2, C3, C4, and C5 are classified as SFS-2 as they harbor radio emitting regions. The other two clumps: C6
and C7 are classified as SFS-1 and QS, respectively. In the case of C6, we would like to comment that although we discern radio (S12) and mid-infrared emission within the clump aperture, this is located nearly $\sim 20''$ (0.6~pc) from the clump peak position, whereas the water maser location is close to the peak ($1''$). The water maser position is fairly accurate from ATCA interferometric observations \citep{2014MNRAS.442.2240W} with a beam~$\sim2''$ and therefore, we grade this clump as SFS-1. There is a possibility that the maser could in reality be associated with S12 in which case, this clump would be categorised as SFS-2. For the present, we proceed with the former premise of classifying C6 as SFS-1. 

\par We next ascertain the massive star forming potential of the clumps C6 and C7. For this, we use the formulation of  \citet{2015arXiv151108810S} to find the mass of the most massive star $\rm{M}_{max}$ likely to be formed in the clump, using a stellar Initial Mass Function (IMF)
of \cite{2001MNRAS.322..231K}. 

$$ \rm{M}_{max} = 20\rm{M}_\odot\left( \frac{\rm{E}_{sf}\,\, \rm{M}_{clump}}{0.3\times1064\rm{M}_\odot} \right)^{1/1.3}$$

Here, $\rm{E}_{sf}$ is the star forming efficiency in the clump that we take as 30\% \citep{2003ARA&A..41...57L}. Taking $\rm{M}_{clump}=300$~M$_\odot$ for C7, we obtain $ \rm{M}_{max} \sim7.6$~M$_\odot$ indicating that C7 is capable of producing intermediate mass stars. A similar exercise with C6 yields  $ \rm{M}_{max} \sim14.5$~M$_\odot$.

\begin{table}[hbt!]
\footnotesize
\caption{The classification of clumps based on multi-wavelength signposts of star formation.}
\begin{center}
\hspace*{-0.5cm}
\begin{tabular}{c c c c c c} \hline \hline
Clump &Radio &IRAC &MIPS &H$_{2}$O& Evolutionary  \\
No.&source& 8~$\mu$m peak&24~$\mu$m source&maser$^a$&Stage\\
\hline
C1 &S10 &\text{\sffamily X}&\checkmark  &\checkmark &SFS-2\\
C2 &S4, S9 &\checkmark &\checkmark &\checkmark &SFS-2\\
C3 &Complex A &\checkmark&\text{\sffamily X}  &\checkmark &SFS-2\\
C4 &S6 &\checkmark &\checkmark &... &SFS-2\\
C5 &S5 &\checkmark &\text{\sffamily X} &\text{\sffamily X} &SFS-2\\
C6 &\text{\sffamily X} &\text{\sffamily X} &\text{\sffamily X} & \checkmark &SFS-1\\
C7 &\text{\sffamily X} &\text{\sffamily X} &\text{\sffamily X} &... &QS\\
\hline 
\end{tabular}
\label{clump_ev}
\\
\checkmark - Detection; \text{\sffamily X} - Non detection; ... - No data available\\
\end{center}
\end{table}


\par
\begin{figure}[hbt!]
\begin{center}
\includegraphics[width=95mm]{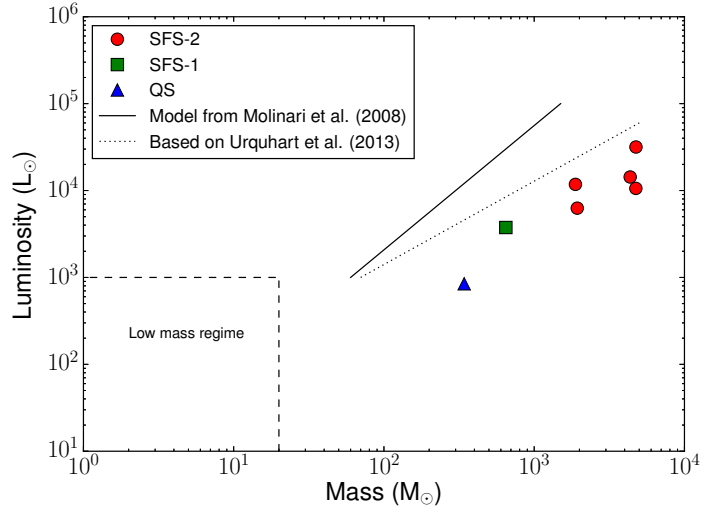}
 \caption{Plot of bolometric luminosity versus mass of the clumps. The locus that distinguishes the final-envelope clearing phase from the accelerating accretion phase according to the models of \citet{2008A&A...481..345M} is shown as a solid line and while the best-fit ZAMS locus to an observational sample by \citet{2013MNRAS.431.1752U}  is represented as a dotted line are shown. The identification of low mass regime is based on \citet{1996A&A...309..827S}. }
 \label{LM}
 \end{center}
\end{figure}

   Another diagnostic tool to visualize the clump evolutionary stages is the mass-luminosity (M - L) diagram. Initially proposed for the evolutionary stages of star formation in the low mass regime by \citet{1996A&A...309..827S}, \citet{2008A&A...481..345M} modified it for high-mass stars as well. Such a plot expresses the relationship between the mass of the YSO envelope and its bolometric luminosity, in which the objects in different evolutionary stages occupy distinct regions in the diagram. The evolution of a massive clump begins in the low luminosity region (lower-right) of the M - L plot. As it evolves, the source proceeds almost vertically, towards regions of higher luminosity. In this stage, the source is said to be in the accelerating accretion phase. 
The solid line separating the accelerating accretion phase from the later envelope clearing phase is shown in Fig.~\ref{LM} \citep{2008A&A...481..345M}, which is based on the turbulent core model of \citet{2003ApJ...585..850M}. Once they reach this line, the (proto)stellar objects move towards the top-left, i.e towards the low mass regime, keeping a nearly constant luminosity, i.e. the final envelope clearing phase. The solid line doesn't represent a ZAMS locus in the true sense. Rather, it represents the location where the accretion driving the protostar assembly will stop. At this time, the more massive objects are already on the ZAMS and corresponds to the stages identified as hot-cores, hypercompact \hii~or ultracompact \hii~regions.
The dotted line in Fig.~\ref{LM} shows the approximate segregation between protostellar evolution and \hii~region evolution and is based on a linear fit to a sample of sources classified as massive YSOs and ultracompact \hii~regions \citep{2013MNRAS.431.1752U}. 

\par The M - L plot for the seven clumps in this region is shown in Fig.~\ref{LM}. As the mass of the clump is derived from dust emission, this is assumed to be the envelope mass. From the plot, it is evident that all the clumps are in the high mass regime and lie in the accelerating accretion phase domain. The
SFS-2 clumps have M $ > 1000$~M$_\odot$ and are grouped towards the right of the M - L plot. The
SFS-1 clump has a larger mass and luminosity compared to the QS clump. We cannot overlook the caveats regarding our assumptions about clump mass and luminosity. The clump luminosity is the based on the integration of fluxes under the best-fit greybody function and does not take into consideration the warm dust component. This leads to an under-estimation of the true luminosity. 
Similarly, the mass estimate does not include the warm dust component. In addition, the clump mass assumes that a single ZAMS star is embedded which need not be the case always, as massive stars are born in rich clusters. These uncertainties are probably the reason why clumps classified as SFS-2 lie in the accelerating accretion phase but close to the envelope clearing phase.

\begin{figure}[hbt!]
\begin{center}
\includegraphics[width=95mm]{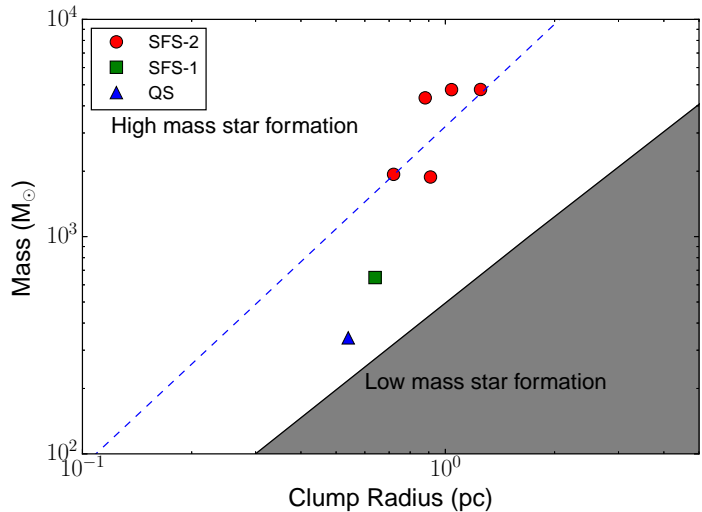}
 \caption{The mass - radius relationship of clumps. The dashed line represents the linear fit to the clumps in the high mass regime as given by \citet{2013MNRAS.431.1752U}. The grey region represents the low mass regime identified by \cite{2010ApJ...723L...7K}. }
\label{RM}
\end{center} 
\end{figure}

\par We also explored the radius - mass (R - M) plot for these clumps based on the models given in \citet{2013MNRAS.431.1752U}. This plot is useful to probe the size - mass relationship in molecular clouds \citep{{2010ApJ...712.1137K},{2010ApJ...716..433K}} and could be used to  differentiate high mass star forming clumps from the low mass ones.  The R versus M values of the clumps are shown in Fig.~\ref{RM}. We observe that all the clumps fall in the high mass region that substantiates the result from the M - L plot. This suggests that even though clump C7 is quiescent, it is capable of forming high mass stars. It is also interesting to note that the more evolved SFS-2 clumps lie in the top-right region of the R - M plot and this agrees well with the linear fit to the resolved clumps of \citet{2013MNRAS.431.1752U} in the high mass regime. 
The SFS-1 and QS clumps show a deviation from this line and are closer to the boundary of low mass star-forming clumps.

\par Thus we see that star formation is active in nearly all clumps. It is difficult to conclude about a collective evolutionary picture based on the evolutionary stage of the clumps although the bright central region (C3) is bustling with star forming activity. In a nutshell, we assert that a majority of the clumps are active and are participating in the massive star formation process.

\section{Conclusions}
\par We examined the star formation activity in the \hii~region complex associated with IRAS~17160--3707 using a range of wavelengths spanning from 3.6~$\mu$m to 870~$\mu$m. Based on our analysis, we arrive at the following conclusions:

\begin{itemize}

\item[(a)]  The radio continuum maps show the presence of twelve compact radio sources in the region with ten of them enveloped in diffuse emission. We have estimated the Lyman continuum fluxes of the individual sources. The 1280 - 610 MHz spectral index map reveals the presence of thermal emission towards compact sources and non-thermal emission in certain pockets of the envelope.

\item[(b)] We identify seven cold dust clumps from the 870~$\mu$m emission map and fitted modified blackbody to fluxes associated with these clumps giving a dust temperature in the range from 24 to 30~K and column density in the range $1.7 - 7.6 \times$~10$^{22}$~cm$^{-2}$. We have also constructed dust temperature, column density and dust emissivity index maps by carrying out modified blackbody fits to the individual pixels in this region.

\item[(c)] The mid-infrared warm dust emission traces the radio morphology and  we see emission associated with the compact sources in the mid-infrared bands. We identify 23 YSOs in the region based on mid-infrared color-color diagrams. Five of them are associated with compact radio sources and the radiative transfer modeling of these sources imply that they are relatively young with best fit age $\le 0.2$~Myr.

\item[(d)] This region harbors an infrared bubble CS-112 likely to be excited by a group of stars located near the geometric centre. We estimated the age of the bubble as 0.7~Myr based on a simple model of photoionized expansion of a \hii~region.

\item[(e)] The dust clumps fall in different classes of evolutionary stages. There are six star-forming clumps and a quiescent clump.  The collective picture indicates that this region is relatively young, currently undergoing a high mass star formation bustle.
\end{itemize}

\acknowledgements
\textit{Acknowledgments}

We thank the referee for a critical reading of
the manuscript and highly appreciate the suggestions,
that markedly contributed to improving the quality of
the paper. We thank N. L. Mart\'{i}n-Hern\'{a}ndez for providing the ATCA maps of this region. G. Nandakumar expresses his thanks to Ishwara Chandra, NCRA-TIFR for conducting an AIPS workshop for reduction of the radio data. 
We thank the staff of the GMRT, who have made the radio observations possible.
GMRT is run by the National Centre for Radio Astrophysics of the Tata Institute
 of Fundamental Research. This research made use of Montage funded by the National 
Science Foundation under Grant Number ACI-1440620, and was previously funded by the 
National Aeronautics and Space Administration's Earth Science Technology Office, 
Computation Technologies Project, under Cooperative Agreement Number NCC5-626 
between NASA and the California Institute of Technology.

\bibliography{AJ_13096_rev2}

\appendix

\section{SED fitting of YSOs}

\begin{figure}[hbt!]
\begin{center}
\hspace*{-1.6cm}
\includegraphics[width=120mm]{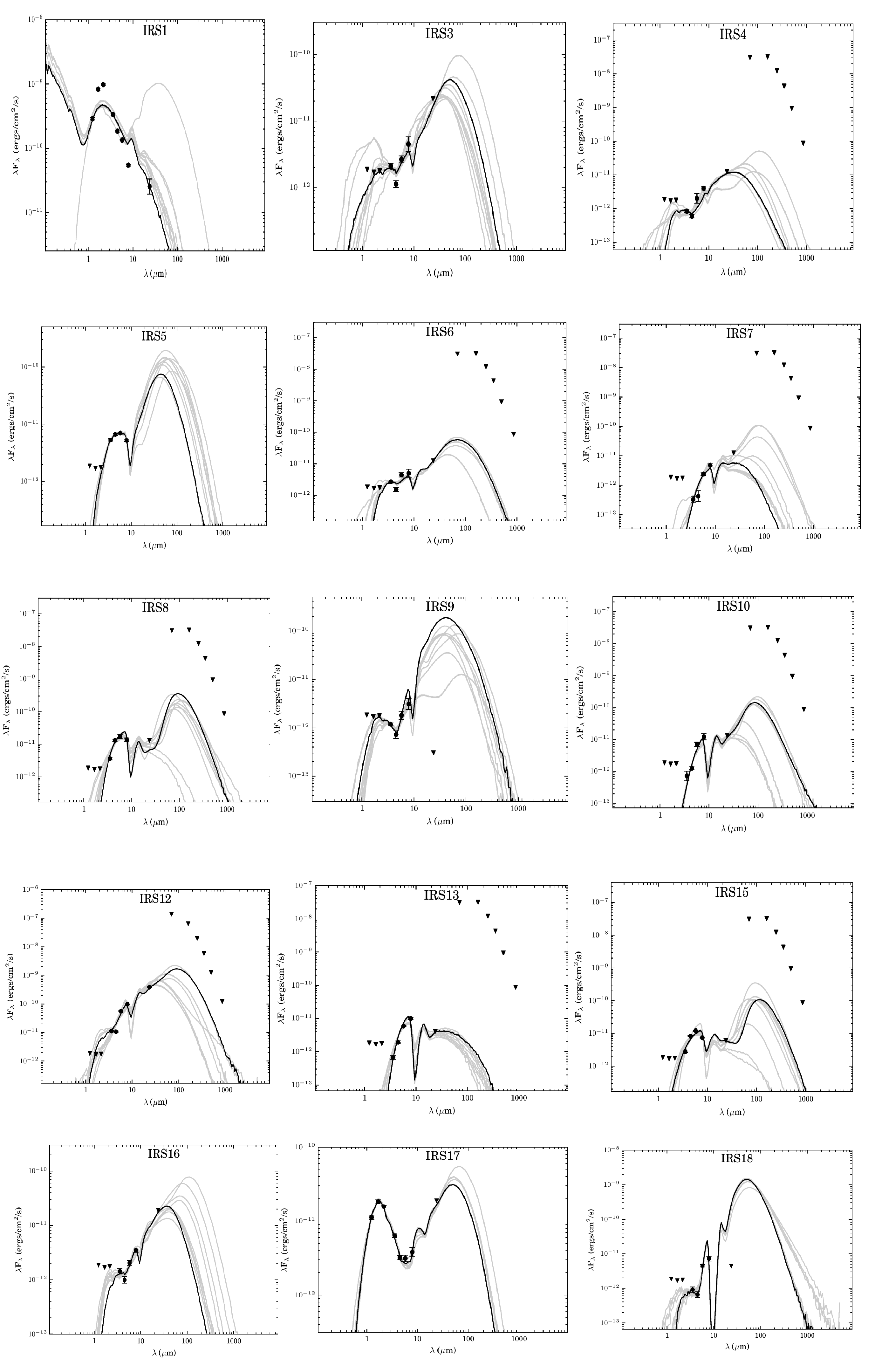} 
 \caption{The infrared SEDs of YSOs fitted with the models of Robitaille
et al. (2007). The filled circles represent the input fluxes while triangles represent upper
limits from the repective clumps. The black line shows the best fitting model, and grey lines show
the subsequent nine good fits.}
\label{IRSmodel1}
\end{center} 
\end{figure}

\begin{figure}[hbt!]
\begin{center}

\hspace*{-1.6cm}
\includegraphics[width=60mm]{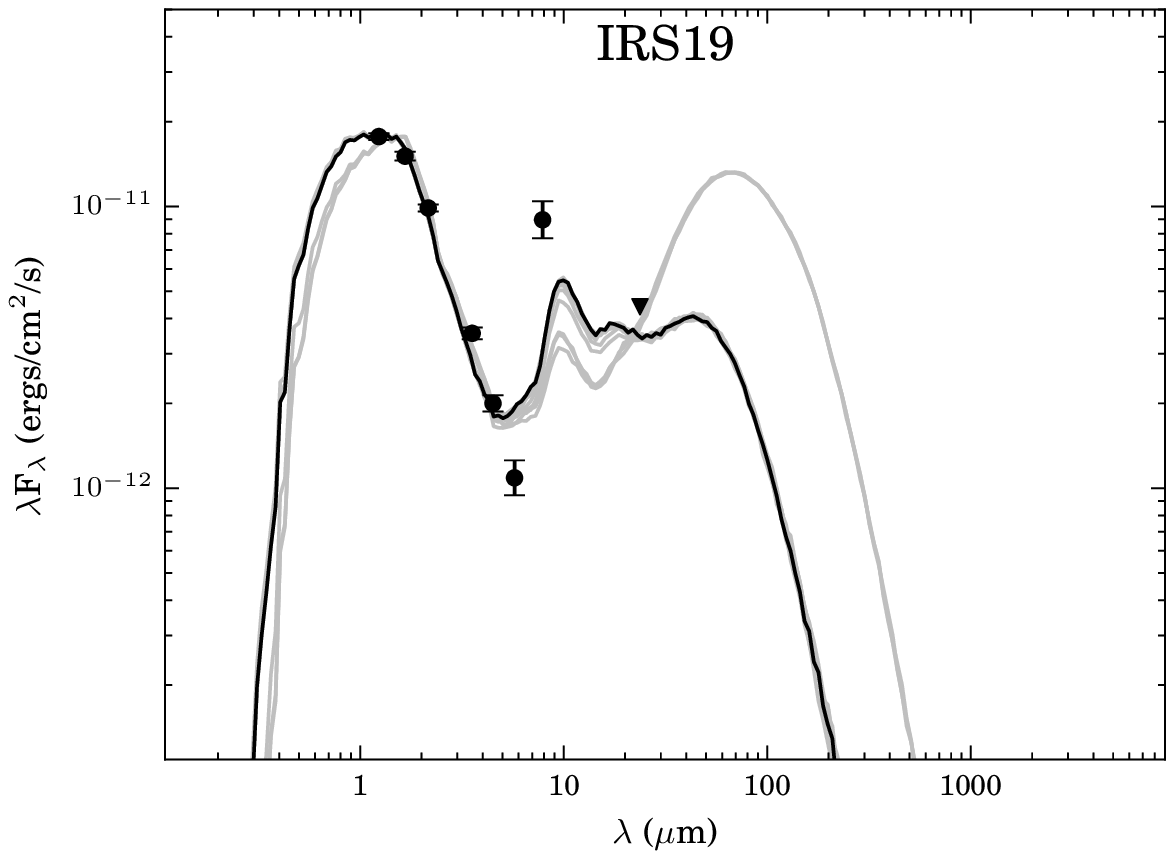} 
\hspace*{-0.2cm} \includegraphics[width=60mm]{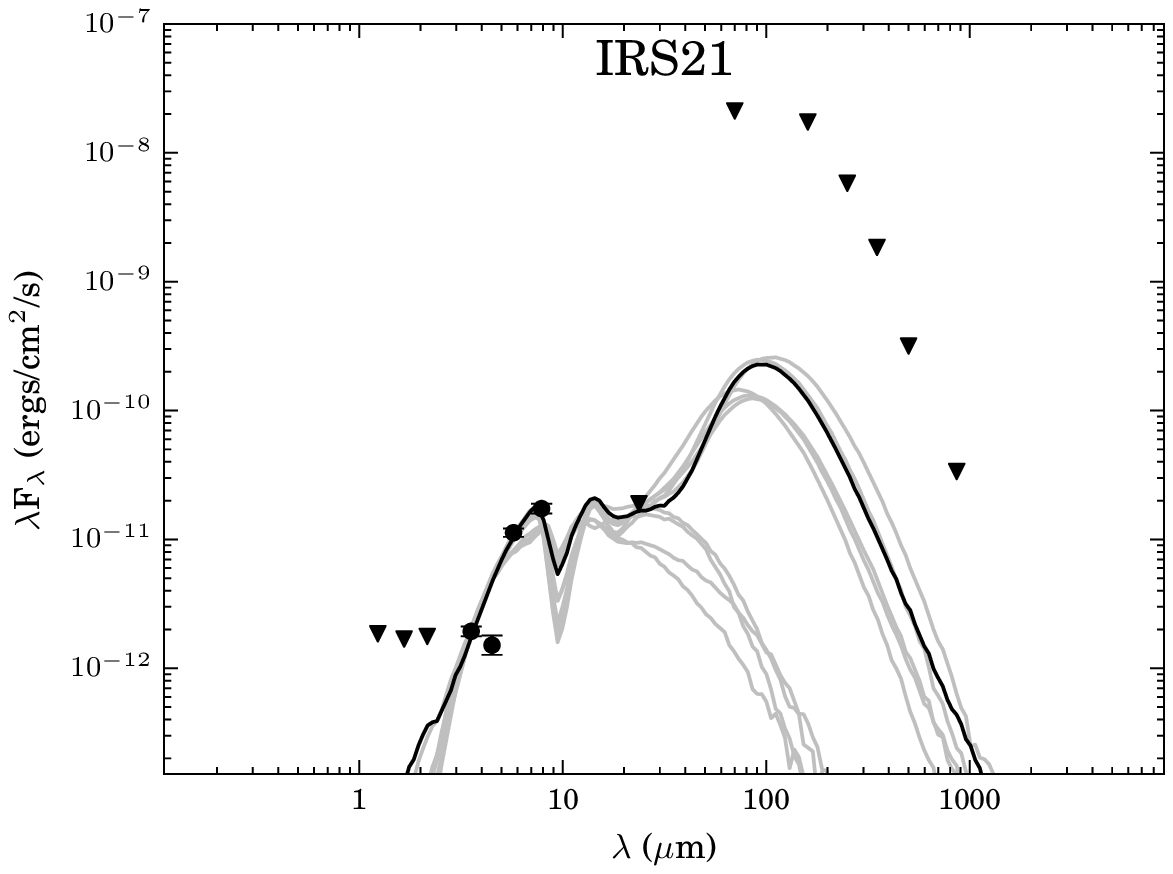} 
\hspace*{-0.2cm} \includegraphics[width=60mm]{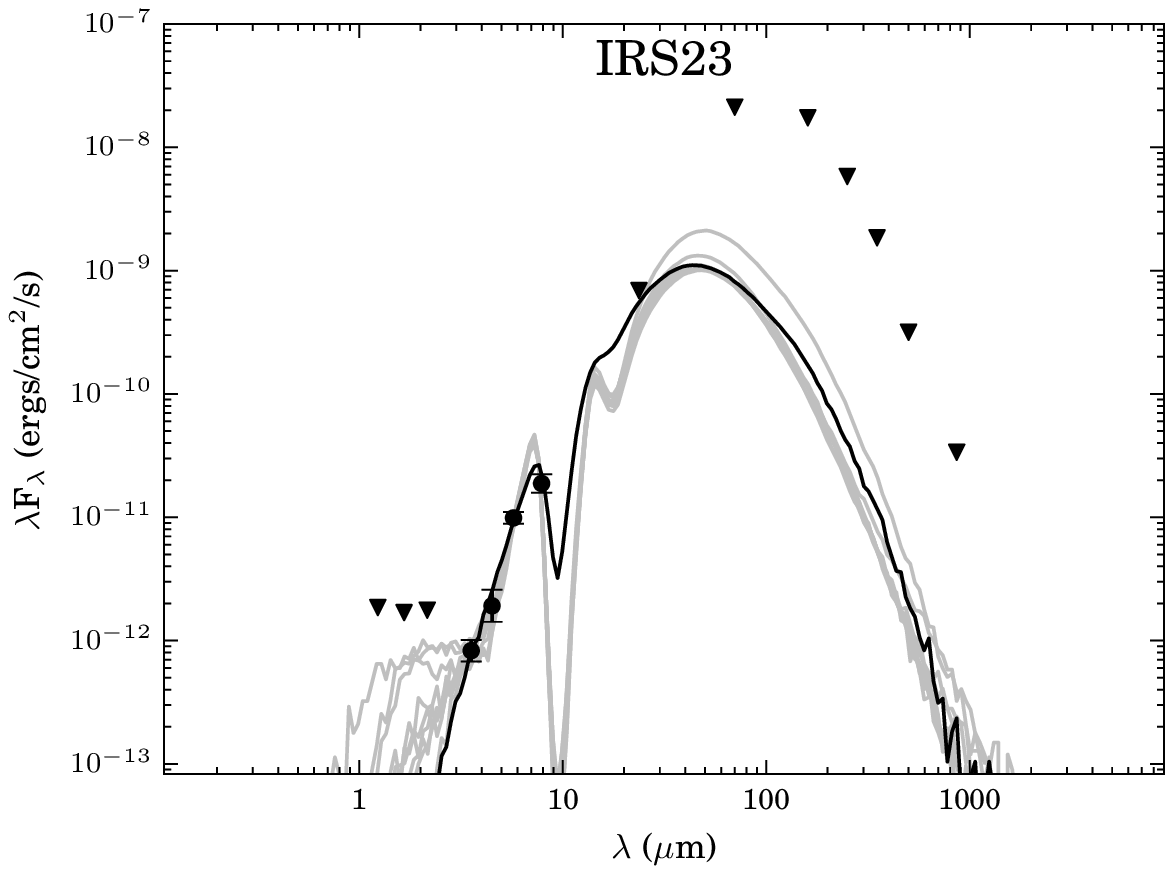}
\hspace*{-1.6cm}
 \includegraphics[width=60mm]{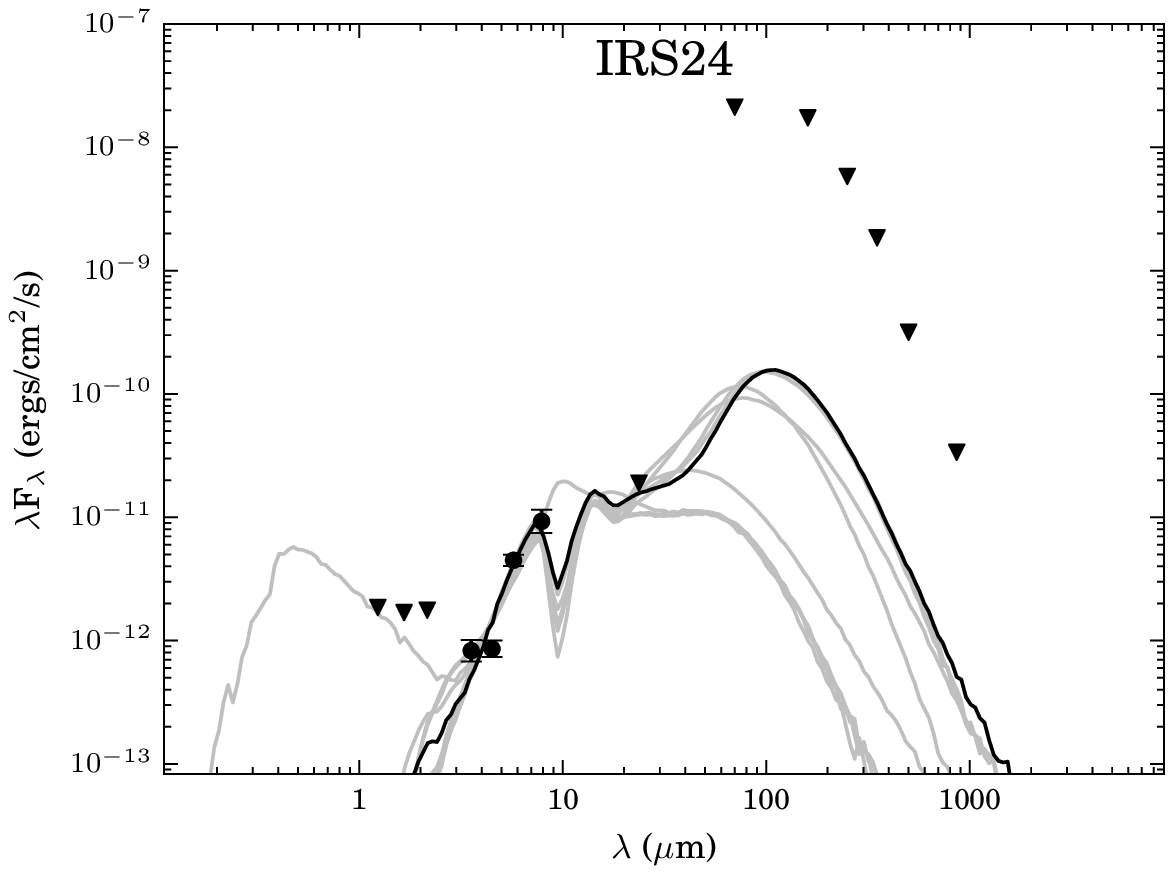} 

 \caption{Same as Fig.~\ref{IRSmodel1}.}
\label{IRSmodel2}
\end{center} 
\end{figure}


\scriptsize
\setlength{\LTleft}{-1cm}
\setlength{\tabcolsep}{2pt}
\begin{longtable}[h!]{c c c c c c c c c c c} 
\caption{Parameters of the models shown in Fig.~\ref{rwiw}. Col. 4 -- 9 give  Mass, Effective Temperature, Luminosity, Inclination angle, Envelope accretion rate, Disk Mass, Extinction and Age, respectively. The parameters listed are for the best fit model and the range of all the ten best fits.}\\
\hline
Source &&$\chi^2$ &Mass  &T$_{eff}$  &Luminosity  &Inc. angle & Env. accretion rate  &Disk mass  & A$_V$   & Age\\ 
& &&(M$_{\odot}$) &(K) &(L$_{\odot}$) & (Deg.) &(M$_{\odot}$/yr) &(M$_{\odot}$)  & (mag) &(Myr)\\
\hline \\
\endfirsthead

\caption{continued}\\
\hline
Source &&$\chi^2$ &Mass  &T$_{eff}$  &Luminosity  &Inc. angle & Env. accretion rate  &Disk mass  & A$_V$   & Age\\ 

& &&(M$_{\odot}$) &(K) &(L$_{\odot}$) & (Deg.) &(M$_{\odot}$/yr) &(M$_{\odot}$)  & (mag) &(Myr)\\

\endhead

\multirow{ 2}{*}{IRS1} & Best fit &2507 & 7.2& 20980 & 2541 & 69.5 &0  & 0.02 & 0 &2.8\\ 
&Range &2507-3060&7.2-10.7&4435-24020&2116-4236&18.2-69.5&$0$-$8.6\times10^{-5}$& $4.8\times10^{-3}$-$0.03$ & 0 & 0.005-3.8 \\ 
\hline  \\

\multirow{ 2}{*}{IRS3} & Best fit &19.9 & 3.9 & 4508 & 89.3 & 56.6 &4.1$\times$10$^{-5}$  &0.02  & 1.4 &0.1 \\ 
&Range &19.9-28.0& 1.2-5.9 &4017-4796 & 34.7-233.1 &18.2-75.5 &  $0.1$-$3.6\times10^{-4}$ &$2.9\times10^{-4}$-$0.04$ &0-5.0 &0.004-0.2\\ 
\hline  \\

\multirow{ 2}{*}{IRS4} & Best fit &11.9 & 1.1 & 4089 & 27.0 & 18.2 &6.9$\times$10$^{-6}$ & 0.02 & 20.2 &0.01 \\ 
&Range & 11.9-24.6 &0.8-1.9&3939-4321&18.2-64.1&18.2-49.5&$4.9\times10^{-6}$-$9.2\times10^{-4}$ &$3.7\times10^{-3}$-0.02&0-28.0&0.003-0.1\\
\hline  \\

\multirow{ 2}{*}{IRS5} & Best fit &0.3 & 5.7 & 4519 & 194.1 & 56.6 &6.1$\times$10$^{-5}$ & 0.05& 15.4 &0.06 \\ 
&Range &0.3-3.8 & 5.0-6.9 &4398-5948 &194.1-424.7 &31.8-63.3 &$0.1$-$3.1\times10^{-4}$&$2.5\times10^{-4}$-$0.08$ &4.0-32.0 &0.03-0.2\\
\hline  \\

\multirow{ 2}{*}{IRS6} & Best fit &22.1 & 5.9 & 4579 & 192.1 & 56.6 &7.2$\times$10$^{-5}$  & 2.3$\times$10$^{-3}$ & 31.4 &0.08 \\ 
&Range &22.1-26.4& 2.9-5.9 &4371-4731 & 76.9-192.1 &18.2-75.5 &  $0.4$-$5.4\times10^{-5}$ &$0.03$-$8.0\times10^{-2}$ &0.5-1.7 &0.19-0.25\\ 
\hline  \\

\multirow{ 2}{*}{IRS7} & Best fit &5.1 & 4.1 & 14420 & 220.9 & 56.6 &0  & 3.9$\times$10$^{-3}$ & 52.6 &3.8 \\ 
&Range &5.0-7.0& 0.6-4.3 &3820-14920 & 56.5-263.6 &18.2-81.4 & $0$-$6.2\times10^{-4}$ &$3.9\times10^{-3}$-$0.1$ &22.4-53.0 &0.01-4.0\\ 
\hline  \\

\multirow{ 2}{*}{IRS8} & Best fit &15.0 & 5.9 & 4250 & 447.1 & 18.2 &1.5$\times$10$^{-3}$  & 0.04 & 24.2 &0.005\\ 
&Range &15.0-65.1& 2.6-6.0 &4107-16210&148.4-447.1 &18.2-75.5 & $0$-$1.9\times10^{-3}$ &$2.0\times10^{-3}$-$0.4$&0-35.2 &0.001-3.4\\ 
\hline  \\

\multirow{ 2}{*}{IRS9} & Best fit &7.4 & 7.5 & 4537 & 474.3 & 87.1 & 4.9$\times$10$^{-5}$  & 8.5$\times$10$^{-3}$ & 5.7 &0.02 \\ 
&Range &7.4-9.6& 1.9-7.5 &4192-4537 & 36.3-474.3 &18.2-87.1 &$0.2$-$4.5\times10^{-4}$ & $2.7\times10^{-3}$-$0.2$ &3.5-19.2 &0.003-0.1\\ 
\hline  \\

\multirow{ 2}{*}{IRS11} & Best fit &21.9 & 1.9 & 4136 & 185.5 & 18.2 &5.2$\times$10$^{-4}$ & 0.1 & 57.5 &0.002 \\ 
&Range &21.9-29.8& 1.9-5.2 &4136-16910 & 170.6-551.0 &18.2-75.5 &  $0$-$1.6\times10^{-3}$ &$2.7\times10^{-4}$-$0.1$ &57.5-73.0 &0.001-5.2\\ 
\hline  \\

\multirow{ 2}{*}{IRS12} & Best fit &240.8 & 10.8 & 4527 & 2199 & 18.2 &2.9$\times$10$^{-3}$ & 0.03 & 18.2 &0.008 \\ 
&Range &240.8-336.5 & 6.8-10.8 &4164-7595 &901.6-2920 &18.2-87.1 & $2.5\times10^{-5}$-$2.9\times10^{-3}$ &0-0.1 &0-20.1 &0.002-0.07\\
\hline  \\

\multirow{ 2}{*}{IRS13} & Best fit &42.8 & 4.9 & 16360 & 488.0 & 75.5 &0  & 0.2 & 102 &1.9 \\ 
&Range &42.8-78.6& 3.7-4.9 &8466-16360 & 149.6-488.0 &56.6-75.5 & 0 &$2.0\times10^{-3}$-$0.2$ &44.1-102.0 &0.9-7.1\\ 
\hline  \\

\multirow{ 2}{*}{IRS15} & Best fit &115.2 & 2.9 & 4175 & 139.1 & 18.2 &1.3$\times$10$^{-3}$  & 8.4$\times$10$^{-3}$ & 11.7 &0.002 \\ 
&Range &115.2-182.5& 2.0-5.9 &4116-15060 & 128.0-447.1 &18.2-49.5 & $0$-$1.5\times10^{-3}$ &$4.2\times10^{-3}$-$0.2$ &11.0-38.0 &0.001-4.1\\ 
\hline  \\

\multirow{ 2}{*}{IRS16} & Best fit &6.1 & 3.7 & 4472 & 95.9 & 75.5 &1.0$\times$10$^{-5}$  & 0.03 & 17.2 &0.1\\ 
&Range &6.1-9.9& 1.0-3.7 &3985-4504&21.5-99.5 &18.2-75.5 &  $1.0\times10^{-5}$-$1.1\times10^{-3}$ & $1.6\times10^{-3}$-$0.1$&8.9-18.0 &0.003-0.2\\ 
\hline \\

\multirow{ 2}{*}{IRS17} & Best fit &8.5 & 5.0 & 4583 & 120.5 & 41.4 &1.4$\times$10$^{-5}$  & 2.4$\times$10$^{-3}$ & 6.0 &0.1 \\ 
&Range &8.5-9.9& 4.1-5.5 &4433-4720 & 119.3-148.7 &18.2-63.3 & $0.1$-$1.2\times10^{-4}$ &$3.1\times10^{-4}$-$0.02$ &2.5-6.8 &0.07-0.1\\ 
\hline \\

\multirow{ 2}{*}{IRS18} & Best fit &6.3 & 8.9 & 4244 & 1968 & 41.4 &3.3$\times$10$^{-4}$ & 0.02 & 1.9 &0.002 \\ 
&Range &6.3-14.5& 6.2-9.1 &4168-4296 & 1280-2167 &31.7-56.6 &  $2.2$-$3.3\times10^{-4}$ &0.02-0.6 &0-10.9 &0.002-0.004\\ 
\hline  \\

\multirow{ 2}{*}{IRS19} & Best fit &85.9 & 4.4 & 5032 & 61.8 & 63.3 &1.6 $\times$10$^{-6}$ & 0.09 & 1.5 &0.4 \\ 
&Range &85.9-115.6 & 2.2-4.4 &4341-5032 &41.4-61.8 &18.2-75.5 & $0.2$-$3.5\times10^{-5}$ &$1.3\times10^{-3}$-$0.09$ &0-1.5 &0.09-0.4\\
\hline  \\

\multirow{ 2}{*}{IRS21} & Best fit &48.1 & 4.5 & 4272 & 276.4 & 18.2 &1.6$\times$10$^{-3}$  & 0.05 & 24.8 &0.009 \\ 
&Range &48.1-67.7& 2.8-5.4 &4114-17280 & 159.7-613.7 &18.2-75.5 &  $0$-$1.7\times10^{-3}$ &$4.6\times10^{-4}$-$0.06$ &24.8-48.0 &0.001-7.9\\ 
\hline  \\

\multirow{ 2}{*}{IRS23} & Best fit &1.5 & 8.3 & 11680 & 3221 & 87.1 &8.9$\times$10$^{-5}$  & 7.9$\times$10$^{-4}$ & 37.0 &0.1 \\ 
&Range &1.5-3.0& 7.8-11.8 &4109-11680 & 1575-3683 &41.4-87.1 & $0.9$-$3.1\times10^{-4}$ &0-0.8 &0-37.0 &0.001-0.1\\ 
\hline  \\

\multirow{ 2}{*}{IRS24} & Best fit &19.9 & 3.9 & 4380 & 192.5 & 18.2 &1.6$\times$10$^{-3}$  & 0.02 & 27.2 &0.02\\ 
&Range &19.9-23.9& 1.2-5.1 &4019-6540&62.9-215.8 &18.2-81.4 &  $4.2\times10^{-7}$-$1.6\times10^{-3}$ & $7.7\times10^{-3}$-$0.2$&0.3-46.7 &0.005-0.5\\
\hline  \\

\label{rwiw_tb}
\end{longtable}


\end{document}